\pdfoutput=1
\documentclass[prd,amsmath,amsfonts,floatfix,nofootinbib,preprintnumbers,letterpaper]{revtex4}

\usepackage{graphicx}
\usepackage{slashed}
\def\la{\langle}
\def\ra{\rangle}

\def\beq{\begin{equation}}
\def\eeq{\end{equation}}
\def\bea{\begin{eqnarray}}
\def\eea{\end{eqnarray}}
\def\barr{\begin{array}}
\def\earr{\end{array}}

\def\op{{\mathcal{O}}}
\def\ampl{{\mathcal{M}}}

\def\Im{\mathop{\mbox{Im}}}

\def\exponential{{\mathrm{e}}}

\def\calD{{\mathcal{D}}}

\def\calF{{\mathcal{F}}}

\def\calH{{\mathcal{H}}}
\def\calI{{\mathcal{I}}}
\def\calJ{{\mathcal{J}}}
\def\calK{{\mathcal{K}}}
\def\calL{{\mathcal{L}}}

\def\calN{{\mathcal{N}}}

\def\calQ{{\mathcal{Q}}}

\def\calS{{\mathcal{S}}}
\def\calT{{\mathcal{T}}}
\def\calU{{\mathcal{U}}}

\def\calW{{\mathcal{W}}}

\begin{document}
\preprint{\vbox{\hbox{JLAB-THY-11-1412}}}
\begin{titlepage}
\begin{flushright}
\end{flushright}
\vskip 0.5cm
\begin{center}
{\Large \bf Axial couplings in heavy hadron chiral perturbation theory\\at the next-to-leading order\\} 
\vskip1cm {\large\bf
William~Detmold$^{a,b}$, C.-J. David~Lin$^{c,d}$, Stefan~Meinel$^{a}$}\\ \vspace{.5cm}
{\normalsize {\sl $^a$ Department of Physics, College of William and Mary, Williamsburg, VA 23187-8795, USA\\
$^{b}$ Jefferson Laboratory, 12000 Jefferson Avenue, Newport News, VA 23606, USA\\
$^{c}$ Institute of Physics, National Chiao-Tung University, Ta-Hsueh Road, Hsinchu 300, Taiwan\\
$^{d}$ Division of Physics, National Centre for Theoretical Sciences, Kuang-Fu Road, Hsinchu 300, Taiwan}}

\vskip1.0cm {\large\bf Abstract:\\[10pt]} \parbox[t]{\textwidth}{{
We present calculations of axial-current matrix elements between various heavy-meson and heavy-baryon states to the next-to-leading order
in heavy hadron chiral perturbation theory in the $p$-regime.  When compared with data from lattice computations or experiments, these results can be
used to determine the axial couplings in the chiral Lagrangian.  Our calculation is performed in partially-quenched chiral perturbation
theory for both SU(4$|$2) and SU(6$|$3).  We incorporate finite-size effects arising from a single
Goldstone meson wrapping around the spatial volume.
Results for full QCD with two and three flavours can be obtained straightforwardly 
by taking the sea-quark masses to be equal to the valence-quark masses.  To illustrate the impact of our chiral perturbation
theory calculation on lattice computations, we analyse the SU(2) full QCD results in detail.
We also study one-loop contributions relevant to
the heavy hadron strong-decay amplitudes involving final-state Goldstone bosons, and demonstrate that the quark-mass dependence
of these amplitudes can be significantly different from that of the axial current matrix elements containing only single hadron external
states.
}}
\end{center}
\end{titlepage}

\section{Introduction}
\label{sec:intro}
The physics of $b$ hadrons is an important and active field of
research, both experimentally and theoretically.
$B$ mesons have played an important role in our understanding of flavour physics in the Standard Model (SM) and 
its possible extension.  The on-going LHCb experiment and possible future $B$ factories will produce
significantly improved experimental information for $B$ mesons
which will, in turn, lead to better constraints on the relevant SM parameters or reveal deviations from the SM.
In addition, a large amount of polarised single-bottom
baryon data will be produced.  This will allow extensive studies
of the spectrum and the decays of these baryons.  Since the baryons carry different spin quantum numbers, they 
may offer additional opportunities for probing the coupling structure of physics beyond the SM.  In performing such
investigations, it is necessary to compare experimental results to precise theoretical calculations in which 
non-perturbative strong-interaction effects are well controlled.  This is becoming achievable because of the progress in Lattice QCD.

Calculations in Lattice QCD are often performed at unphysical light-quark masses due to the limited computing
resources.  In order to obtain high-precision theoretical predictions for spectral quantities and matrix elements,
it is essential to use chiral perturbation theory ($\chi$PT) to extrapolate to the physical quark masses.  For 
systems of hadrons containing a single valence $b$ or $\overline{b}$ quark, the relevant chiral effective field theory is 
heavy-hadron $\chi$PT (HH$\chi$PT)~\cite{Burdman:1992gh,
  Wise:1992hn,Yan:1992gz,Cho:1992gg,Cho:1992cf}. 
 In addition to the low-energy constants in the chiral Lagrangian of the 
Goldstone boson sector, there are three unknown coupling constants
in this effective theory at the leading order (LO).  These constants, defined explicitly as $g_{1,2,3}$ in
Eq.~(\ref{eq:HHChPT_lagrangian_LO}) in Sec.~\ref{sec:HHChPT},
accompany axial couplings of heavy hadrons to the Goldstone boson sector
and appear in all chiral extrapolations using HH$\chi$PT.  Therefore, the accurate determination of  $g_{1,2,3}$ is one 
of the most important tasks in the Lattice QCD calculations for $b$-physics phenomenology.

In this work, we compute the matrix elements of the quark-level axial currents,
\beq
\label{eq:quark_level_currents}
 \calJ_{ud,\mu} = \bar{d}\gamma_{\mu}\gamma_{5} u , \,\, {\mathrm{and}} \,\,
 \calJ_{us,\mu} = \bar{s}\gamma_{\mu}\gamma_{5} u ,
\eeq
between various heavy-light meson and single${-}b$ baryon states to the next-to-leading order (NLO) in HH$\chi$PT.
In particular, we calculate the relevant one-loop contributions to these 
matrix elements.  When compared with data from lattice calculations or experiments, our results
can be used to extract the above-mentioned three axial couplings in HH$\chi$PT.
Our calculation is performed in partially-quenched chiral
perturbation theory (PQ$\chi$PT) using the supersymmetric formulation~\cite{Bernard:1994sv}, for both SU(4$|$2) and 
SU(6$|$3). The ``full-QCD" limit can be taken straightforwardly from our results by setting the sea-quark masses
to be equal to the valence-quark masses.  Our one-loop computation is carried out for finite spatial volume in the 
$p$-regime~\footnote{Studies of the heavy-meson systems in the $\epsilon$-regime can be found in Ref.~\cite{Bernardoni:2009sx,Briceno:2011rz}.}, 
following the same method as in Refs.~\cite{Arndt:2004bg,Detmold:2006gh}.  As pointed out in Ref.~\cite{Colangelo:2010ba}, 
in heavy-light meson systems, finite-volume effects arising from higher-order terms in the chiral expansion can be estimated.  
This requires high-precision information on the $B^{\ast}{-}B{-}\pi$ coupling beyond that which is currently available.
Nevertheless, such higher-order effects are insignificant for current and future lattice calculations,
since computations with small pion masses in large volumes are becoming standard.

In this paper, we present our results in the isospin limit.  However, in the 
case of SU(6$|$3), we include the SU(3) breaking\footnote{More
  precisely, we consider identical SU(3) breaking effects in the sea,
  valence and ghost sectors of SU(6$|$3) but for simplicity, refer to this as SU(3)
breaking.} effects, both in the external states and in the axial currents.  
At NLO in HH$\chi$PT, the axial-current matrix elements for heavy hadrons can be written in the general form
\beq
\label{eq:chiral_extrap_generic}
 g \left ( 1 + g^{2} L + g^{\prime 2} L^{\prime} + L^{\prime\prime} \right ) + {\mathrm{analytic}}\mbox{ }{\mathrm{terms}},
\eeq
where $g$ and $g^{\prime}$ are variously $g_{1}$, $g_{2}$ and $g_{3}$ in 
Eq.~(\ref{eq:HHChPT_lagrangian_LO}), and $L$, $L^{\prime}$ and $L^{\prime\prime}$
are the contributions from one-loop diagrams.  The determination of $g_{1}$ using lattice QCD has been 
attempted by various groups~\cite{deDivitiis:1998kj,Abada:2003un,Ohki:2008py,Becirevic:2009yb,Bulava:2010ej,Bulava:2011yz}.  
However, the correct quark mass dependence (based on the symmetries of
QCD) of the axial matrix elements was not previously known.
Using the current work, extrapolations to the physical quark masses
can be made rigorously.

This paper is organised in the following way.  Section~\ref{sec:HHChPT} contains an introduction to HH$\chi$PT. 
In Sec.~\ref{sec:NLO}, we first present the general structure of the one-loop contributions to the axial-current matrix 
elements, before giving the results in the case of SU(2) in
Sec.~\ref{sec:one-loop-contr-su2}. Results for SU(4$|$2) and SU(6$|$3)
HH$\chi$PT are presented in Sec.~\ref{sec:one-loop-contr-su42-su63}, emphasising the quark flavour flow picture. 
In Sec.~\ref{strong-decays}, the strong-decay amplitudes involving
final-state Goldstone mesons are also computed before we conclude.
Technical details of the results are included in the appendices.

\section{Heavy hadron chiral perturbation theory}
\label{sec:HHChPT}
The partially quenched (PQ) chiral Lagrangian\footnote{In this paper, we only address situations where there are 
no multi-particle thresholds involved in loops. 
Therefore, in spite of the sickness pointed out in 
Ref.~\cite{Bernard:1996ez}, we can still
use the Minkowski formalism of PQ chiral perturbation theory.} 
for the Goldstone mesons is
\beq
\label{eq:goldstone_chiral_lagrangian}
 {\mathcal{L}}_{\mathrm{G}} =
 \frac{f^{2}}{8} {\mathrm{str}} \left [ 
 \big (\partial_{\mu} \Sigma^{\dagger}\big )\big (\partial^{\mu} \Sigma \big)
 + \Sigma^{\dagger} \chi + \chi^{\dagger} \Sigma\right ]
 + \left [
\alpha (\partial_{\mu}\Phi_{0})(\partial^{\mu} \Phi_{0})
  - M_{0}^{2} \Phi^{2}_{0} 
        \right ] ,
\eeq
where
$\Sigma= {\mathrm{exp}}(2 i \Phi/f)$ is the non-linear Goldstone particle
field, with
$\Phi$ being the matrix containing the standard Goldstone 
fields in the quark-flavour basis.
We use $f=132$~MeV. 
In this work, we
follow the supersymmetric formulation of PQ chiral perturbation theory 
(PQ$\chi$PT)~\cite{Bernard:1994sv}.
Therefore under
${\mathrm{SU}}(4|2)_{{\mathrm{L}}}\otimes {\mathrm{SU}}(4|2)_{{\mathrm{R}}}$
or ${\mathrm{SU}}(6|3)_{{\mathrm{L}}}\otimes {\mathrm{SU}}(6|3)_{{\mathrm{R}}}$, $\Sigma$ transforms as
\beq
\label{eq:Sigma_transform_rule}
\Sigma \longrightarrow U_{{\mathrm{L}}} \: \Sigma \: U_{{\mathrm{R}}}^{\dagger},
\eeq
where 
\bea
 && U_{{\mathrm{L}}} \in {\mathrm{SU}}(4|2)_{{\mathrm{L}}} \mbox{ }{\mathrm{or}}\mbox{ } {\mathrm{SU}}(6|3)_{{\mathrm{L}}},\nonumber\\
&&\nonumber\\
 && U_{{\mathrm{R}}} \in {\mathrm{SU}}(4|2)_{{\mathrm{R}}} \mbox{ }{\mathrm{or}}\mbox{ } {\mathrm{SU}}(6|3)_{{\mathrm{R}}}.
\eea
The symbol ``str'' in the above equation
means ``supertrace''.
The variable $\chi$ is defined as
\beq
\label{eq:chi_definition}
 \chi \equiv 2 B_{0} {\mathcal{M}}_{q},
\eeq
where $B_0$ is a low energy constant related to the chiral condensate
and, in the isospin limit, the quark mass matrix, $\ampl_{q}$ is
\beq
\label{eq:PQ_SU42_mass_matrix}
 \ampl_{q} = {\mathrm{diag}} 
 (\underbrace{m_{u},m_{u}}_{{\mathrm{valence}}},
  \underbrace{m_{u^{\prime}},m_{u^{\prime}}}_{{\mathrm{sea}}},
  \underbrace{m_{u},m_{u}}_{{\mathrm{ghost}}}) ,
\eeq
in the SU(4$|$2) theory, and is
\beq
\label{eq:PQ_SU63_mass_matrix}
 \ampl_{q} = {\mathrm{diag}} 
 (\underbrace{m_{u},m_{u},m_{s}}_{{\mathrm{valence}}},
  \underbrace{m_{u^{\prime}},m_{u^{\prime}},m_{s^{\prime}}}_{{\mathrm{sea}}},
  \underbrace{m_{u},m_{u},m_{s}}_{{\mathrm{ghost}}}) ,
\eeq
in the SU(6$|$3) theory.
We keep the strange quark mass different from that 
of the up and down quarks
in the valence, sea and ghost sectors.  
Notice that the flavour singlet state $\Phi_0={\mathrm{str}}(\Phi)/\sqrt{6}$ 
is rendered heavy by the
$U(1)_A$ anomaly in PQQCD \cite{Sharpe:2001fh,Sharpe:2000bc}
and can be integrated out, resulting in residual "hairpin" structures.

%
%

The inclusion of the heavy-light mesons in 
chiral perturbation theory was first proposed in
Refs.~\cite{Burdman:1992gh, Wise:1992hn,Yan:1992gz}, with the
generalisation to quenched and partially quenched theories given in 
Ref.~\cite{Sharpe:1996qp,Savage:2001jw}.  The $1/M_{P}$ and
chiral corrections were studied by Boyd and Grinstein 
\cite{Boyd:1995pa}.  The $B$ and $B^{\ast}$ meson fields
appear in this effective theory through the ``superfield"
\beq
\label{eq:H_field}
 H^{(\bar b)}_{i} = \left ( B^{\ast}_{i,\mu} \gamma^{\mu} - B_{i}\gamma_{5}\right ) \frac{1 - \slashed{v}}{2} ,
\eeq
where $v_{\mu}$ is the 4-velocity of the meson fields, 
$B_{i}$ and $B^{\ast}_{i,\mu}$ annihilate 
pseudoscalar and vector
mesons containing an anti-$b$ quark\footnote{We follow the standard
  notation~\cite{Nakamura:2010zzi} for the flavour content of $B$ mesons, so that e.g. $B_u=B^+=u\bar b$.} and a light quark of flavour $i$.
Under the heavy-quark spin transformation $S_{h}$ and the unbroken light-flavour transformation $U(x)$, the field $H^{(\bar b)}$ transforms as
\beq
\label{eq:H_transformation_properties}
 H^{(\bar b)}_{i}(x) \longrightarrow U_{i}^{\mbox{ }j}(x)\mbox{ }H^{(\bar b)}_{j}(x)\mbox{ }S_{h}^{-1}.
\eeq
Also, the conjugate field, which creates heavy-light mesons containing an anti-$b$
quark and a light quark of flavour $i$, is defined as
\beq
 \bar{H}^{(\bar b)}_{i} = \gamma^{0} H^{(\bar b)\dagger}_{i} \gamma_{0},
\eeq
and transforms under $S_{h}$ and $U(x)$ as
\beq
\label{eq:Hbar_transformation_properties}
 \bar{H}^{(\bar b)}_{i}(x) \longrightarrow S_{h}\mbox{ } \bar{H}^{(\bar b)}_{j}(x) \mbox{ }
   \big ( U^{\dagger} \big )^{j}_{\mbox{ }i}(x) .
\eeq
The introduction of the single${-}b$ baryons to $\chi$PT was pioneered by
authors of Refs.~\cite{Yan:1992gz,Cho:1992gg,Cho:1992cf}, and the effective theory was generalised to the PQ scenario
in Ref.~\cite{Arndt:2003vx}.  Since the two valence light quarks in such baryons may carry total spin quantum 
numbers\footnote{The total spin of the
light degrees of freedom is a conserved quantum number because of the heavy quark symmetry.}  
$s_{l}=0$ or $s_{l}=1$, there are two types of heavy baryons.  At the quark level, these two types of baryons carrying
light flavours $i$ and $j$ are described by the interpolating fields
\bea
 \calT^{\gamma}_{ij} &\sim& b^{\gamma,c} \left [ q^{\alpha,a}_{i} q^{\beta,b}_{j}
     + q^{\beta,b}_{j} q^{\alpha,a}_{i}\right ] \epsilon_{abc}
       \left ( C \gamma_{5}\right )_{\alpha\beta} \mbox{ }{\mathrm{for}}\mbox{ }s_{l} = 0 ,
\nonumber\\
&& \nonumber\\
\label{eq:quark_level_baryon_interpolating_fields}
 \calS^{\gamma,\mu}_{ij} &\sim& b^{\gamma,c} \left [ q^{\alpha,a}_{i} q^{\beta,b}_{j}
     - q^{\beta,b}_{j} q^{\alpha,a}_{i}\right ] \epsilon_{abc}
       \left ( C \gamma^{\mu}\right )_{\alpha\beta} \mbox{ }{\mathrm{for}}\mbox{ }s_{l} = 1 ,
\eea
where $C$ is the charge-conjugation matrix, $\alpha$, $\beta$ and $\gamma$ are the Dirac indices and
$a$, $b$ and $c$ are colour indices.  In full QCD, the $\calT$ fields are anti-symmetric and the $\calS$ fields are
symmetric under the exchange of the light flavour indices.  In the PQ theory, the flavour structure of these interpolating fields 
has the properties
\bea
 \calT_{ij} &=& (-1)^{\eta_{i}\eta_{j}} \calT_{ji} , \nonumber\\
&& \nonumber\\
\label{eq:PQ_baryon_flavour_properties}
 \calS^{\mu}_{ij} &=& (-1)^{1+\eta_{i}\eta_{j}} \calS^{\mu}_{ji},
\eea
where 
\beq
\label{eq:eta_definition}
\eta_{i} = \left \{
       \begin{array}{l}
         1 \mbox{ }{\mathrm{when}}\mbox{ }i \in {\mathrm{valence}}\mbox{ }{\mathrm{and}}\mbox{ }
                  {\mathrm{sea}}, \\
         0 \mbox{ }{\mathrm{when}}\mbox{ }i \in {\mathrm{ghost}} , \\
       \end{array}
     \right .
\eeq
accounts for different statistics of quarks in PQQCD.  These fields
transform as {\bf 39}- and {\bf 42}-plets under 
the SU(6$|$3) flavour rotation, while they transform as {\bf 17}- and
{\bf 19}-plets under 
the SU(4$|$2) flavour rotation. The baryon fields are included in heavy-hadron
chiral perturbation theory (HH$\chi$PT) according to the flavour properties in 
Eq.~(\ref{eq:PQ_baryon_flavour_properties}).  In the case of $N_{f} = 3$ ($N_{f}$ refers to the number of sea-quark 
flavours), 
the pure valence-valence sector of the $s_{l}=0$ baryons
is related to the physical states of $\Lambda_{b}$ and $\Xi^{\pm 1/2}_{b}$ via
\beq
\label{eq:valence_T_baryons}
 T_{({\mathrm{valence}}{-}{\mathrm{valence}}}) = \frac{1}{\sqrt{2}}\left (
  \begin{array}{ccc}
   0                &     \Lambda_{b}    &    \Xi^{+1/2}_{b}    \\
   -\Lambda_{b}     &         0          &    \Xi^{-1/2}_{b}    \\
   -\Xi^{+1/2}_{b}  &   -\Xi^{-1/2}_{b}  &        0             \\
  \end{array}
  \right ),
\eeq
where the superscript indicates the 3-component of the isospin. Since the light-light di-quark is of spin-1 in the $\calS^{\mu}_{ij}$ fields, such baryons can be
in spin $1/2$ or $3/2$ states which are degenerate in the heavy quark limit.  
Therefore they are best described by the ``superfield" 
\beq
\label{eq:S_superfield}
 S^{\mu}_{ij} = \sqrt{\frac{1}{3}} (v^{\mu} + \gamma^{\mu}) \gamma_{5} B_{ij}
       + B^{\ast \mu}_{ij} ,
\eeq
where $B_{ij}$ and $B^{\ast\mu}_{ij}$ are spin-$1/2$ and $3/2$ baryons.  In the pure valence-valence
sector,
\beq
\label{eq:valence_S_baryons}
 B_{({\mathrm{valence}}{-}{\mathrm{valence}}}) = \left (
  \begin{array}{ccc}
   \Sigma^{+1}_{b}   &  \frac{1}{\sqrt{2}}\Sigma^{0}_{b} &  \frac{1}{\sqrt{2}}\Xi^{\prime +1/2}_{b}  \\
   \frac{1}{\sqrt{2}}\Sigma^{0}_{b} & \Sigma^{-1}_{b}    &  \frac{1}{\sqrt{2}}\Xi^{\prime -1/2}_{b}  \\
  \frac{1}{\sqrt{2}}\Xi^{\prime +1/2}_{b} & \frac{1}{\sqrt{2}}\Sigma^{0}_{b} & \Omega_{b} \\
  \end{array}
  \right ) ,
\eeq
and similarly for the $B^{\ast\mu}_{ij}$ fields.  The $S^{\mu}_{ij}$ and $T_{ij}$ fields have the same property 
as $\bar{H}^{(\bar{b})}_{i}$ under the heavy-quark spin transformation $S_{h}$.  For the unbroken 
light-flavour transformation,
\bea
 S^{\mu}_{ij}(x) &\longrightarrow& U^{\mbox{ }k}_{i}(x) \: U^{\mbox{ }l}_{j}(x) \: S^{\mu}_{kl}(x), \nonumber\\
&&\nonumber\\
 T_{ij}(x) &\longrightarrow& U^{\mbox{ }k}_{i}(x) \: U^{\mbox{ }l}_{j}(x) \: T_{kl}(x),
\eea
with the flavour indices satisfying Eq.~(\ref{eq:PQ_baryon_flavour_properties}).  
These $S_{ij}^{\mu}$ and $T_{ij}$ are annihilation field operators and we denote the corresponding creation
fields by $\bar{S}^{\mu}_{ij}$ and $\bar{T}_{ij}$.

The Goldstone mesons couple to the above heavy meson and baryon fields in the HH$\chi$PT Lagrangian 
via the non-linear realisation
\beq
 \xi \equiv \exponential^{i\Phi/f} = \sqrt{\Sigma} ,  \label{eq:xitransformation1}
\eeq
which transforms as
\beq
 \xi(x) \longrightarrow U_{\mathrm{L}} \:\:
  \xi(x) \:\: U^{\dagger}(x) = U(x) \:\: \xi(x) \:\:
   U^{\dagger}_{\mathrm{R}}.  \label{eq:xitransformation2}
\eeq
The $\xi$ field can be used to construct vector and 
axial-vector fields
\bea
 V^{\mu} &=& \frac{1}{2} \left ( \xi^{\dagger} \partial^{\mu}\xi
       + \xi\partial^{\mu}\xi^{\dagger}\right ) ,\nonumber\\
&&\nonumber\\
 A^{\mu} &=& \frac{i}{2} \left ( \xi^{\dagger} \partial^{\mu}\xi
       - \xi\partial^{\mu}\xi^{\dagger}\right ) .
\eea 
The vector field can then serve as the gauge field in defining the chiral covariant derivative which
acts on the heavy hadrons,
\bea
 \calD^{\mu} H^{(\bar b)}_{i} 
      &=& \partial^{\mu} H^{(\bar b)}_{i} + \left ( V^{\mu} \right )^{\mbox{ }j}_{i} H^{(\bar b)}_{j} , \nonumber\\
&& \nonumber\\
 \calD^{\mu} T_{ij} &=& \partial^{\mu} T_{ij} + \left ( V^{\mu} \right )_{i}^{\mbox{ }k} T_{kj}
         + (-1)^{\eta_{i}(\eta_{j} + \eta_{k})} \left ( V^{\mu} \right )_{j}^{\mbox{ }k}
            T_{ik}, \nonumber\\
&& \nonumber\\
 \calD^{\mu} S_{ij}^\nu &=& \partial^{\mu} S_{ij}^\nu + \left ( V^{\mu} \right )_{i}^{\mbox{ }k} S_{kj}^\nu
         + (-1)^{\eta_{i}(\eta_{j} + \eta_{k})} \left ( V^{\mu} \right )_{j}^{\mbox{ }k}
            S_{ik}^\nu.
\eea
The leading-order HH$\chi$PT Lagrangian is then
\bea
 {\mathcal{L}}^{({\mathrm{LO}})}_{\mathrm{HH\chi PT}} 
 &=& -i \, {\mathrm{tr_{D}}}
  \left [
   \bar{H}^{(\bar b)i} v_{\mu} \calD^{\mu} H^{(\bar b)}_{i}
  \right ] +  i \left ( \bar{T} v_{\mu} \calD^{\mu} T \right )_{{\mathrm{f}}}
   - i \big ( \bar{S}^{\nu} v_{\mu} \calD^{\mu} S_{\nu} \big )_{{\mathrm{f}}}
    + \Delta^{(B)} \big ( \bar{S}^{\nu} S_{\nu} \big )_{{\mathrm{f}}} \nonumber\\
\label{eq:HHChPT_lagrangian_LO}
& &  + \, g_{1}\,
   {\mathrm{tr_{D}}}
  \left [
  \bar{H}^{(\bar b)}_{i} \gamma^\mu \gamma_{5}\: H^{(\bar b)}_{j} 
   A_\mu^{ij}
  \right ] 
    - i g_{2}\, \epsilon_{\mu\nu\sigma\rho} 
         \big ( \bar{S}^{\mu} v^{\nu} A^{\sigma} S^{\rho} \big )_{{\mathrm{f}}}
    + \sqrt{2}\, g_{3}\, \left [ 
       \left ( \bar{T} A^{\mu} S_{\mu} \right )_{{\mathrm{f}}}
       + \left ( \bar{S}_{\mu} A^{\mu} T \right )_{{\mathrm{f}}}
      \right ]  ,
\eea
where $v_{\mu}$ is the velocity of the heavy hadrons, 
${\mathrm{tr_{D}}}[\mbox{ }]$ means taking the trace in
Dirac space, and $\big (\mbox{ }\big )_{{\mathrm{f}}}$ is the implementation of the 
PQ-theory flavour contraction rules~\cite{Arndt:2003vx}
\bea
 \big ( \bar{T} Y T \big )_{{\mathrm{f}}} &=& \bar{T}^{ji} Y^{\mbox{ }l}_{i} T_{lj} , \nonumber\\
&&\nonumber\\
 \big ( \bar{S}^{\mu} Y S_{\mu} \big )_{{\mathrm{f}}} &=& \bar{S}^{\mu,ji} Y^{\mbox{ }l}_{i} S_{\mu,lj} , \nonumber\\
&&\nonumber\\
\label{eq:baryon_PQ_flavour_contraction}
 \big ( \bar{T} Y^{\mu} S_{\mu} \big )_{{\mathrm{f}}} &=& \bar{T}^{ji} \left ( Y^{\mu}\right )^{\mbox{ }l}_{i} S_{\mu,lj} .
\eea
The parameter $\Delta^{(B)}$ is the mass difference between the $S$ and $T$ fields with the same light
flavour indices,
\bea
 \Delta^{(B)} = M_{S_{i,j}} - M_{T_{i,j}} . \nonumber
\eea
It is grouped together with the definition of other mass parameters in Eq.~(\ref{eq:mass_param}) in Appendix~\ref{app:masses_in_loops}.  
This mass difference is of
$O(\Lambda_{{\mathrm{QCD}}})$, and does not vanish either in the chiral limit or in 
the heavy-quark limit.

The LO Lagrangian for HH$\chi$PT contains terms of $O(p)$ and no light-quark mass dependence.
To generate the flavour SU(3) breaking effects in heavy-meson and baryon spectrum, which give rise to the mass 
differences $\delta^{(M)}_{i,j}$ and $\delta^{(B)}_{ij,kl}$ in Eq.~(\ref{eq:mass_param}),
one introduces
\bea
  {\mathcal{L}}^{(\chi)}_{\mathrm{HH\chi PT}} &=&\lambda_{1}\mbox{ }{\mathrm{tr_{D}}} \left [
  \bar{H}^{(\bar b)}_{i} \chi_{\xi}^{ij} H^{(\bar b)}_{j}
 \right ]
+ \lambda_{2}\mbox{ }{\mathrm{tr_{D}}} \left [ 
  \bar{H}^{(\bar b)i} 
  H^{(\bar b)}_{i}
 \right ]
 {\mathrm{str}}\left ( \chi_{\xi}
  \right )\nonumber\\
\label{eq:HHChPT_lagrangian_chi}
 && + \, \lambda_{3} \big ( \bar{S}^{\mu} \chi_{\xi} S_{\mu}\big )_{{\mathrm{f}}}
    + \lambda_{4} \big ( \bar{S}^{\mu}  S_{\mu}\big )_{{\mathrm{f}}}\, {\mathrm{str}}\left (
   \chi_{\xi}
  \right ) + \, \lambda_{5} \big ( \bar{T} \chi_{\xi} T \big )_{{\mathrm{f}}}
    + \lambda_{6} \big ( \bar{T}  T\big )_{{\mathrm{f}}}\, {\mathrm{str}}\left (
   \chi_{\xi}
  \right ) ,
\eea
where
\beq
\label{eq:chi_xi_definition}
\chi_{\xi} = \xi \chi \xi + \xi^{\dagger} \chi \xi^{\dagger} .
\eeq
In the computation of the axial-current matrix elements, 
the flavour breaking effects in Eq.~(\ref{eq:HHChPT_lagrangian_chi}) are formally sub-leading compared to
those encoded in the pure Goldstone Lagrangian, Eq.~(\ref{eq:goldstone_chiral_lagrangian}).  Nevertheless,
we keep them in our calculation as they can be numerically significant.

In this work, we also include the heavy quark spin symmetry breaking term
\beq
\label{eq:HQ_spin_breaking_term}
\frac{\bar{\lambda}_{2}}{M_{B}}
{\mathrm{tr_{D}}} \left [
 \bar{H}^{(\bar b)i}\sigma_{\mu\nu} H^{(\bar b)}_{i} \sigma^{\mu\nu}
\right ] ,
\eeq
where $M_{B}$ is the $B$ meson mass.  
This counterterm leads to the mass difference between the $B^{\ast}$ and $B$ mesons with the same light flavour,
\bea
 \Delta^{(M)} &=& M_{B^{\ast}_{i}} - M_{B_{i}} ,\nonumber
\eea
which vanishes in the heavy-quark limit.
This mass difference is also grouped together with other mass parameters 
in Eq.~(\ref{eq:mass_param}) in Appendix~\ref{app:masses_in_loops}. In principle,
there are also such heavy quark spin breaking terms in the baryon sector, resulting in mass differences
between $B_{ij}$ and $B^{\ast}_{ij}$ baryons in
Eq.~(\ref{eq:S_superfield}). However, these mass differences are
numerically much smaller than $\Delta^{(M)}$~\cite{Aaltonen:2007rw}.

\section{Axial current matrix elements at the next-to-leading order}
\label{sec:NLO}
Applying the Noether theorem to the chiral Lagrangian in the previous section, one can derive the leading-order axial currents
corresponding to their quark-level counterparts in Eq.~(\ref{eq:quark_level_currents}).
For matrix elements involving external states of single heavy hadrons, the relevant LO currents are
\beq
\label{eq:axial_current}
 J^{(N_{f})}_{ij,\mu} = g_{1}\, {\mathrm{tr}}_{{\mathrm{D}}} \left [ \bar{H}^{(\bar b)}_{k}  \gamma_{\mu}\gamma_{5}\: H^{(\bar b)}_{l}\left ( \tau^{(+)}_{ij,\xi} \right )^{kl}\right ] 
- i g_{2}\, \epsilon_{\mu\nu\sigma\rho} \big ( \bar{S}^{\nu} v^{\sigma} \tau^{(+)}_{ij,\xi} S^{\rho}\big )_{{\mathrm{f}}}
+ \sqrt{2} \, g_{3}\, \left [ \big ( \bar{S}_{\mu} \tau^{(+)}_{ij,\xi} T\big
)_{{\mathrm{f}}} + \big (\bar{T} \tau^{(+)}_{ij,\xi} S_{\mu} \big )_{{\mathrm{f}}} \right ] ,
\eeq
where the subscript $ij$ means the current changes the light quark flavours from $i$ to $j$, and 
\beq
\label{eq:tau_xi_ij}
  \tau^{(+)}_{ij,\xi} = \frac{1}{2} \left ( \xi^{\dagger} \tau_{ij} \xi + \xi \tau_{ij} \xi^{\dagger} \right ),
\eeq
with the matrices $\tau_{ij}$ defined as
\beq
 \left ( \tau_{ij} \right )_{kl} = \delta_{il} \delta_{jk}, 
\eeq
where $k$ and $l$ run through all the light-quark flavours in PQQCD.  The superscript
$N_{f}$ is the number of sea quark flavours and $N_{f} = 2,3$
represent the cases of SU(4$|$2) and SU(6$|$3) respectively. These leading-order axial currents generate the LO terms, 
as well as the NLO contributions via one-loop 
corrections, in the matrix elements studied in this work\footnote{In addition to the terms in Eq.~(\ref{eq:axial_current}),
there are other operators in the LO currents arising from the chiral Lagrangian introduced in the previous section.  Nevertheless,
these terms do not appear in the matrix elements to the order we work at.  That is, their contributions to the one-loop corrections
vanish.}.

\begin{figure}[!t]
  \centering
\includegraphics[width=0.35\columnwidth]{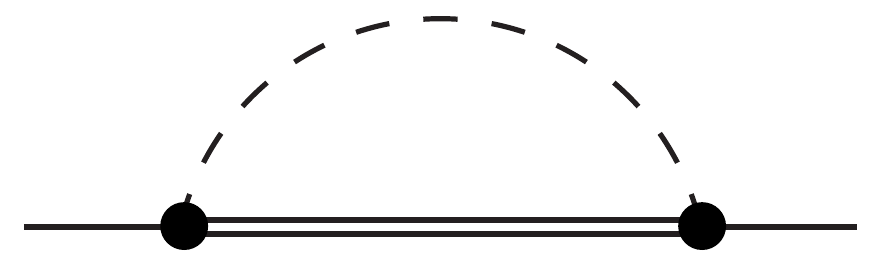}
\hspace*{1mm}
\includegraphics[width=0.225\columnwidth]{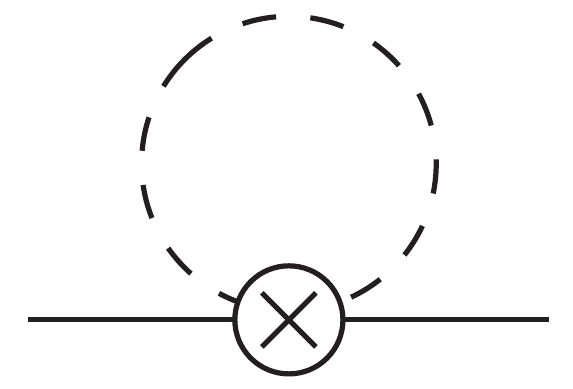}
\hspace*{1mm}
\includegraphics[width=0.35\columnwidth]{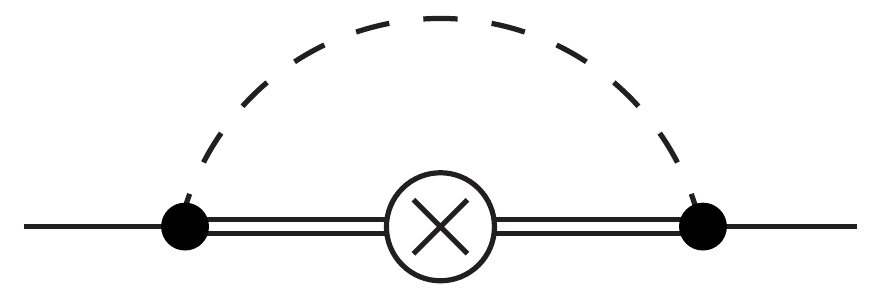} 
\\
(a)\hspace*{5.1cm}(b)\hspace*{5.1cm}(c)
  \caption{One-loop diagrams contributing to the matrix elements of axial currents between heavy hadrons.
   The dashed lines are the Goldstone meson propagators, including the possible ``hairpin" structures.  The single
   solid lines denote generically the external heavy hadrons, while the double solid lines are the internal
   heavy hadrons.  They can be $B$, $B^{\ast}$ mesons or $T_{ij}$, $S_{ij}$ baryons. The circled crosses are
   the insertions of the LO axial current $J^{(N_{f})}_{ij,\mu}$ given in Eq.~(\ref{eq:axial_current}), while the 
   other vertices are from the strong chiral Lagrangian in Eq.~(\ref{eq:HHChPT_lagrangian_LO}).  
   Diagram (a) is the self-energy of the heavy hadron and it leads to the wavefunction renormalisation
   contribution to the matrix elements.  Diagrams (b) and (c) are the ``tadpole" and ``sunset" types respectively.}
  \label{fig:oneloop}
\end{figure}

There is a significant increase in the number of terms in the next-to-leading order axial currents in HH$\chi$PT, and we
postpone the detailed investigation of these NLO currents to Subsection~\ref{sec:NLO_analytic} below.  Here we first write down 
the generic form of the chiral expansion of the axial-current matrix elements to the NLO,
\beq
\label{eq:NLO-master}
  \la H_{j} |\calJ_{ij,\mu}| H_{i}\ra_{{\mathrm{QCD}}}
   = \la H_{j} |J^{(N_{f})}_{ij,\mu}| H_{i}\ra_{\rm LO} \times
    \left [ 1 + \frac{1}{f^{2}}\left ( \calT^{(N_{f})}_{ij} + \calW^{(N_{f})}_{H_{i}}
      + \calW^{(N_{f})}_{H_{j}} + \calQ^{(N_{f})}_{H_{i}\rightarrow H_{j}} \right )
      + \calN^{(N_{f})}_{H_{i}\rightarrow H_{j}}  \right ] ,
\eeq
where the equality symbol means the matching between (PQ)QCD and the chiral effective theory,
$\mu$ is the Lorentz index, and $\calJ_{ij,\mu}$ are the quark-level currents given in Eq.~(\ref{eq:quark_level_currents}).  
The flavour indices are denoted by $i$ and $j$ which are {\it not} summed in the above expression, and $H_{i}$ is a heavy hadron state
(meson or baryon) containing the light flavour $i$.
The symbols 
$\calT^{(N_{f})}_{ij}$, $\calW^{(N_{f})}_{H_{i}}$ and 
$\calQ^{(N_{f})}_{H_{i}\rightarrow H_{j}}$ are results from the tadpole, 
wavefunction-renormalisation and
sunset diagrams at one-loop, as depicted in Fig~\ref{fig:oneloop}, where single and double solid lines represent
generically the external and internal heavy hadrons while the dashed lines are the Goldstone propagators.  The
circled crosses in Fig~\ref{fig:oneloop} are the insertions of the LO axial current $J^{(N_{f})}_{ij,\mu}$ given
in Eq.~(\ref{eq:axial_current}).
The tadpole contributions $\calT^{(N_{f})}_{ij}$ are independent of the external states, since they emerge completely 
from the flavour structure of the currents.

In Eq.~(\ref{eq:NLO-master}),  we have written the NLO analytic terms as the LO matrix elements times 
$\calN^{(N_{f})}_{H_{i}\rightarrow H_{j}}$.  In Subsection~\ref{sec:NLO_analytic} below, we will study
these NLO analytic terms, and show that they can be presented in this manner.
In this section, we examine the analytic terms (polynomials in the Goldstone masses) in 
the matrix elements in Eq.~(\ref{eq:NLO-master}) for various external states.  These are encoded in
\bea
&& \la H_{j} |J^{(N_{f})}_{ij,\mu}| H_{i}\ra_{\rm LO} \mbox{ }{\mathrm{and}}\mbox{ }
      \calN^{(N_{f})}_{H_{i}\rightarrow H_{j}} . \nonumber
\eea
The non-analytic contributions
arising from the one-loop diagrams will be discussed in Sections~\ref{sec:one-loop-contr-su2} and
\ref{sec:one-loop-contr-su42-su63}.

\subsection{Leading-order matrix elements}
\label{eq:LO_ME}

Lattice computations are often performed using the baryon interpolating fields in
Eq.~(\ref{eq:quark_level_baryon_interpolating_fields}).  Therefore we carry out the $\chi$PT calculation for
the $T_{ij}$ and $S_{ij}^{\mu}$ external states.  From our results, it is straightforward to obtain matrix elements for
physical external baryon states using Eqs.~(\ref{eq:valence_T_baryons}) and (\ref{eq:valence_S_baryons}).
The leading-order HH$\chi$PT predictions for the matrix elements studied in this work are
\bea
\nonumber  \la B^*_d  | J^{(N_f)}_{ud,\mu} | B_u  \ra_{\rm LO}
         = \la B^*_s  | J^{(3)}_{us,\mu}   | B_u  \ra_{\rm LO}
&=& - 2\:g_1 \: \varepsilon^*_\mu, \\
\nonumber  \la S_{dd} | J^{(N_f)}_{ud,\mu} | T_{du} \ra_{\rm LO}
= \sqrt{2} \la S_{sd} | J^{(3)}_{ud,\mu}   | T_{su} \ra_{\rm LO}
= \sqrt{2} \la S_{ds} | J^{(3)}_{us,\mu}   | T_{du} \ra_{\rm LO}
=          \la S_{ss} | J^{(3)}_{us,\mu}   | T_{su} \ra_{\rm LO}
&=& - g_3 \:  \overline{U}_\mu \: \calU, \\
\nonumber  \la S_{dd} | J^{(N_f)}_{ud,\mu} | S_{du} \ra_{\rm LO}
= \sqrt{2} \la S_{sd} | J^{(3)}_{ud,\mu}   | S_{su} \ra_{\rm LO}
= \sqrt{2} \la S_{ds} | J^{(3)}_{us,\mu}   | S_{du} \ra_{\rm LO}
=          \la S_{ss} | J^{(3)}_{us,\mu}   | S_{su} \ra_{\rm LO}
&=& - \frac{i}{\sqrt{2}}\:g_2 \:v^\sigma \: \epsilon_{\sigma \mu \nu\rho}\: \overline{U}^\nu U^\rho, \\ \label{eq:LOmatrixelts}
\eea
where $\varepsilon^\mu$ is the polarisation vector of the $B^*$ meson,
$\calU$ is the Dirac spinor of the $T$ baryon, and the $U^\mu$'s are the ``superfield
spinors'' of the $S$ baryons. The basis polarisation vectors and spinors satisfy the spin sums
\bea
\nonumber \sum_{s=1}^3 \varepsilon_\mu(v,s) \varepsilon^*_\nu(v,s) &=& -g_{\mu\nu}+v_\mu v_\nu, \\
\nonumber \sum_{s=1}^2 \calU(v,s) \overline{\calU}(v,s) &=& \frac{1+\slashed{v}}{2}, \\
\sum_{s=1}^6 U^\mu (v,s) \overline{U}^\nu (v,s) &=& -(g^{\mu\nu}-v^\mu v^\nu) \frac{1+\slashed{v}}{2}.
\eea
Note that $U^\mu$ is \emph{not} a Rarita-Schwinger spinor; instead it contains the degrees of freedom of both the spin-$1/2$ and spin-$3/2$ components of the superfield.
In Eq.~(\ref{eq:LOmatrixelts}), the states are normalized as
\bea
\nonumber \langle B_i(v,\mathbf{k}) | B_i(v,\mathbf{k}') \rangle &=& 2v^0 (2\pi)^3 \delta^3(\mathbf{k}-\mathbf{k}'), \\
\nonumber \langle B_i^*(v,\mathbf{k},s) | B_i^*(v,\mathbf{k}',s') \rangle &=& 2v^0 (2\pi)^3 \delta_{ss'}\delta^3(\mathbf{k}-\mathbf{k}'), \\
\nonumber \langle T_{ij}(v,\mathbf{k},s) | T_{ij}(v,\mathbf{k}',s') \rangle &=& v^0 (2\pi)^3 \delta_{ss'}\delta^3(\mathbf{k}-\mathbf{k}'), \\
\langle S_{ij}(v,\mathbf{k},s) | S_{ij}(v,\mathbf{k}',s') \rangle &=& v^0 (2\pi)^3 \delta_{ss'}\delta^3(\mathbf{k}-\mathbf{k}').
\eea
%

%
\subsection{Next-to-leading order analytic terms}
\label{sec:NLO_analytic}
In this subsection, we investigate the NLO counterterms in the axial currents.  Their matrix elements between 
single heavy hadron states are written as
\bea
&& \la H_{j} |J^{(N_{f})}_{ij,\mu}| H_{i}\ra_{\rm LO}\times
      \calN^{(N_{f})}_{H_{i}\rightarrow H_{j}} , \nonumber
\eea
in Eq.~(\ref{eq:NLO-master}).  These NLO counterterms play a significant role in the chiral expansion, since they
have to be included to renormalise the one-loop contributions from the LO axial currents to matrix elements.
  
First we notice that the chiral Lagrangian in
Eq.~(\ref{eq:HHChPT_lagrangian_chi}) does not contain any space-time derivative, therefore it does not 
lead to new terms in the axial currents upon applying the Noether theorem.  To obtain the NLO axial currents,
we introduce additional operators in the chiral Lagrangian,
\bea
\nonumber \calL^{({\mathrm{NLO,\:axial}})}_{{\mathrm{HH}}\chi{\mathrm{PT}}} &=&
     \kappa_{1}^{(H)}\, {\mathrm{tr}}_{{\mathrm{D}}} \left [ \bar{H}^{(\bar b)}_{i} \gamma^\mu\gamma_5\: H^{(\bar b)}_{j} 
        (A_\mu)^{i}_{\mbox{ }k}\:\chi_{\xi}^{kj}\right ] + 
     \kappa_{2}^{(H)}\, {\mathrm{tr}}_{{\mathrm{D}}} \left [ \bar{H}^{(\bar b)}_{i} \gamma^\mu\gamma_5\: H^{(\bar b)}_{j} 
        \chi_{\xi}^{ik}(A_\mu)_{k}^{\mbox{ }j}\right ] \\
&& \hspace{-0.2cm} + \,    \kappa_{3}^{(H)}\, {\mathrm{tr}}_{{\mathrm{D}}} \left [ \bar{H}^{(\bar b)}_{i} \gamma^\mu\gamma_5\: H^{(\bar b)}_{j} 
        (A_\mu)^{ij}\right ]{\rm str}(\chi_{\xi}) +
    \kappa_{4}^{(H)}\, {\mathrm{tr}}_{{\mathrm{D}}} \left [ \bar{H}^{(\bar b)}_{i} \gamma^\mu\gamma_5\: H^{(\bar b)}_{i} 
        (A_\mu)_{kl}\:\chi_{\xi}^{lk}\right ] \nonumber \\
&& \hspace{-0.2cm} + \,  \kappa_{1}^{(S)}\, \epsilon_{\mu\nu\sigma\rho} \big ( \bar{S}^{\mu} v^{\nu} A^{\sigma} \chi_{\xi} S^{\rho} \big )_{{\mathrm{f}}} 
  +  \kappa_{2}^{(S)}\, \epsilon_{\mu\nu\sigma\rho} \big ( \bar{S}^{\mu} v^{\nu} \chi_{\xi} A^{\sigma} S^{\rho} \big )_{{\mathrm{f}}} \nonumber \\
&& \hspace{-0.2cm}  +\, \kappa_{3}^{(S)}\, \epsilon_{\mu\nu\sigma\rho} \big ( \bar{S}^{\mu} v^{\nu} A^{\sigma}  S^{\rho} \big )_{{\mathrm{f}}}\, {\mathrm{str}} \left ( \chi_{\xi}\right )
   + \kappa_{4}^{(S)}\, \epsilon_{\mu\nu\sigma\rho} \big ( \bar{S}^{\mu} v^{\nu} S^{\rho} \big )_{{\mathrm{f}}}\, {\mathrm{str}} \left ( A^{\sigma} \chi_{\xi}\right )\nonumber\\
\label{eq:HHChPT_lagrangian_NLO}
&& \hspace{-0.2cm}+\, \kappa_{1}^{(T)}\, \left [ \big ( \bar{T} A^{\mu} \chi_{\xi} S_{\mu} \big )_{{\mathrm{f}}} + \big ( \bar{S}_{\mu} A^{\mu} \chi_{\xi} T \big )_{{\mathrm{f}}} \right ] 
 + \kappa_{2}^{(T)}\, \left [ \big ( \bar{T} \chi_{\xi} A^{\mu} S_{\mu} \big )_{{\mathrm{f}}} + \big ( \bar{S}_{\mu} \chi_{\xi} A^{\mu} T \big )_{{\mathrm{f}}} \right ] \nonumber\\
&& \hspace{-0.2cm} +\, \kappa_{3}^{(T)}\, \left [ \big ( \bar{T} A^{\mu}  S_{\mu} \big )_{{\mathrm{f}}} + \big ( \bar{S}_{\mu} A^{\mu} T \big )_{{\mathrm{f}}} \right ]\, {\mathrm{str}} \left ( \chi_{\xi} \right ) ,
\eea
where $\chi_{\xi}$ is defined in Eq.~(\ref{eq:chi_xi_definition}).  
The mesonic sector of the above Lagrangian was already introduced in Refs.~\cite{Boyd:1995pa,Stewart:1998ke}.
Upon applying the Noether theorem to Eq.~(\ref{eq:HHChPT_lagrangian_NLO}), one 
obtains the currents
which lead to the NLO analytic terms $\calN^{(N_{f})}_{H_{i}\rightarrow H_{j}}$ in Eq.~(\ref{eq:NLO-master}),
\bea
\nonumber J^{({\mathrm{NLO,\:analytic}})}_{ij,\mu} &=&
     \kappa_{1}^{(H)}\, {\mathrm{tr}}_{{\mathrm{D}}} \left [ \bar{H}^{(\bar b)}_{k}\gamma_{\mu}\gamma_{5}\: H^{(\bar b)}_{l} 
        \left ( \tau^{(+)}_{ij,\xi}\mbox{ }\chi_{\xi}\right )^{kl}\right ] + 
     \kappa_{2}^{(H)}\, {\mathrm{tr}}_{{\mathrm{D}}} \left [ \bar{H}^{(\bar b)}_{k} \gamma_{\mu}\gamma_{5}\: H^{(\bar b)}_{l} 
          \left ( \chi_{\xi}\mbox{ }\tau^{(+)}_{ij,\xi}\right )^{kl}\right ] \nonumber\\
&& \hspace{-0.2cm} + \,
     \kappa_{3}^{(H)}\, {\mathrm{tr}}_{{\mathrm{D}}} \left [ \bar{H}^{(\bar b)}_{k} \gamma_{\mu}\gamma_{5}\: H^{(\bar b)}_{l} 
        \left ( \tau^{(+)}_{ij,\xi} \right )^{kl}\right ]{\rm str}(\chi_{\xi}) +
    \kappa_{4}^{(H)}\, {\mathrm{tr}}_{{\mathrm{D}}} \left [ \bar{H}^{(\bar b)}_{k} \gamma_{\mu}\gamma_{5}\: H^{(\bar b)}_{k} 
        \left ( \tau^{(+)}_{ij,\xi} \mbox{ }\chi_{\xi} \right )^{ll}\right ] \nonumber\\
&& \hspace{-0.2cm} + \, 
     \kappa_{1}^{(S)}\, \epsilon_{\mu\nu\sigma\rho} \big ( \bar{S}^{\nu} v^{\sigma}  \tau^{(+)}_{ij,\xi}\mbox{ }\chi_{\xi} S^{\rho} \big )_{{\mathrm{f}}} 
  +  \kappa_{2}^{(S)}\, \epsilon_{\mu\nu\sigma\rho} \big ( \bar{S}^{\nu} v^{\sigma} \chi_{\xi}\mbox{ }\tau^{(+)}_{ij,\xi}  S^{\rho} \big )_{{\mathrm{f}}} \nonumber \\
&& \hspace{-0.2cm}  +\, \kappa_{3}^{(S)}\, \epsilon_{\mu\nu\sigma\rho} \big ( \bar{S}^{\nu} v^{\sigma} \tau^{(+)}_{ij,\xi} S^{\rho} \big )_{{\mathrm{f}}}\, {\mathrm{str}} \left ( \chi_{\xi}\right )
   + \kappa_{4}^{(S)}\, \epsilon_{\mu\nu\sigma\rho} \big ( \bar{S}^{\nu} v^{\sigma} S^{\rho} \big )_{{\mathrm{f}}}\, {\mathrm{str}} \left ( \tau^{(+)}_{ij,\xi} \mbox{ }\chi_{\xi}\right )\nonumber\\
\label{eq:axial_current_NLO}
&& \hspace{-0.2cm}+\, \kappa_{1}^{(T)}\, \left [ \big ( \bar{T} \tau^{(+)}_{ij,\xi}\mbox{ } \chi_{\xi} S_{\mu} \big )_{{\mathrm{f}}} + 
    \big ( \bar{S}_{\mu}  \tau^{(+)}_{ij,\xi} \mbox{ } \chi_{\xi} T \big )_{{\mathrm{f}}} \right ] 
 + \kappa_{2}^{(T)}\, \left [ \big ( \bar{T} \chi_{\xi}\mbox{ }  \tau^{(+)}_{ij,\xi}  S_{\mu} \big )_{{\mathrm{f}}} + 
  \big ( \bar{S}_{\mu} \chi_{\xi} \mbox{ } \tau^{(+)}_{ij,\xi} T \big )_{{\mathrm{f}}} \right ] \nonumber\\
&& \hspace{-0.2cm} +\, \kappa_{3}^{(T)}\, \left [ \big ( \bar{T} \tau^{(+)}_{ij,\xi}  S_{\mu} \big )_{{\mathrm{f}}} + 
    \big ( \bar{S}_{\mu} \tau^{(+)}_{ij,\xi}  T \big )_{{\mathrm{f}}} \right ]\, {\mathrm{str}} \left ( \chi_{\xi} \right ) ,
\eea
where $\tau^{(+)}_{ij,\xi}$ is defined in Eq.~(\ref{eq:tau_xi_ij}). Although it is not explicitly shown in the above
equation, these NLO currents depend on $N_{f}$.

Comparing the currents $J^{({\mathrm{NLO,\:analytic}})}_{ij,\mu}$ to their leading-order counterparts,
$J^{(N_{f})}_{ij,\mu}$ in Eq.~(\ref{eq:axial_current}), one observes that they share the similar feature in the 
combination of the heavy-hadron fields with the flavour matrices $\tau^{(+)}_{ij,\xi}$.  The complication in 
$J^{({\mathrm{NLO,\:analytic}})}_{ij,\mu}$ results completely from the insertion of
$\chi_{\xi}$, which contains one power of the quark-mass matrix.  This shows that one can write the 
NLO matrix elements as
\bea
\la H_{j} |J^{({\mathrm{NLO,\:analytic}})}_{ij,\mu}| H_{i}\ra_{{\mathrm{NLO}}} &=& \la H_{j} |J^{(N_{f})}_{ij,\mu}| H_{i}\ra_{\rm LO}\times
      \calN^{(N_{f})}_{H_{i}\rightarrow H_{j}} , \nonumber
\eea
and
\bea
 && \calN^{(N_{f})}_{H_{i}\rightarrow H_{j}} \sim O(m_{q}) \sim O(M^{2}_{{\mathrm{Goldstone}}}), \nonumber
\eea
where $m_{q}$ is the light-quark mass.

\section{One-loop contributions in SU(2) HH$\chi$PT}
\label{sec:one-loop-contr-su2}
We now turn to the discussion of the one-loop results for the axial-current matrix elements.  In this section, we
first present a simple case, namely SU(2) $\chi$PT in the infinite-volume limit, and use it to illustrate the
 main features of these one-loop contributions.  Details of the SU(4$|$2) and SU(6$|$3) PQ$\chi$PT results are 
addressed in the next section.

We start by reducing the leading-order matrix elements in 
Eq.~(\ref{eq:LOmatrixelts}) to a simpler form.  Notice that all these matrix elements are proportional to the
axial couplings, $g_{1,2,3}$.  Therefore, from the generic form of the 
chiral expansion for the axial-current matrix elements given in Eq.~(\ref{eq:NLO-master}), we can define the
``effective" axial couplings 
\bea
   \left ( g_{1} \right )_{{\mathrm{eff}}}
   &=& g_{1} \times
    \left [ 1 + \frac{1}{f^{2}}\left ( \calT^{(2)}_{ud} + \calW^{(2)}_{B_{u}}
      + \calW^{(2)}_{B^{\ast}_{d}} + \calQ^{(2)}_{B_{u}\rightarrow B^{\ast}_{d}} \right )
      + \calN^{(2)}_{B_{u}\rightarrow B^{\ast}_{d}}  \right ] , \nonumber\\
&& \nonumber\\
  \left ( g_{2} \right )_{{\mathrm{eff}}}
   &=& g_{2} \times
    \left [ 1 + \frac{1}{f^{2}}\left ( \calT^{(2)}_{ud} + \calW^{(2)}_{T_{du}}
      + \calW^{(2)}_{S_{dd}} + \calQ^{(2)}_{T_{du}\rightarrow S_{dd}} \right )
      + \calN^{(2)}_{T_{du}\rightarrow S_{dd}}  \right ] , \nonumber\\
&& \nonumber\\
\label{eq:effective_couplings}
  \left ( g_{3} \right )_{{\mathrm{eff}}}
   &=& g_{3} \times
    \left [ 1 + \frac{1}{f^{2}}\left ( \calT^{(2)}_{ud} + \calW^{(2)}_{S_{du}}
      + \calW^{(2)}_{S_{dd}} + \calQ^{(2)}_{S_{du}\rightarrow S_{dd}} \right )
      + \calN^{(2)}_{S_{du}\rightarrow S_{dd}}  \right ] ,
\eea
with the wavefunction renormalisation ($\calW$), tadpole ($\calT$) and sunset ($\calQ$) 
diagram contributions from Fig.~\ref{fig:oneloop} (a), (b) and (c). 

The result for the tadpole
diagram is particularly simple. In the infinite-volume limit, it is 
\bea
  && \calT^{(2)}_{ud} \stackrel{{\mathrm{infinite}}{-}V}{\longrightarrow}
     -2 I(M_{\pi}) = - \frac{2}{16 \pi^{2}} M^{2}_{\pi} {\mathrm{log}}\left ( \frac{M^{2}_{\pi}}{\mu^{2}}\right ), \nonumber
\eea
following the definition of the function $I(m)$ in Eq.~(\ref{eq:IandF_infinite_volume}) in Appendix~\ref{app:integral}.
Here $M_{\pi}$ is the pion mass, and $\mu$ is the renormalisation scale.  The dependence on $\mu$ is cancelled by the NLO
counterterm contributions $\calN^{(2)}$ in the above expression for the effective axial couplings.

In this SU(2) full QCD case, the infinite-volume limit of the wavefunction renormalisation and sunset diagrams can be 
written in two functions
\bea
 H(m,\Delta) &=& \frac{\partial F(m,\Delta)}{\partial \Delta} ,\nonumber\\
&& \nonumber\\
\label{eq:HandK_infinite_volume}
 K(m,\Delta_{1},\Delta_{2}) &=& \frac{F(m,\Delta_{1}) - F(m,\Delta_{2})}{\Delta_{1} - \Delta_{2}} ,
\eea
with the function $F$ defined in Eq.~(\ref{eq:IandF_infinite_volume}).  The scale $\Delta$ in these functions 
results from the mass difference between the external and the internal heavy hadrons.  In the heavy quark 
and the isospin limits, we have
\bea
  && M_{B^{\ast}_{d}} - M_{B_{u}} = M_{S_{dd}} - M_{S_{du}} = 0. \nonumber
\eea
Therefore the only relevant heavy-hadron mass difference in these limits is
\bea
 && \Delta^{(B)} = M_{S_{du}} - M_{T_{du}} = M_{S_{dd}} - M_{T_{du}} \sim 200 \mbox{ }{\mathrm{MeV}},\nonumber
\eea

and the effective couplings in Eq.~(\ref{eq:effective_couplings}) are
\bea
 \left ( g_{1}\right )_{{\mathrm{eff}}} &=& g_{1} \left [ 1 - \frac{2}{f^{2}} I (M_{\pi}) 
      + \frac{4 g^{2}_{1}}{f^{2}} H(M_{\pi},0)  + {\mathrm{analytic}}\mbox{ }{\mathrm{terms}} \right ], \nonumber\\
&& \nonumber\\
 \left ( g_{2}\right )_{{\mathrm{eff}}} &=& g_{2} \left [ 1 - \frac{2}{f^{2}} I (M_{\pi}) 
      + \frac{3 g^{2}_{2}}{2 f^{2}} H(M_{\pi},0)  
      + \frac{g^{2}_{3}}{f^{2}} \left ( H(M_{\pi},-\Delta^{(B)}) - 2 K(M_{\pi},-\Delta^{(B)},0) \right )  
         + {\mathrm{analytic}}\mbox{ }{\mathrm{terms}}\right ]
      , \nonumber\\
&& \nonumber\\
 \left ( g_{3}\right )_{{\mathrm{eff}}} &=& g_{3} \bigg [ 1 - \frac{2}{f^{2}} I (M_{\pi}) 
      + \frac{g^{2}_{2}}{f^{2}} \left ( -2 H(M_{\pi},\Delta^{(B)}) + H(M_{\pi},0)\right ) \nonumber\\
\label{eq:g_eff_su2_results}
   && \hspace{2.65cm} + \frac{g^{2}_{3}}{2 f^{2}} \left ( H(M_{\pi},-\Delta^{(B)}) 
            + 9 H(M_{\pi},\Delta^{(B)}) -2 K(M_{\pi},\Delta^{(B)},0)\right ) 
       + {\mathrm{analytic}}\mbox{ }{\mathrm{terms}}\bigg ] ,
\eea
with the analytic terms resulting from $\calN^{(2)}$ in Eq.~(\ref{eq:effective_couplings}).
Here we stress that the tadpole diagram  is the dominant one-loop contribution to the chiral expansion 
of $(g_{1})_{{\mathrm{eff}}}$.  This is because the typical value of the coupling, $g_{1}^{2} \sim 0.25$, 
is small, leading to the suppression of
other diagrams in the above equation\footnote{Since $g_{2,3} \sim O(1)$, this suppression is not present in
the chiral expansion of $(g_{2,3})_{{\mathrm{eff}}}$.}. A numerical comparison of the individual 
contributions from different types of Feynman diagrams will be given in Sec.~\ref{sec:individual-contribs}.

Before proceeding with further discussion of the formulae in Eq.~(\ref{eq:g_eff_su2_results}), we notice that
the function $H(m,\Delta)$ can be related to $I(m)$ when $\Delta = 0$,
\beq
\label{eq:H_to_I}
 H(m,0) = -I(m) = -\frac{m^{2}}{16 \pi^{2}} {\mathrm{log}}\left ( \frac{m^{2}}{\mu^{2}} \right ).
\eeq
This leads to the simplification of the chiral expansion of $(g_{1})_{{\mathrm{eff}}}$,
\beq
\label{eq:g_1_eff_simple_form}
  \left ( g_{1}\right )_{{\mathrm{eff}}} = g_{1} \bigg [ 1 - \frac{2}{(4 \pi f)^{2}} 
     M^{2}_{\pi} \: \log \left( \frac{M^{2}_{\pi}}{\mu^2}\right) 
      - \frac{4 g^{2}_{1}}{(4 \pi f)^{2}} M_\pi^2\: \log \left( \frac{M^{2}_{\pi}}{\mu^2}\right)   
     + c(\mu)\:M_\pi^2 \bigg ].
\eeq
The renormalisation-scale dependence from the loop diagrams is cancelled by the coefficient, $c(\mu)$, of the analytic term
which also encodes the contributions from the NLO Lagrangian.

In the following two subsections, we first address an issue related to the chiral limit of the formulae presented
above, and then present an estimation for the numerical size of the one-loop corrections.

\subsection{Wavefunction renormalisation and sunset diagrams in the chiral limit}
\label{sec:chiral_limit_subtraction}
As pointed out in Eqs.~(\ref{eq:HandK_infinite_volume}) and (\ref{eq:g_eff_su2_results}),
the infinite-volume one-loop contributions from the wavefunction renormalisation and sunset
diagrams can be written in terms of the functions $H$ and $K$, which are obtained by taking derivatives of the function 
defined in Eq.~(\ref{eq:IandF_infinite_volume}),
\bea
F(m,\Delta) &=& \frac{-1}{16\pi^{2}} 
 \left [ \left ( m^{2}- \frac{2\Delta^{2}}{3} \right )\Delta 
     {\log\left ( \frac{m^{2}}{\mu^{2}}\right )}  
   + \left ( \frac{10\Delta^{2}}{9} - \frac{4 m^{2}}{3}\right ) \Delta
   + \frac{2 (\Delta^{2}-m^{2})}{3} m R\left ( \frac{\Delta}{m}\right )
\right ], \mbox{ }{\mathrm{where}}\nonumber\\
&& \nonumber\\
R(x) &\equiv& 
 \sqrt{x^{2}-1}\mbox{ }\left [ 
       {\mathrm{log}}\left (  x - \sqrt{x^{2}-1+i\epsilon} \right )
     - {\mathrm{log}}\left (  x + \sqrt{x^{2}-1+i\epsilon} \right ) \right ] \nonumber.
\eea
This function is obtained by regularising the loop integrals with the subtraction scheme defined in 
Eq.~(\ref{eq:lambdabar}) in Appendix~\ref{app:integral}.  
Implementing this scheme is a common practice in $\chi$PT calculations~\cite{Gasser:1985gg}.
It leads to the result that $F(m,\Delta)$ does not vanish in the limit $m\rightarrow 0$ unless $\Delta=0$.  
Such behaviour does not cause any conceptual problem in the effective theory, 
since the axial couplings, $g_{1,2,3}$, can undergo finite renormalisation depending on the subtraction scheme used to
regulate one-loop integrals.  Various subtraction schemes always lead to 
the same physical quantities, such as the hadronic masses and axial transition amplitudes, which are scheme-independent.  
On the other hand, it would be desirable and natural to choose a scheme in which the one-loop contributions
decouple in the chiral limit.
As pointed out in Refs.~\cite{Tiburzi:2005na,WalkerLoud:2006sa}, it is possible to find a scheme such that the real part of $F$ vanishes in the chiral limit.  It is implemented by simply rewriting $F$ as
\beq
\label{eq:F_prime}
 F^{\rm (sub)}(m,\Delta) = \frac{-1}{16\pi^{2}} 
 \left [ m^{2}\Delta 
     {\log\left ( \frac{m^{2}}{\mu^{2}}\right )}  
         - \frac{2 \Delta^{3}}{3} {\log\left ( \frac{m^{2}}{4 \Delta^{2}}\right )}
     - \frac{4 m^{2}\Delta}{3} 
   + \frac{2 (\Delta^{2}-m^{2})}{3} m R\left ( \frac{\Delta}{m}\right )
\right ] ,
\eeq
and appropriately modifying the counterterms to absorb the difference
(a finite polynomial in $\Delta$).  It is straightforward to demonstrate that when $\Delta+m > 0$, in which case
the external heavy hadrons are stable particles, this function is real and
\beq
\label{eq:chiral_limit_F_prime}
 \mathop{{\mathrm{lim}}}_{m\rightarrow 0} F^{\rm (sub)}(m,\Delta) = 0.
\eeq

In the case $\Delta + m < 0$ which corresponds to the situation that the external heavy hadron becomes unstable, the functions $F$ and $F^{\rm (sub)}$ are complex.
Although the real part of $F^{\rm (sub)}$ vanishes in the chiral limit, the imaginary part remains non-zero.
This occurs when
\beq
\label{eq:unstable_threshold}
  M_{\pi} < |M_{T_{ij}} - M_{S^{\mu}_{ij}}| = \Delta^{(B)} \sim 200 \mbox{ }{\mathrm{MeV}} .
\eeq
Below this threshold, one cannot define matrix elements containing external $S^{\mu}_{ij}$ hadrons.  In principle, more complicated matrix elements can be used to determine the couplings $g_{2}$ and $g_{3}$ for the pion masses in the regime of Eq.~(\ref{eq:unstable_threshold}), but this is beyond the scope of this work.  See
Refs.~\cite{Lellouch:2000pv,Bernard:2007cm} for related discussions.  Here we stress that one can perform lattice
calculations in the regime where the pion mass is larger than
$\Delta^{(B)}$ but small compared to the chiral symmetry breaking scale, such that the external hadrons are all stable
and the chiral expansion is still valid. These calculations enable the extraction the axial couplings, $g_{1,2,3}$, which
can then be used to perform chiral extrapolations and make predictions for other quantities.

\subsection{Evaluation of individual contributions}
\label{sec:individual-contribs}
\begin{figure}[!t]
  \centering
  \includegraphics[width=0.5\columnwidth]{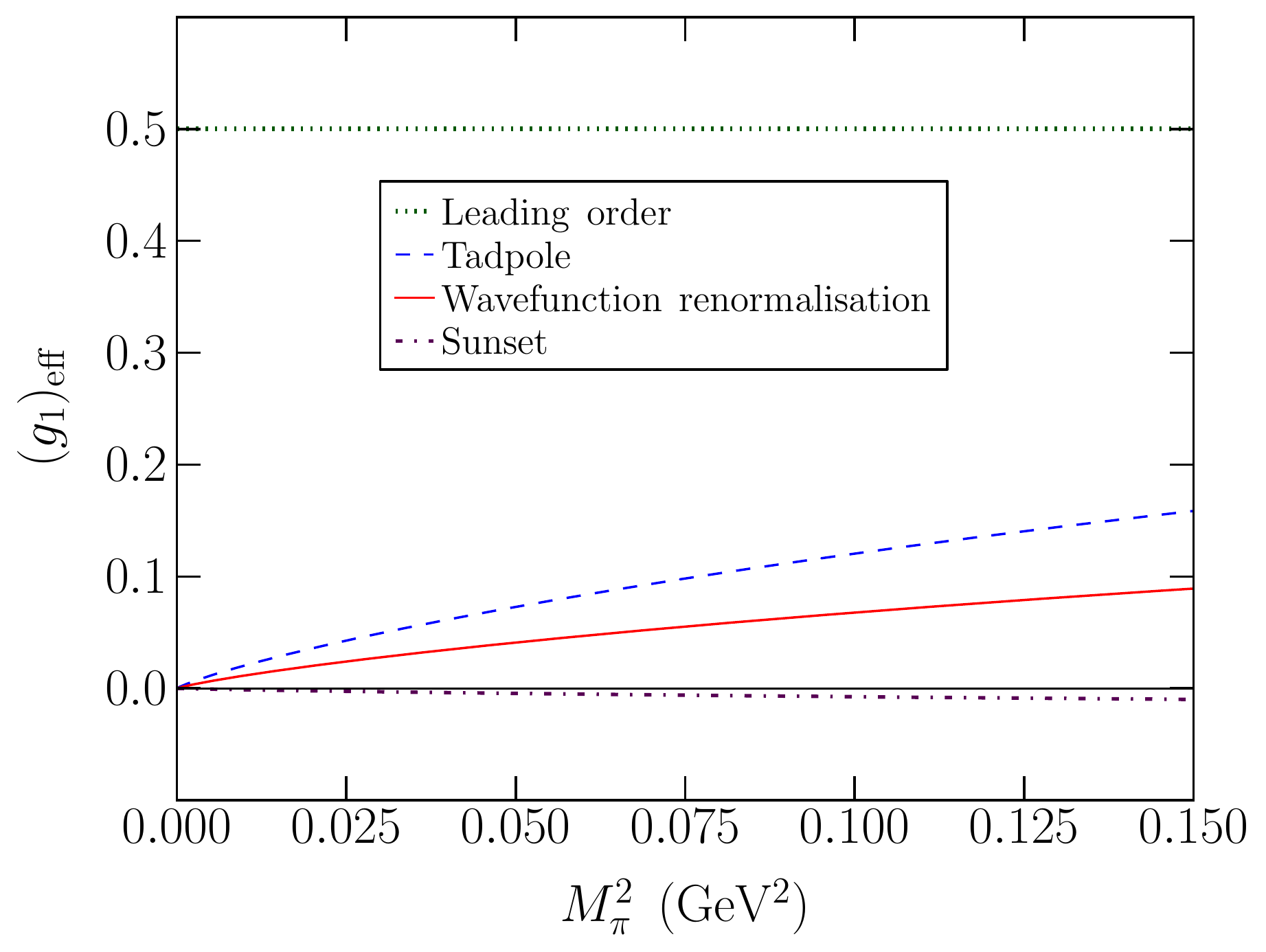}
  \includegraphics[width=0.5\columnwidth]{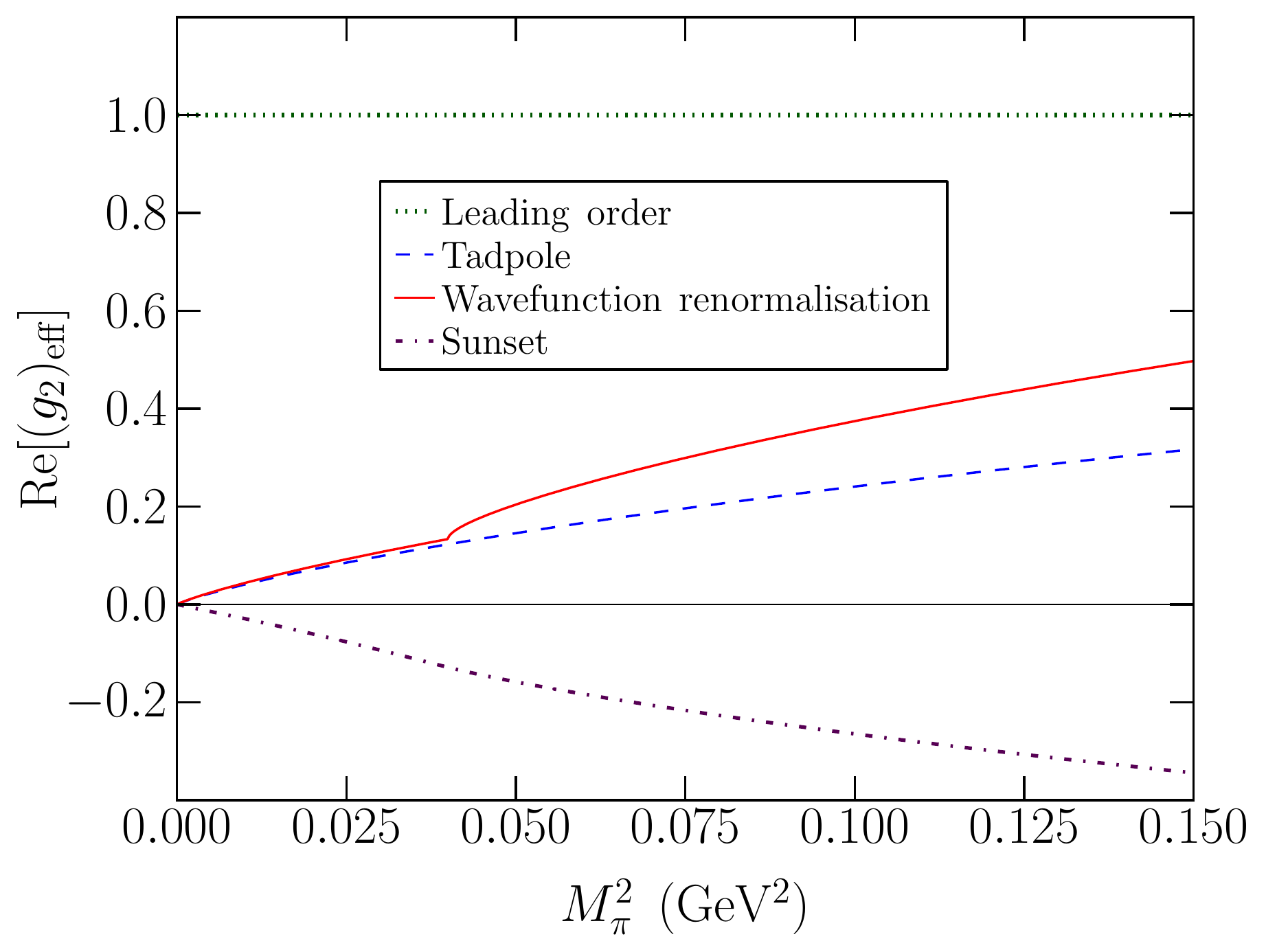}
  \includegraphics[width=0.5\columnwidth]{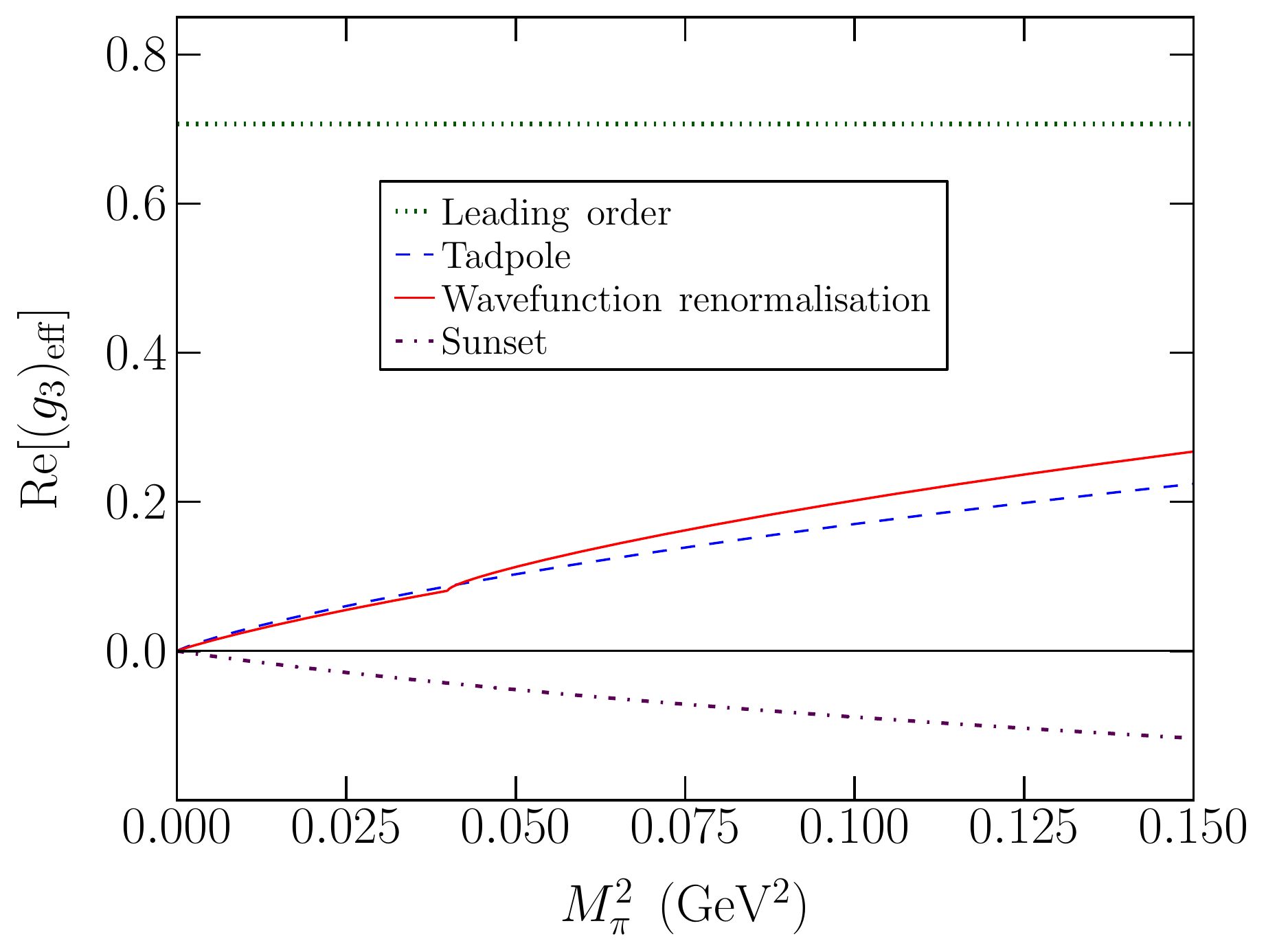} 
  \caption{Comparison of the individual infinite-volume one-loop contributions to the
    pion mass dependence of the (real part of the) various effective couplings,
    $\left ( g_{1,2,3} \right )_{{\mathrm{eff}}}$ evaluated using the values of the LECs given in the text.
The kinks in the wavefunction renormalisation and sunset contributions to the baryonic couplings
arise from the $S\rightarrow T\pi$ threshold at $M_\pi=\Delta^{(B)}$. Below this threshold, the curves 
lose their physical interpretation.  The
subtraction scheme is that presented in Eq.~(\ref{eq:F_prime}).}
  \label{fig:indiviual-contribs}
\end{figure}

In this subsection, we use the simple infinite volume, SU(2) case to explore the typical 
size of the one-loop contributions.  This can be best summarised by the plots in Fig.~\ref{fig:indiviual-contribs}.
In these plots, the pion mass dependence of the
loop contributions to
three effective axial couplings [their real part in the case of $\left ( g_{2,3}\right )_{{\mathrm{eff}}}$] is shown for
exemplary  values of the various low energy constants. 
These results are obtained using the subtraction scheme defined in Eq.~(\ref{eq:F_prime}) 
in Sec.~\ref{sec:chiral_limit_subtraction}.  The leading
order contribution is also shown.  We take $g_1=0.5$, a value consistent
with recent determinations
\cite{deDivitiis:1998kj,Abada:2003un,Ohki:2008py,Becirevic:2009yb,Bulava:2010ej,Bulava:2011yz}
and then 
use the quark model expectations for the other couplings, $g_2=2g_1$ and $g_3=\sqrt{2}g_1$ (in
our normalisation) \cite{Yan:1992gz}\footnote{These values are also consistent with preliminary lattice QCD results~\cite{our_work:2011}.}
which are far less constrained. We work in the heavy-quark limit so that $\Delta^{(M)}=0$,
and we have set the $S-T$ mass differences to $\Delta^{(B)}=200$ MeV, consistent with
experiment~\cite{Aaltonen:2007rw}.
The renormalization scale used here is $\mu=4\pi f$.

It is clear from these figures that the tadpole contributions provide an
important part of the chiral non-analytic behaviour of the axial
couplings. Furthermore, in the range of pion masses considered here, $M_\pi\alt$ 400 MeV,
the NLO contributions from loops are numerically small corrections to the leading-order
results.\footnote{The imaginary parts of $(g_{2,3})_{\rm eff}$ that arise for $M_\pi<\Delta^{(B)}$ are also small, $|\Im(g_{2,3})_{\rm eff}|<0.05$.}
 This indicates that in this range the $SU(2)$ chiral expansion of the axial-current matrix elements
is well-behaved. Variations of the low-energy constants, $g_1$, $g_2$, $g_3$, $\Delta^{(B)}$, and the renormalisation
scale, $\mu$, within reasonable ranges do not substantially alter the behaviour shown in Fig.~\ref{fig:indiviual-contribs}.

\section{One-loop contributions in SU(4$|$2) and SU(6$|$3) HH$\chi$PT}
\label{sec:one-loop-contr-su42-su63}
In this section, we study the one-loop contributions in Eq.~(\ref{eq:NLO-master}) in SU(4$|$2) and SU(6$|$3) partially quenched HH$\chi$PT
in finite volume.
These results are complicated because we keep the SU(3) light-flavour breaking effects from Eq.~(\ref{eq:HHChPT_lagrangian_chi}) in our calculation.
Here we investigate the structure of the one-loop computation via analysing the quark flavour flow picture~\cite{Sharpe:1992ft}.  The details of 
the results are given in Appendices~\ref{app:waverenorm_tables} and \ref{app:sunset_tables}.

\subsection{The tadpole diagrams}
\label{sec:one_loop_tadpole}
First we present the contributions from the tadpole diagrams.  These take the simple form,
\bea
 \calT^{(2)}_{ud} &=& -2 \calI(M_{u,u^{\prime}}) ,\nonumber\\
&& \nonumber\\
 \calT^{(3)}_{ud} &=& -2 \calI(M_{u,u^{\prime}}) - \calI(M_{u,s^{\prime}}) ,\nonumber\\
&&\nonumber\\
\label{eq:all_tadpole}
 \calT^{(3)}_{us} &=& -\calI(M_{u,u^{\prime}}) - \frac{1}{2} \calI(M_{u,s^{\prime}}) - \calI(M_{s,u^{\prime}})
          -\frac{1}{2} \calI(M_{s,s^{\prime}}) + \frac{1}{6} \tilde{\calI}_{3}(M_{u,u})
          - \frac{1}{3} \tilde{\calI}_{3}(M_{u,s}) + \frac{1}{6} \tilde{\calI}_{3}(M_{s,s}),
\eea
where the functions $\calI$ and $\tilde{\calI}_{3}$ are defined in (\ref{eq:I}) and (\ref{eq:SU63_DP_IHK}), respectively. The tadpole diagram
results are completely determined by the structure of the axial currents.

\subsection{The self-energy diagrams}
\label{sec:one_loop_waverenorm}

In this subsection, we present the heavy hadron wavefunction renormalisation, resulting from 
the self-energy diagrams. These are more complicated 
than the tadpole-diagram results in
our calculation, since we keep track of the flavour SU(3) breaking effects from both the Goldstone 
masses [Eq.~(\ref{eq:goldstone_chiral_lagrangian})] and the 
heavy-meson and baryon spectrum [Eq.~(\ref{eq:HHChPT_lagrangian_chi})].  It is helpful to analyse the quark 
flavour flow diagrams~\cite{Sharpe:1992ft} to understand the structure of the results.  
To investigate this structure, we first assign a ``direction" to each flavour flow line:
\begin{itemize}
 \item The flow following the direction of a line means a quark with that flavour, while the flow against the
       direction means its anti-quark.
\end{itemize}
For the analysis of the heavy meson wavefunction renormalisation, we follow the nomenclature for the 
coefficients in front of the sum (integral) in a loop diagram:
\begin{itemize}
 \item The ``tilded'' coefficients accompany the ``hairpin'' contributions from the light flavour-singlet mesons. 
 \item The ``primed'' coefficients multiply the sums in which a $B$ meson appears in the loop, while the ``unprimed"
       coefficients are for the cases involving an internal $B^{\ast}$ meson.
\end{itemize}

\begin{figure}[!t]
  \centering
\includegraphics[width=0.3\columnwidth]{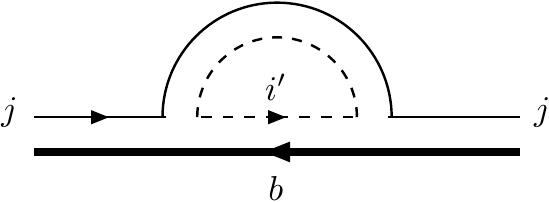}
\hspace*{13mm}
\includegraphics[width=0.3\columnwidth]{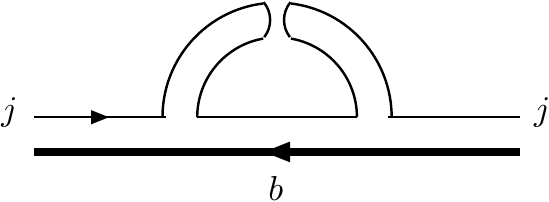} 
\\
\vspace{0.5cm}
\hspace*{-0.00cm}(a)\hspace*{6.50cm}(b)
 \caption{Quark flavour flow structure for the meson self-energy diagrams. 
   The thick line represents the anti-$b$ quark, and the thin 
   lines are the valence light quarks, while the dashed line is the sea light quark.
   Diagram (a) contributes to the $w$ and $w^{\prime}$ terms in Eq.~(\ref{eq:meson_waverenorm_master})
   when the internal heavy-light meson is $B^{\ast}_{i^{\prime}}$ and $B_{i^{\prime}}$ respectively, while
   the Goldstone meson is composed of a $j$ valence quark and an $i^{\prime}$ sea anti-quark.
   Diagram (b) is the ``hairpin" structure and results in terms containing $\tilde{w}$ 
   (internal $B^{\ast}_{j}$) and
   $\tilde{w}^{\prime}$ (internal $B_{j}$) in Eq.~(\ref{eq:meson_waverenorm_master}).
  \label{fig:qFlow_waverenorm_meosns} }
\end{figure}

The quark flow picture for the heavy meson wavefunction renormalisation diagrams is presented in Fig.~\ref{fig:qFlow_waverenorm_meosns}.
Since there is only one valence light quark involved, and the internal valence-quark loops are cancelled by the
ghost-quark loops, the only possible non-hairpin structure is from the sea-quark contributions.  This is depicted in
Fig.~\ref{fig:qFlow_waverenorm_meosns} (a), where the Goldstone meson is composed of a $j$ valence quark, and an $i^{\prime}$ sea anti-quark.  The hairpin contribution is presented in Fig.~\ref{fig:qFlow_waverenorm_meosns} (b).

Following the above nomenclature and the quark flavour flow picture in Fig.~\ref{fig:qFlow_waverenorm_meosns}, 
the results for the heavy-meson wavefunction renormalisation can be written as,
\bea
 \calW^{(N_{f})}_{B_{j}} &=& g_{1}^{2} \sum_{a} \bigg [ w^{(N_{f})}_{B_{j},a} \mbox{ }\calH(M_{j,a}, \Delta^{(M)} + \delta^{(M)}_{a,j})
                 + \tilde{w}^{(N_{f})}_{B_{j},a} \mbox{ }\tilde{\calH}_{N_{f}}(M_{j,a}, \Delta^{(M)}) \bigg ]
          ,\nonumber\\
&&\nonumber\\
 \calW^{(N_{f})}_{B_{j}^{\ast}} &=& g_{1}^{2}  \sum_{a} \bigg [ w^{\prime (N_{f})}_{B_{j}^{\ast},a} 
                                 \mbox{ }\calH(M_{j,a}, - \Delta^{(M)} + \delta^{(M)}_{a,j})
                 + \tilde{w}^{\prime (N_{f})}_{B_{j}^{\ast},a} \mbox{ }\tilde{\calH}_{N_{f}}(M_{j,a}, - \Delta^{(M)})\nonumber\\
\label{eq:meson_waverenorm_master}
    &&   \hspace{0.85cm} + \mbox{ }w^{(N_{f})}_{B_{j}^{\ast},a} 
                                 \mbox{ }\calH(M_{j,a}, \delta^{(M)}_{a,j})
                 + \tilde{w}^{(N_{f})}_{B_{j}^{\ast},a} \mbox{ }\tilde{\calH}_{N_{f}}(M_{j,a}, 0) \bigg ] ,
\eea
where the summations are over the flavours $u$ and $u^{\prime}$ in the SU(4$|$2) theory, and are over the flavours $u$, $s$, $u^{\prime}$ 
and $s^{\prime}$ in the SU(6$|$3) theory.
The functions $\calH$ and $\tilde{\calH}_{N_{f}}$ are results of the sums (integrals) involved in the loops, and are 
defined in Eqs.~(\ref{eq:def_calH}),
(\ref{eq:DP_IHK}), (\ref{eq:SU42_DP_IHK}) and (\ref{eq:SU63_DP_IHK}) in Appendix~\ref{app:integral}.  
The mass parameters $M_{j,a}$, $\Delta^{(M)}$ and
$\delta^{(M)}_{a,j}$ are defined in Eq.~(\ref{eq:mass_param}) in Appendix~\ref{app:masses_in_loops}.
The coefficients $w$, $w^{\prime}$, $\tilde{w}$ and $\tilde{w}^{\prime}$ are presented in
Table~\ref{tab:meson_waverenorm_table} in Appendix~\ref{app:waverenorm_tables}.

\begin{figure}[!t]
  \centering
\includegraphics[width=0.4\columnwidth]{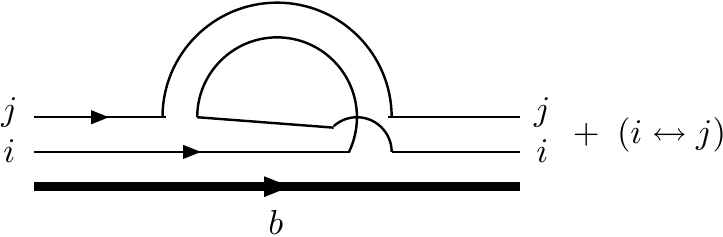}
\hspace*{13mm}
\includegraphics[width=0.4\columnwidth]{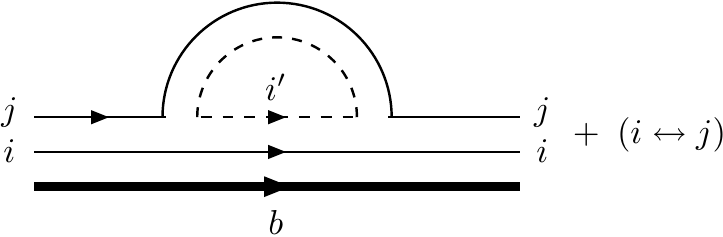} 
\\
\vspace{0.5cm}
\hspace*{-1.80cm}(a)\hspace*{8.30cm}(b)
 \caption{Quark flavour flow structure for the baryon self-energy diagrams without 
   the ``hairpin" structure.  The thick line represents the $b$ quark, and the thin 
   lines are the valence light quarks, while the dashed line is the sea light quark. 
   Diagram (a) is the "crossing" type which does not involve sea quark contributions.  
   The explicitly-shown diagrams give rise to the terms multiplied by $w$ and $w^{\prime}$ in 
   Eq.~(\ref{eq:baryon_waverenorm_master}) when the internal baryons are $S_{ai}$ and 
   $T_{ai}$ respectively [$a=i$ in diagram (a) and $a=i^{\prime}$ in diagram (b)].  
   Interchanging the flavour indices $i$ and $j$ 
   leads to the corresponding $u$ and $u^{\prime}$ terms in the same equation.
  \label{fig:qFlow_waverenorm_baryons} }
\end{figure}

\begin{figure}[!t]
  \centering
\includegraphics[width=0.3\columnwidth]{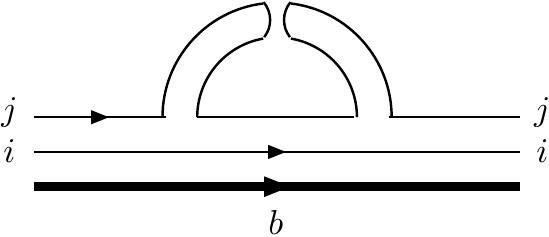}
\hspace*{5mm}
\includegraphics[width=0.3\columnwidth]{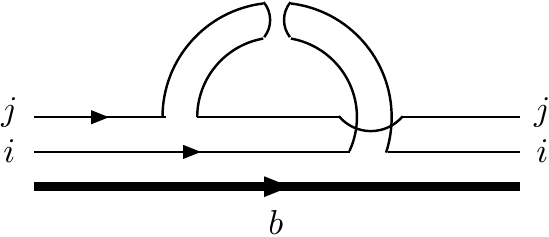} 
\hspace*{5mm}
\includegraphics[width=0.3\columnwidth]{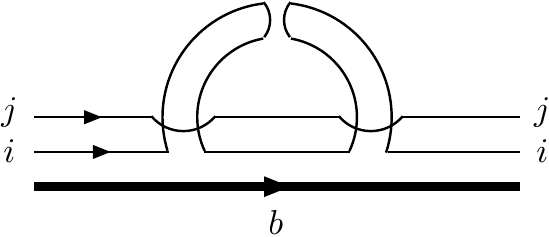} 
\\
\vspace{0.5cm}
\hspace*{-0.00cm}(a)\hspace*{5.7cm}(b)\hspace*{5.7cm}(c)
  \caption{Quark flavour flow structure for the baryon self-energy diagrams involving the ``hairpin" structure.  The thick line represents the $b$ quark, and the thin 
   lines are the valence light quarks.  Diagram (a) contributes to the $\tilde{w}$ (internal $S_{ij}$ baryon) 
   and $\tilde{w}^{\prime}$ (internal $T_{ij}$ baryon) terms in
   Eq.~(\ref{eq:baryon_waverenorm_master}), while Diagrams (b) and (c) result in the $\tilde{u}$ (internal $S_{ij}$ baryon) and $\tilde{u}^{\prime}$
   (internal $T_{ij}$ baryon) terms in the same equation.}
  \label{fig:qFlow_waverenorm_baryons_hairpin}
\end{figure}

Next, we discuss the structure of the baryon self-energy diagrams.  We start by modifying the above rule for
assigning the ``primed" coefficients,
\begin{itemize}
 \item The ``primed'' coefficients multiply the sums in which the $T$ baryon appears in the loop, 
       while the ``unprimed" coefficients are for the cases involving the internal $S$ baryon.
\end{itemize}
These diagrams are further complicated by the presence of two light
valence quarks.  To keep track of the flavour flow of these two quarks, we introduce an additional 
rule to our notation,
\begin{itemize}
 \item For a baryon ($T_{ij}$ or $S_{ij}$) of light flavour indices $i$ and $j$, we assign the 
       coefficient $u$ to the diagram if the quark carrying flavour $i$ appears in the Goldstone meson.  
       For all the other
       cases, including the appearance of the anti-$i$ in the Goldstone meson, they are accompanied by
       the coefficient $w$.
\end{itemize}
The flavour flow structure for the baryon self-energy diagrams can be summarised in 
Figs.~\ref{fig:qFlow_waverenorm_baryons} and \ref{fig:qFlow_waverenorm_baryons_hairpin}.
For the diagrams explicitly shown in Fig.~\ref{fig:qFlow_waverenorm_baryons} (a) and (b),
the Goldstone mesons are composed of ($j$, anti-$i$) and ($j$, anti-$i^{\prime}$) respectively.  Therefore, they
are accompanied by the $w$-type coefficients ($w$ for the internal $S$ baryon and $w^{\prime}$ for the 
internal $T$ baryon).  Terms with the $u$-type
coefficients are obtained by exchanging the flavours $i$ and $j$, as also indicated in this figure.
Notice that the  ``non-hairpin" valence-valence Goldstone contributions appear via the 
"crossing" configuration in Fig~\ref{fig:qFlow_waverenorm_baryons} (a).
The ``hairpin" structure of the baryon self-energy diagrams is presented in 
Fig.~\ref{fig:qFlow_waverenorm_baryons_hairpin}.  From the above rules, it is clear that
the diagram in Fig.~\ref{fig:qFlow_waverenorm_baryons_hairpin} (a) leads to a 
$\tilde{w}$-type term, while those in Fig.~\ref{fig:qFlow_waverenorm_baryons_hairpin} (b) and (c)
are multiplied by $\tilde{u}$-type coefficients.

Following the above discussion, we obtain the results for the baryon wavefunction renormalisation,
\bea
 \calW^{(N_{f})}_{T_{ij}} &=&  g_{3}^{2}  \sum_{a} \bigg [ w^{(N_{f})}_{T_{ij},a} \mbox{ }\calH(M_{j,a}, \Delta^{(B)} + \delta^{(B)}_{ai,ij})
                 + \tilde{w}^{(N_{f})}_{T_{ij},a} \mbox{ }\tilde{\calH}_{N_{f}}(M_{j,a}, \Delta^{(B)}) \nonumber\\
    &&   \hspace{0.85cm} + \mbox{ } u^{(N_{f})}_{T_{ij},a} \mbox{ }\calH(M_{i,a}, \Delta^{(B)} + \delta^{(B)}_{aj,ij})
                 + \tilde{u}^{(N_{f})}_{T_{ij},a} \mbox{ }\tilde{\calH}_{N_{f}}(M_{i,a}, \Delta^{(B)})
           \bigg ] , \nonumber\\
&&\nonumber\\
 \calW^{(N_{f})}_{S_{ij}} &=&  g_{2}^{2}  \sum_{a} \bigg [ w^{(N_{f})}_{S_{ij},a} \mbox{ }\calH(M_{j,a}, \delta^{(B)}_{ai,ij})
                 + \tilde{w}^{(N_{f})}_{S_{ij},a} \mbox{ }\tilde{\calH}_{N_{f}}(M_{j,a}, 0) \nonumber\\
    &&   \hspace{0.85cm} + \mbox{ }  u^{(N_{f})}_{S_{ij},a} \mbox{ }\calH(M_{i,a}, \delta^{(B)}_{aj,ij})
                 + \tilde{u}^{(N_{f})}_{S_{ij},a} \mbox{ }\tilde{\calH}_{N_{f}}(M_{i,a}, 0)
           \bigg ] \nonumber\\
    &&  \hspace{-0.25cm}   +  g_{3}^{2}  \sum_{a} \bigg [ w^{\prime (N_{f})}_{S_{ij},a} \mbox{ }\calH(M_{j,a}, -\Delta^{(B)} + \delta^{(B)}_{ai,ij})
                 + \tilde{w}^{\prime (N_{f})}_{S_{ij},a} \mbox{ }\tilde{\calH}_{N_{f}}(M_{j,a}, -\Delta^{(B)}) \nonumber\\
\label{eq:baryon_waverenorm_master}
    &&   \hspace{0.85cm} + \mbox{ }  u^{\prime (N_{f})}_{S_{ij},a} \mbox{ }\calH(M_{i,a}, -\Delta^{(B)} + \delta^{(B)}_{aj,ij})
                 + \tilde{u}^{\prime (N_{f})}_{S_{ij},a} \mbox{ }\tilde{\calH}_{N_{f}}(M_{i,a}, -\Delta^{(B)})
           \bigg ]  ,
\eea
where the summations are over the flavours $u$ and $u^{\prime}$ in the SU(4$|$2) theory, and are over the flavours $u$, $s$, $u^{\prime}$ 
and $s^{\prime}$ in the SU(6$|$3) theory.
The relevant coefficients, $w$, $u\ldots$ are presented in Tables~\ref{tab:T_waverenorm_table} 
and \ref{tab:S_waverenorm_table} in Appendix~\ref{app:waverenorm_tables}. The
mass parameters $M_{a,b}$, $\Delta^{(B)}$ and $\delta^{(B)}_{ab,cd}$ are defined in Eq.~(\ref{eq:mass_param}) in
Appendix~\ref{app:masses_in_loops}.  These results  agree with those in the literature~\cite{Tiburzi:2004kd}\footnote{The SU(3)
breaking effects arising from Eq.~(\ref{eq:HHChPT_lagrangian_chi}) 
are not included in the results in Ref.~\cite{Tiburzi:2004kd}. We have also checked these wavefunction renormalisation diagrams 
against the full-QCD, SU(3)-limit results at $\Delta^{(B)}=0$ in Ref.~\cite{Cho:1992cf}, and found agreement. }.

\subsection{The sunset diagrams}
\label{sec:one_loop_sunset}

In this subsection, we discuss the structure of the sunset diagrams.  
The Lorentz indices carried by the hadronic states are completely absorbed
into the tree-level contribution in Eq.~(\ref{eq:NLO-master}), therefore they are omitted in the
notation below. In order to organise the results, we follow the same convention in assigning
the ``flow direction" to a quark line and the ``tilded" coefficients to the terms involving
the ``hairpin" structure, as that in the self-energy diagrams.

\begin{figure}[!t]
  \centering
\includegraphics[width=0.3\columnwidth]{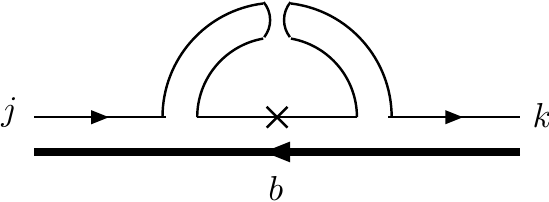} 
\\
 \caption{Quark flavour flow structure for the meson sunset diagrams with external $B_{j}$ and 
   $B^{\ast}_{k}$ states. 
   The thick line represents the anti-$b$ quark, and the thin 
   line is the valence light quark.  The cross is the current $J_{jk}$.  This diagram results in the 
   $\tilde{y}$ (internal $B^{\ast}_{j}$ and $B^{\ast}_{k}$) and $\tilde{y}^{\prime}$ 
   (internal $B^{\ast}_{j}$ and $B_{k}$) terms in Eq.~(\ref{eq:meson_sunset_master}). This is the
   only possible quark flavour flow configuration for the meson sunset diagrams.
  \label{fig:qFlow_sunset_meosns} }
\end{figure}

First we study the sunset diagram for the axial-current matrix element between the $B_{j}$ and
$B^{\ast}_{k}$ mesons. Because of the flavour-changing structure of the currents that we consider in this work, 
it is straightforward
to demonstrate that in this case the Goldstone meson must involve the ``hairpin" contribution. This
is depicted in Fig.~\ref{fig:qFlow_sunset_meosns}.  
Furthermore, the internal heavy meson with the light flavour $j$ must be a $B^{\ast}_{j}$ since there is
no $B{-}B{-}$Goldstone coupling in the Lagrangian or the current.  On the other hand, the
internal heavy meson involving the light flavour $k$ can be either $B_{k}$ or $B^{\ast}_{k}$.
These two cases are distinguished by the ``primed" and the ``unprimed" coefficients in the results.
We then obtain the sunset-diagram contribution to this matrix element as
\beq
\label{eq:meson_sunset_master}
 \calQ^{(N_{f})}_{B_{j}\rightarrow B^{\ast}_{k}} =  g_{1}^{2} \bigg [
       \tilde{y}^{(N_{f})}_{B_{j}, B^{\ast}_{k}} \mbox{ }\tilde{\calK}_{N_{f}} (M_{j,k}, \Delta^{(M)}, \Delta^{(M)}+\delta^{(M)}_{k,j})
       + \tilde{y}^{\prime (N_{f})}_{B_{j}, B^{\ast}_{k}} \mbox{ } \tilde{\calK}_{N_{f}} (M_{j,k}, \Delta^{(M)}, \delta^{(M)}_{k,j}) 
            \bigg ] ,
\eeq
where $\calK$ and $\tilde{\calK}_{N_{f}}$ are the sums (integrals) involved in the loops, and are 
defined in Eqs.~(\ref{eq:def_calH}),
(\ref{eq:DP_IHK}), (\ref{eq:SU42_DP_IHK}) and (\ref{eq:SU63_DP_IHK}) in Appendix~\ref{app:integral}.  
The mass parameters 
$M_{j,k}$, $\Delta^{(M)}$ and
$\delta^{(M)}_{k,j}$ are defined in Eq.~(\ref{eq:mass_param}) in Appendix~\ref{app:masses_in_loops}.
The coefficients $\tilde{y}$ and $\tilde{y}^{\prime}$ are presented in
Table~\ref{tab:y_tilde} in Appendix~\ref{app:sunset_tables}.

\begin{figure}[!t]
  \centering
\includegraphics[width=0.3\columnwidth]{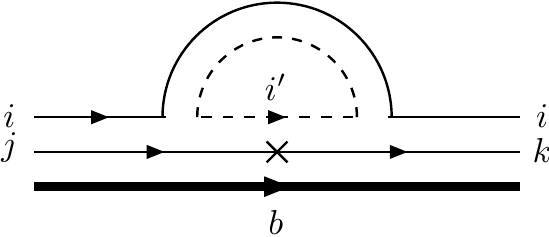} 
\hspace*{5mm}
\includegraphics[width=0.3\columnwidth]{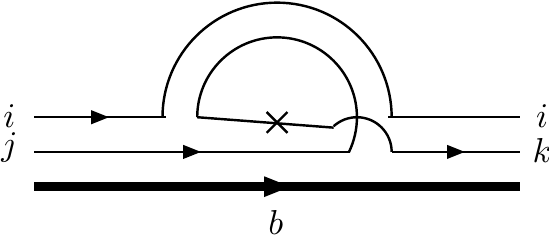} 
\hspace*{5mm}
\includegraphics[width=0.3\columnwidth]{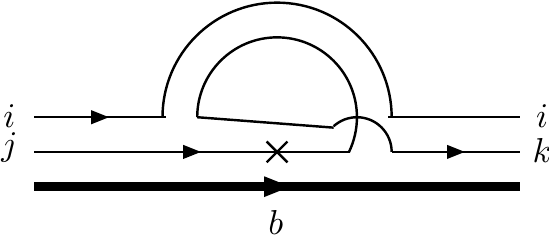} 
\\
\vspace{0.5cm}
\hspace*{-0.00cm}(a)\hspace*{5.7cm}(b)\hspace*{5.7cm}(c)
 \caption{Quark flavour flow structure for the baryon sunset diagrams without the ``hairpin" structure. 
   The thick line represents the $b$ quark, and the thin 
   lines are the valence light quarks, while the dashed line is the sea light quark.  The cross is the current $J_{jk}$.
These diagrams lead to the $x$ (internal $S_{aj}$ and $S_{ak}$ baryons with $a = i^{\prime}, j, k$ in diagram (a), (b), (c) respectively), 
  $x^{\prime}$ (internal $T_{aj}$ and $S_{ak}$ baryon ) and $x^{\prime\prime}$
(internal $S_{aj}$ and $T_{ak}$ baryons) terms in Eq.~(\ref{eq:baryon_sunset_master}).
  \label{fig:qFlow_sunset_baryon} }
\end{figure}

\begin{figure}[!t]
  \centering
\includegraphics[width=0.3\columnwidth]{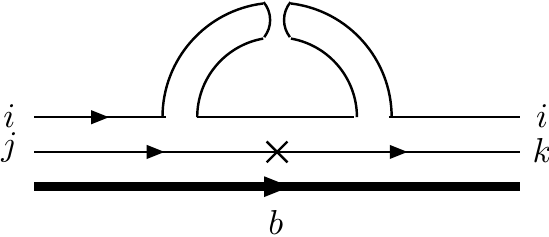} 
\hspace*{5mm}
\includegraphics[width=0.3\columnwidth]{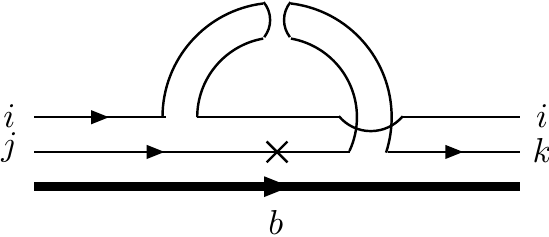} 
\hspace*{5mm}
\includegraphics[width=0.3\columnwidth]{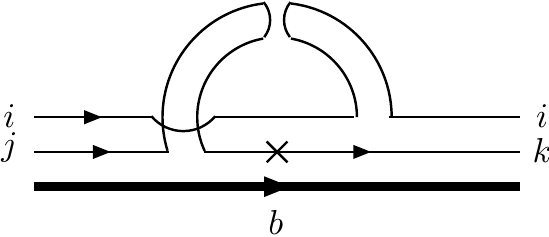} 
\\
\vspace{0.5cm}
\hspace*{-0.00cm}(a)\hspace*{5.7cm}(b)\hspace*{5.7cm}(c)
 \caption{Quark flavour flow structure for the baryon sunset diagrams with the ``hairpin" structure involving the spectator quark (flavour $i$). 
   The thick line represents the $b$ quark, and the thin 
   lines are the valence light quarks.  The cross is the current $J_{jk}$.   These diagrams lead to the $\tilde{x}$ (internal $S_{ij}$ and $S_{ik}$ baryons),
$\tilde{x}^{\prime}$ (internal $T_{ij}$ and $S_{ik}$ baryon ) and $\tilde{x}^{\prime\prime}$ (internal $S_{ij}$ and $T_{ik}$ baryons) terms in Eq.~(\ref{eq:baryon_sunset_master}).
  \label{fig:qFlow_sunset_baryon_hairpin_with_spectator} }
\end{figure}

\begin{figure}[!t]
  \centering
\includegraphics[width=0.3\columnwidth]{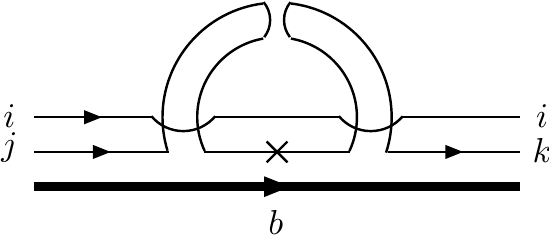} 
\\
 \caption{Quark flavour flow structure for the baryon sunset diagrams with the ``hairpin" structure in which the spectator quark (flavour $i$) is absent. 
   The thick line represents the $b$ quark, and the thin 
   lines are the valence light quarks.  The cross is the current $J_{jk}$.   These diagrams lead to the $\tilde{y}$ (internal $S_{ij}$ and $S_{ik}$ baryons),
$\tilde{y}^{\prime}$ (internal $T_{ij}$ and $S_{ik}$ baryon ) and $\tilde{y}^{\prime\prime}$ (internal $S_{ij}$ and $T_{ik}$ baryons) terms in Eq.~(\ref{eq:baryon_sunset_master}).
  \label{fig:qFlow_sunset_baryon_hairpin_without_spectator} }
\end{figure}

Next, we investigate the sunset diagrams for the following axial-current transitions involving baryons,
\bea
 T_{ij} \longrightarrow S^{\mu}_{ik}\nonumber\\
&&\nonumber\\
\label{eq:baryon_transitions}
 S^{\mu}_{ij} \longrightarrow S^{\nu}_{ik} ,
\eea
where the spectator quark carries the flavour index $i$.  The quark flavour flow configurations are
shown in Figs.~\ref{fig:qFlow_sunset_baryon}, ~\ref{fig:qFlow_sunset_baryon_hairpin_with_spectator} 
and \ref{fig:qFlow_sunset_baryon_hairpin_without_spectator}.  Again, we use the ``tilded" coefficients
to denote terms in which the ``hairpin" structure appears. Because of parity, there are no axial couplings
amongst even number of Goldstone mesons, therefore the flavour indices $j$ and 
$k$ must appear in the internal baryons. These internal
baryons can be $T$ or $S$ type.  Denoting the other flavour index in the loop by $a$, we adopt the following convention 
to distinguish various possibilities for the internal $S$ and $T$ contributions:
\begin{itemize}
 \item If the internal baryons are $S_{aj}$ and $S_{ak}$, then the coefficient for the diagram is ``unprimed".
 \item If the internal baryons are $T_{aj}$ (left in the loop) and $S_{ak}$ (right in the loop), then the coefficient for the diagram is ``primed".  Such terms
are absent in the $T\rightarrow S$ transition amplitudes.
 \item If the internal baryons are $S_{aj}$ (left in the loop) and $T_{ak}$ (right in the loop), then the coefficient for the diagram is ``double-primed".
\end{itemize}

To keep track of the flow of the spectator quark $i$ in these processes, we follow the rules:
\begin{itemize}
 \item If the spectator quark flavour is present in the Goldstone meson, then the diagram corresponds to a term with 
       $x{-}$type coefficient ($x$, $x^{\prime}$, $x^{\prime\prime}$, $\tilde{x}$, 
       $\tilde{x}^{\prime}$ or $\tilde{x}^{\prime\prime}$).
 \item If the spectator quark flavour is absent in the Goldstone meson, then the diagram corresponds to a term with 
       $y{-}$type coefficient.
\end{itemize}

In Fig.~\ref{fig:qFlow_sunset_baryon}, we show the quark flavour flow diagrams containing no
``hairpin" structure.  In such flow configurations, the spectator quark flavour always appears in the 
Goldstone meson.  Therefore they will only be accompanied by the $x{-}$type coefficients.  Notice that
the valence-valence Goldstone mesons also appear in these diagrams via the ``crossing" 
configurations in Figs.~\ref{fig:qFlow_sunset_baryon} (b) and (c).
The ``hairpin" contributions to the quark flavour flow configurations for the processes in 
Eq.~(\ref{eq:baryon_transitions}) are shown in Figs.~\ref{fig:qFlow_sunset_baryon_hairpin_with_spectator} 
and \ref{fig:qFlow_sunset_baryon_hairpin_without_spectator}.  These two figures are distinguished by the
presence/absence of the spectator quark flavour in the Goldstone propagator.  Therefore they correspond to
terms with $\tilde{x}{-}$ and $\tilde{y}{-}$type coefficients respectively.

Following the above rules in analysing the quark flavour flow structure, the results for the
baryon sunset diagrams can be written as
\bea
 \calQ^{(N_{f})}_{T_{ij}\rightarrow S_{ik}} &=& g_{2}^{2} \bigg \{ \sum_{a} \bigg [
              x^{(N_{f})}_{T_{ij},S_{ik},a}\mbox{ }\calK (M_{i,a},\Delta^{(B)}+\delta^{(B)}_{aj,ij},\Delta^{(B)}+\delta^{(B)}_{ak,ij})
               + \tilde{x}^{(N_{f})}_{T_{ij},S_{ik},a}\mbox{ }\tilde{\calK}_{N_{f}}(M_{i,a},\Delta^{(B)},\Delta^{(B)}+\delta^{(B)}_{ik,ij})\bigg ] \nonumber\\
  && \hspace{0.6cm} + \mbox{ }\tilde{y}^{(N_{f})}_{T_{ij},S_{ik}}\mbox{ }\tilde{\calK}_{N_{f}}(M_{j,k},\Delta^{(B)},\Delta^{(B)}+\delta^{(B)}_{ik,ij})
            \bigg \}\nonumber\\
  && \hspace{-0.25cm} +  g_{3}^{2} \bigg \{  \sum_{a} \bigg [
              x^{\prime\prime (N_{f})}_{T_{ij},S_{ik},a}\mbox{ }\calK (M_{i,a},\Delta^{(B)}+\delta^{(B)}_{aj,ij},\delta^{(B)}_{ak,ij})
             + \tilde{x}^{\prime\prime (N_{f})}_{T_{ij},S_{ik},a}\mbox{ }\tilde{\calK}_{N_{f}}(M_{i,a},\Delta^{(B)},\delta^{(B)}_{ik,ij}) \bigg ] \nonumber\\
  && \hspace{0.6cm} + \mbox{ }\tilde{y}^{\prime\prime (N_{f})}_{T_{ij},S_{ik}}\mbox{ }\tilde{\calK}_{N_{f}}(M_{j,k},\Delta^{(B)},\delta^{(B)}_{ik,ij})     
            \bigg \} , \nonumber\\
&&\nonumber\\
 \calQ^{(N_{f})}_{S_{ij}\rightarrow S_{ik}} &=& g_{2}^{2} \bigg \{ \sum_{a} \bigg [ 
              x^{(N_{f})}_{S_{ij},S_{ik},a}\mbox{ }\calK (M_{i,a},\delta^{(B)}_{aj,ij},\delta^{(B)}_{ak,ij})
               + \tilde{x}^{(N_{f})}_{S_{ij},S_{ik},a}\mbox{ }\tilde{\calK}_{N_{f}}(M_{i,a}, 0, \delta^{(B)}_{ik,ij})
      \bigg ] 
        + \mbox{ }\tilde{y}^{(N_{f})}_{S_{ij},S_{ik}}\mbox{ }\tilde{\calK}_{N_{f}}(M_{j,k}, 0,\delta^{(B)}_{ik,ij})
             \bigg \}\nonumber\\
  && \hspace{-0.25cm} +  g_{3}^{2} \bigg \{ \sum_{a} \bigg [
              x^{\prime (N_{f})}_{S_{ij},S_{ik},a}\mbox{ }\calK (M_{i,a},-\Delta^{(B)}+\delta^{(B)}_{aj,ij},\delta^{(B)}_{ak,ij})
               + \tilde{x}^{\prime (N_{f})}_{S_{ij},S_{ik},a}\mbox{ }\tilde{\calK}_{N_{f}}(M_{i,a}, -\Delta^{(B)}, \delta^{(B)}_{ik,ij})
      \bigg ] \nonumber\\
  && \hspace{0.6cm} + \mbox{ }\tilde{y}^{\prime (N_{f})}_{S_{ij},S_{ik}}\mbox{ }\tilde{\calK}_{N_{f}}(M_{j,k}, -\Delta^{(B)},\delta^{(B)}_{ik,ij})\nonumber\\
 && \hspace{0.6cm}
             \sum_{a} \bigg [
              x^{\prime\prime (N_{f})}_{S_{ij},S_{ik},a}\mbox{ }\calK (M_{i,a},\delta^{(B)}_{aj,ij},-\Delta^{(B)}+\delta^{(B)}_{ak,ij})
               + \tilde{x}^{\prime\prime (N_{f})}_{S_{ij},S_{ik},a}\mbox{ }\tilde{\calK}_{N_{f}}(M_{i,a}, 0, -\Delta^{(B)}+\delta^{(B)}_{ik,ij})
      \bigg ] \nonumber\\
\label{eq:baryon_sunset_master}
  && \hspace{0.6cm} + \mbox{ }\tilde{y}^{\prime\prime(N_{f})}_{S_{ij},S_{ik}}\mbox{ }\tilde{\calK}_{N_{f}}(M_{j,k}, 0, -\Delta^{(B)}+\delta^{(B)}_{ik,ij})
     \bigg \},
\eea
where the summations are over the flavours $u$ and $u^{\prime}$ in the SU(4$|$2) theory, and are over the flavours $u$, $s$, $u^{\prime}$ 
and $s^{\prime}$ in the SU(6$|$3) theory.  The relevant coefficients are presented in 
Tables~\ref{tab:baryon_x}, \ref{tab:baryon_x_prime}, \ref{tab:baryon_x_double_prime}, \ref{tab:y_tilde},
\ref{tab:baryon_tilde_x}, \ref{tab:baryon_tilde_x_prime}, and \ref{tab:baryon_tilde_x_double_prime} in
Appendix~~\ref{app:sunset_tables}.

\section{$H_{1}\rightarrow H_{2}\mbox{ }\pi(K)$ transition amplitudes}
\label{strong-decays}
The axial-current matrix elements, 
presented in the previous sections, are closely related to those in the strong-decay amplitudes, such as
\bea
 && B^{\ast}_{d} \rightarrow B_{u}\pi, \mbox{ }
 B^{\ast}_{s} \rightarrow B_{u} K, \nonumber\\
&&\nonumber\\
 && \Sigma^{(\ast)}_{b} \rightarrow \Lambda_{b} \pi,\nonumber\\
&&\nonumber\\
\label{eq:strong_decays}
 && \Sigma^{\ast }_{b} \rightarrow \Sigma_{b} \pi.
\eea
Note that, with the exception of $\Sigma^{(\ast)}_{b} \rightarrow \Lambda_{b}\pi$, for bottom hadrons the above
decays are kinematically forbidden in nature.  In HH$\chi$PT, the LO and NLO analytic terms for these decay amplitudes have 
the same structure as the matrix elements in Eq.~(\ref{eq:NLO-master}).  That is, the LO contributions are 
all proportional to the axial couplings $g_{1,2,3}$, while the NLO results are polynomials in the Goldstone
masses.  Therefore we only address the one-loop diagrams for these decays.

\begin{figure}[!t]
  \centering
\includegraphics[width=0.35\columnwidth]{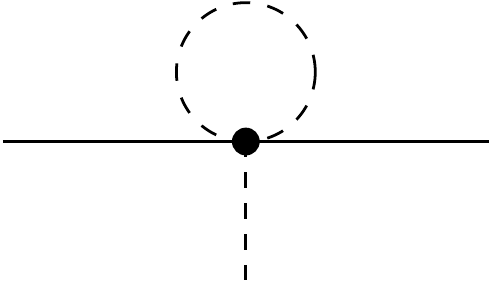}
\hspace*{10mm}
\includegraphics[width=0.35\columnwidth]{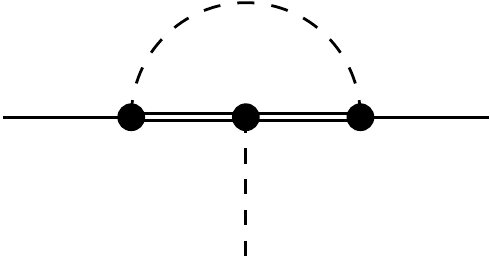} 
\\
\vspace*{4mm}
(a)\hspace*{7.0cm}(b)
  \caption{Diagrams contributing to the decay amplitudes in Eq.~(\ref{eq:strong_decays}). 
   The self-energy diagrams leading to wavefunction renormalisation
   of the external particles are not shown in this figure.
   The dashed lines are the Goldstone mesons.  The single
   solid lines denote generically the external heavy hadrons, while the double solid lines are the internal
   heavy hadrons.  They can be $B$, $B^{\ast}$ mesons or $T_{ij}$, $S_{ij}$ baryons. The vertices are all from the 
   axial-coupling terms proportional to $g_{1,2,3}$ in the strong chiral Lagrangian in Eq.~(\ref{eq:HHChPT_lagrangian_LO}).  
   Diagrams (a) and (b) are the ``tadpole" and ``sunset" types respectively.}
  \label{fig:strong_decay_oneloop}
\end{figure}

To compute the one-loop amplitudes for the processes in Eq.~(\ref{eq:strong_decays}), one has to calculate 
the wavefunction renormalisation of the Goldstone bosons and the heavy hadrons, as well as the tadpole and sunset 
diagrams in Fig.~\ref{fig:strong_decay_oneloop}. The Goldstone boson
wavefunction renormalisation can be found in standard references such as~\cite{Bernard:1994sv}
and \cite{Sharpe:2000bc}, and the heavy hadron wavefunction renormalisation is presented in 
Eqs.~(\ref{eq:meson_waverenorm_master}) and (\ref{eq:baryon_waverenorm_master}).
The amplitudes from the sunset diagram in Fig.~\ref{fig:strong_decay_oneloop} (b) are identical to those from 
the corresponding diagram in Fig.~\ref{fig:oneloop} (c).  Therefore they are equal to the results presented in
Eqs.~(\ref{eq:meson_sunset_master}) and (\ref{eq:baryon_sunset_master}).
The tadpole diagram in Fig.~\ref{fig:strong_decay_oneloop} (a) differs from that of the axial-current matrix elements
in Fig.~\ref{fig:oneloop} (b) by a factor of one-third.  That is, one can take the results in Eq.~(\ref{eq:all_tadpole}),
and multiply them by $1/3$ to obtain the corresponding tadpole-diagram contributions to the decay amplitudes in Eq.~(\ref{eq:strong_decays}).
It turns out that the contribution from the tadpole diagram is exactly cancelled by
the contribution from the wavefunction renormalization of the external Goldstone boson \cite{Cheng:1993kp}.
As is shown in Fig.~\ref{fig:indiviual-contribs}, the tadpole
diagrams provide significant contributions to the axial current matrix
elements and will lead to significant differences between the
quark-mass dependence of axial current matrix elements and that of strong decay amplitudes.

These decay amplitudes have also been computed in Ref.~\cite{Cheng:1993kp},
to one-loop order in SU(3) HH$\chi$PT  in the infinite-volume limit
with $\Delta^{(M)} = \Delta^{(B)} = 0$, and without the
SU(3) breaking effects from the Lagrangian in Eq.~(\ref{eq:HHChPT_lagrangian_chi}).  Our results agree with those 
presented in Ref.~\cite{Cheng:1993kp} in the same limits.

\section{Conclusion}
\label{sec:conclusion}
With the expectation of precise data from the LHCb collaboration and
from the potential SuperB experiment, accurate QCD calculations of quantities involving $B$ mesons
and single-$b$ baryons will be important in further constraining flavour physics and in looking for physics beyond the SM. 
This is a challenging but necessary task.
In this paper, we have presented calculations for axial-current matrix elements involving single
heavy hadron external states in HH$\chi$PT at the NLO.
We have performed these computations in partially quenched $\chi$PT for both $N_{f}=2$ and $N_f=2+1$,
including finite volume effects.
Our results are essential for extracting the axial couplings in HH$\chi$PT from experimental data or
lattice QCD. These axial couplings are central quantities in $b$ physics, as they control the light quark mass
dependence of $b$-hadron observables and determine the strong decay widths of heavy hadrons.

We have discussed the $SU(2)$ case in detail, numerically analysing the behaviour of the various loop contributions
for natural values of the low-energy constants.
Based on our study, we conclude that the $SU(2)$ chiral expansion of the axial current matrix elements is
well-behaved for $M_\pi \alt$ 400 MeV. This implies that lattice calculations that are 
performed in this regime can be used to determine the axial couplings reliably.

\section*{Acknowledgments}
We warmly thank Brian Tiburzi, Andr\'{e} Walker-Loud and Matthew Wingate for helpful discussions.
This work is supported by U.S. DoE contract number DE-AC05-06OR-23177, grant number DE-SC000-1784, Jeffress Memorial Trust
grant number J-968, and Taiwanese NSC grant 
number 99-2112-M-009-004-MY3.  We acknowledge the hospitality of Academia Sinica Taipei, The College of William and Mary, Thomas Jefferson 
National Accelerator Facility, National Centre of Theoretical Sciences and National Chiao-Tung University.

\appendix
\section{Mass parameters}
\label{app:masses_in_loops}
In this appendix, we define various quantities appearing in our results.  First we present the hadron masses,
\bea
 && M_{a,b}^{2} = B_{0} (m_{a} + m_{b}) , \,\,\,\,\,
 \tilde{\delta}_{VS}^{2} = M_{u,u}^{2} - M_{u,u^{\prime}}^{2} , \,\,\,\,\,
 \tilde{\delta}_{VSs}^{2} = M_{s,s}^{2} - M_{s,s^{\prime}}^{2} , \,\,\,\,\,
 M_{X}^{2} = \frac{1}{3} \left ( M_{u,u}^{2} + 2 M_{s,s}^{2} 
   - 2\tilde{\delta}_{VS}^{2} - 4 \tilde{\delta}_{VSs}^{2} \right ) , \nonumber\\
&& \nonumber\\
&&  \Delta^{(M)} = M_{B^{\ast}_{a}} - M_{B_{b}} , \,\,\,\,\,
 \delta^{(M)}_{a,b} = M_{B_{a}} - M_{B_{b}} 
           = M_{B^{\ast}_{a}} - M_{B^{\ast}_{b}}, \nonumber\\
&& \nonumber\\
\label{eq:mass_param}
 && \Delta^{(B)} = M_{S_{ab}} - M_{T_{ab}} , \,\,\,\,\,
 \delta^{(B)}_{ab,cd} = M_{T_{ab}} - M_{T_{cd}} = M_{S_{ab}} - M_{S_{cd}} ,
\eea
where $B_{0}$ is defined in Eq.~(\ref{eq:chi_definition}). As explained in the main text, $\Delta^{(M)}$ vanishes in the heavy quark limit, while $\Delta^{(B)}$ remains non-zero and is of $O(\Lambda_{{\mathrm{QCD}}})$.  In this paper, we work in the isospin limit, and denote the pion mass as $M_{u,u}$.

It is useful to define the following quantities which appear in the ``hairpin'' contributions
to the flavour-singlet meson propagators in the SU(6$|$3) theory.
\bea
 &&  A_{u,u} = \frac{2 \left(\tilde{\delta}_{VS}^2-M_{u,u}^2+M_X^2\right)
 \tilde{\delta}_{VS}^2}{\left(M_{u,u}^2-M_X^2\right)^2}+\frac{3}{2} ,  \,\,\,\,\,\,\,\,
  A_{s,s} = \frac{3 \left( 8 \tilde{\delta}_{VSs}^{4} + \left ( 
         2 \tilde{\delta}_{VS}^{2} - M_{u,u}^{2} + M_{s,s}^{2} \right )^{2} \right ) }
         {\left ( 2 \tilde{\delta}_{VS}^{2} + 4 \tilde{\delta}_{VSs}^{2} 
            - M_{u,u}^{2} + M_{s,s}^{2} \right )^{2} }\,, \nonumber\\
&&\nonumber\\
 && C_{u,u} = 3 \tilde{\delta}_{VS}^2-\frac{2 \tilde{\delta}_{VS}^4}{M_{u,u }^2-M_X^2} , \,\,\,\,\,\,\,\,
  C_{s,s} = \frac{6 \tilde{\delta}_{VSs}^{2}\left ( 2 \tilde{\delta}_{VS}^{2} 
         - M_{u,u}^{2} + M_{s,s}^{2}\right ) }
          { 2 \tilde{\delta}_{VS}^{2} + 4 \tilde{\delta}_{VSs}^{2} 
              - M_{u,u}^{2} + M_{s,s}^{2} } , \nonumber\\
&&\nonumber\\
&&  D^{(u)}_{u,s} = \frac{2 \tilde{\delta}_{VS}^{2} 
               \left ( M_{u,u}^{2} - M_{s,s}^{2} + 2 \tilde{\delta}_{VSs}^{2} \right ) }
                {\left ( M_{u,u}^{2} - M_{s,s}^{2} \right ) \left ( M_{u,u}^{2} - M_{X}^{2} \right )} , \,\,\,\,\,\,\,\,
  D^{(s)}_{u,s} = \frac{ 2 \tilde{\delta}_{VSs}^{2} 
                \left ( M_{u,u}^{2} - M_{s,s}^{2} - 2 \tilde{\delta}_{VS}^{2}\right ) }
               {\left ( M_{u,u}^{2} - M_{s,s}^{2}\right ) \left ( M_{s,s}^{2} - M_{X}^{2} \right ) } , 
\nonumber\\
&&\nonumber\\
&&\nonumber\\
\label{eq:ACD}
&&  D^{(X)}_{u,s} = \frac{\left ( M_{u,u}^{2} - M_{X}^{2} -2\tilde{\delta}_{VS}^{2}\right ) 
                          \left ( M_{s,s}^{2} - M_{X}^{2} -2\tilde{\delta}_{VSs}^{2}\right )}
              {\left ( M_{u,u}^{2} - M_{X}^{2} \right ) \left ( M_{s,s}^{2} - M_{X}^{2} \right )} .
\eea
In the full QCD limit, where $m_{u^\prime} = m_{u}$ and $m_{s^\prime} = m_{s}$,
\bea
 && A^{{\mathrm{QCD}}}_{u,u} = \frac{3}{2} , \mbox{ }\,\,  A^{{\mathrm{QCD}}}_{s,s} = 3 , \mbox{ }\,\,
  C^{{\mathrm{QCD}}}_{u,u} = 0 , \mbox{ }\,\, C^{{\mathrm{QCD}}}_{s,s} = 0 , \nonumber\\
&&\nonumber\\
 && D^{(u){\mathrm{QCD}}}_{u,s} = 0, \mbox{ }\,\, D^{(s){\mathrm{QCD}}}_{u,s} = 0, \mbox{ }\,\,
    D^{(X){\mathrm{QCD}}}_{u,s} = 1 .
\eea

\section{Integrals and sums}
\label{app:integral}
In this appendix, we present results of loop integrals and sums
using dimensional
regularisation, with the ultra-violet divergences removed by subtracting the term,
\beq
\label{eq:lambdabar}
 \bar{\lambda} = \frac{2}{4 - d} - \gamma_{E} + {\mathrm{log}}(4\pi) + 1,
\eeq
where $d$ is the number of space-time dimensions.  This is a commonly-used scheme in $\chi$PT calculations~\cite{Gasser:1985gg}.
It is different from the 
$\overline{{\mathrm{MS}}}$ scheme by the constant ``1" on the right-hand side of the above equation.  It can also be
changed into the scheme discussed in Sec.~\ref{sec:chiral_limit_subtraction} straightforwardly.
Finite volume effects in the 
limit $m L \gg 1$ ($m$ is a generic Goldstone mass and L is the spatial lattice 
volume) are 
computed by replacing the momentum integrals by sums in the spatial directions. 
The one-loop contributions appearing in this work can all be obtained by investigating
the following sums/integrals
\bea
 \calI(m) &\equiv&
 \mu^{4 - d} \sum \hspace{-0.45cm} \int \frac{d^{d}k}{( 2\pi )^{d}} \frac{i}{k^{2}-m^{2}+ i\epsilon}
    - \frac{ m^{2}}{16 \pi^{2}} \bar{\lambda}, \\
 & &\nonumber\\
\label{eq:IandF}
 \calF(m,\Delta) &\equiv& 
 \left ( g^{\rho\nu} - v^{\rho} v^{\nu}\right ) \left [ \frac{\mu^{4 - d}}{(d - 1)} 
 \sum \hspace{-0.45cm} \int
 \frac{d^{d}k}{( 2\pi )^{d}} 
  \frac{i k_{\rho} k_{\nu}}{(k^{2}-m^{2}
+ i\epsilon)(v\cdot k - \Delta + i\epsilon)} 
  + \frac{g_{\rho\nu}}{16 \pi^{2}}\bar{\lambda}\left ( \frac{2 \Delta^{2}}{3} - m^{2}\right ) \Delta
\right ] , \nonumber
\eea
where $\mu$ is the renormalisation scale, and the symbol 
\bea
 && \sum \hspace{-0.45cm} \int d^{d}k \nonumber
\eea
means performing the sums in three spatial directions using the Poisson summation formula, 
followed by dimensionally regularising the infinite-volume integrals.

We can further separate the infinite-volume limit of $\calI$ and $\calF$ from the finite-volume contributions,
\bea
 \calI(m) &=& I(m) + I_{{\mathrm{FV}}}(m) , \nonumber \\
 & &\nonumber\\
 \calF(m) &=& F(m,\Delta) + F_{{\mathrm{FV}}}(m). \label{eq:I}
\eea
The functions $I$ and $F$ are results from the ordinary one-loop integrals,
\bea
 I(m) &=& \frac{m^{2}}{16 \pi^{2}} {\log\left ( \frac{m^{2}}{\mu^{2}}\right )} ,
\nonumber\\
&&\nonumber\\
\label{eq:IandF_infinite_volume}
 F(m,\Delta) &=& \frac{-1}{16\pi^{2}} 
 \left [ \left ( m^{2}- \frac{2\Delta^{2}}{3} \right )\Delta 
     {\log\left ( \frac{m^{2}}{\mu^{2}}\right )}  
   + \left ( \frac{10\Delta^{2}}{9} - \frac{4 m^{2}}{3}\right ) \Delta
   + \frac{2 (\Delta^{2}-m^{2})}{3} m R\left ( \frac{\Delta}{m}\right )
\right ] ,
\eea
with
\beq
\label{eq:R}
R(x) \equiv 
 \sqrt{x^{2}-1}\mbox{ }\left [ 
       {\mathrm{log}}\left (  x - \sqrt{x^{2}-1+i\epsilon} \right )
     - {\mathrm{log}}\left (  x + \sqrt{x^{2}-1+i\epsilon} \right ) \right ] .
\eeq
The function $F(m,\Delta)$ does not vanish in the $m\rightarrow 0$ limit unless $\Delta=0$.  
One can adopt the scheme discussed in Sec.~\ref{sec:chiral_limit_subtraction} by simply
rewriting $F$ as $F^{({\mathrm{sub}})}$ defined in Eq.~(\ref{eq:F_prime}), and the real part of
the function $F^{({\mathrm{sub}})}$ is zero in the chiral limit for arbitrary $\Delta$.

For the case in which the external hadrons are stable particles, the finite-volume pieces can be
shown to be~\cite{Arndt:2004bg,Detmold:2005pt}
\footnote{Similar formulae for finite-volume effects are also obtained in Ref.~\cite{Beane:2004tw}.}
\bea
 I_{\mathrm{FV}}(m) 
  &=& \frac{m}{4 \pi^{2}} \sum_{\vec{u}\not=\vec{0}}
 \frac{1}{u L} K_{1}\left (u m L\right )\nonumber\\
& &\nonumber\\
 &\stackrel{m L \gg 1}{\longrightarrow}&
 \frac{1}{4\pi^{2}} \sum_{\vec{u}\not=\vec{0}}
 \sqrt{\frac{m\pi}{2 u L}}
  \left (\frac{1}{u L}\right ) \exponential^{-u m L}
  \times \left \{
  1 + \frac{3}{8 u m L} 
 - \frac{15}{128 (u m L)^{2}}
 + \op\left ( \left [\frac{1}{u m L}\right ]^{3}\right )
 \right \} , \nonumber\\
& &\nonumber\\
\label{eq:FandI_FV}
 F_{\mathrm{FV}}(m,\Delta) &=& \frac{-1}{12\pi^{2}} \sum_{\vec{u}\not=
   \vec{0}} \frac{1}{u\, L}
 \int_{0}^{\infty} d |\vec{k}| \,
 \frac{|\vec{k}|\,{\mathrm{sin}}(u |\vec{k}| L)}{
     \sqrt{|{\vec k}|^2+m^2} + \Delta}
 \left ( \Delta +\frac{m^2}{\sqrt{|{\vec k}|^2+m^2}} \right ) 
 \nonumber\\&&\nonumber\\
 &\stackrel{m L \gg 1}{\longrightarrow}& 
 \frac{- m^2}{24\pi}\sum_{\vec{u}\not=
   \vec{0}}\frac{{\mathrm{e}}^{-u m L}}{ u \,L} 
   {\mathcal{A}} \,,
\eea
where $\vec{u}=(u_1,u_2,u_3)$ with $u_i\in {\mathbb{Z}}$, $u \equiv
|\vec{u}|$ and 
\begin{eqnarray}
\label{eq:calA_def}
 {\mathcal{A}} &=&\exponential^{(z^{2})} \big [ 
1 - {\mathrm{Erf}}(z)\mbox{ }\big ]
+\left (\frac{1}{u m L} \right ) \bigg [
 \frac{1}{\sqrt{\pi}} \left ( \frac{9z}{4} - 
\frac{z^{3}}{2}\right )
 + \left(\frac{z^{4}}{2}-2\,z^2\right)\exponential^{(z^{2})} 
 \big [ 1 - {\mathrm{Erf}}(z)\mbox{ }\big ]
\bigg ]\\
& &
-\left (\frac{1}{u m L} \right )^{2}\bigg [
\frac{1}{\sqrt{\pi}}\left ( -\frac{39z}{64} + 
\frac{11z^{3}}{32}
  -\frac{9z^{5}}{16} + \frac{z^{7}}{8} \right )
-\left ( -\frac{z^{6}}{2} + \frac{z^{8}}{8}\right )
\exponential^{(z^{2})} \big [ 1 - {\mathrm{Erf}}(z)\mbox{ }\big ]
\bigg ]
+\op\left ( \frac{1}{(u m L)^{3}}\right ) ,
\nonumber
\end{eqnarray}
with
\begin{equation}
 z = \left ( \frac{\Delta}{m} \right )
  \sqrt{\frac{u m L}{2}} .
\end{equation}
Higher order terms in the $1/(u m L)$ expansion in
Eq.~(\ref{eq:calA_def}) can be easily calculated.  The integer $u_{i}$ can be interpreted as the number of times that the pion wraps around the spatial volume in the $i{-}$the direction.  

The functions appearing in our one-loop results are
\bea
 && \calH(m,\Delta) \equiv \frac{\partial \calF(m,\Delta)}{\partial \Delta}
\qquad {\rm and} \qquad
 \calK (m,\Delta_{1},\Delta_{2}) \equiv \frac{\calF(m,\Delta_{1})-\calF(m,\Delta_{2})}{\Delta_{1} - \Delta_{2}} \,,
\nonumber\\
&&\nonumber\\
\label{eq:def_calH}
 &&\calI_{\eta^{\prime}}(m) \equiv \frac{\partial \calI(m)}{\partial m^{2}} , \qquad
  \calH_{\eta^{\prime}}(m,\Delta) \equiv \frac{\partial \calH(m,\Delta)}{\partial m^{2}}
\qquad {\rm and} \qquad
 \calK_{\eta^{\prime}}(m,\Delta_{1},\Delta_{2}) \equiv \frac{\partial \calK(m,\Delta_{1},\Delta_{2})}{\partial m^{2}} .
\eea
Notice that 
\beq
\label{eq:K_to_H}
 \calK \left ( m,\Delta,\Delta \right ) \equiv \mathop{{\mathrm{lim}}}_{\Delta^{\prime}\rightarrow\Delta}
     \calK(m,\Delta,\Delta^{\prime})  =  \calH \left ( m,\Delta \right ).
\eeq
To present the residual flavour-singlet ``hairpin'' contributions in a compact form, we define three functions
\beq
\label{eq:DP_IHK}
\tilde{\calI}_{N_{f}}(m), \mbox{ } \tilde{\calH}_{N_{f}}(m,\Delta), \mbox{ } {\mathrm{and}}\mbox{ }
    \tilde{\calK}_{N_{f}}(m,\Delta_{1},\Delta_{2}) .
\eeq
They take the explicit form
\bea
 \tilde{\calI}_{2}(m) &=& \calI(m) + 2 \tilde{\delta}^{2}_{VS} \mbox{ }\calI_{\eta^{\prime}}(m) ,\nonumber\\
&& \nonumber\\
 \tilde{\calH}_{2}(m,\Delta) &=& \calH(m,\Delta) + 2 \tilde{\delta}^{2}_{VS}\mbox{ } \calH_{\eta^{\prime}}(m,\Delta) ,
   \nonumber\\
&& \nonumber\\
\label{eq:SU42_DP_IHK}
  \tilde{\calK}_{2}(m,\Delta_{1},\Delta_{2}) &=& \calK(m,\Delta_{1},\Delta_{2}) 
    + 2 \tilde{\delta}^{2}_{VS}\mbox{ } \calK_{\eta^{\prime}}(m,\Delta_{1},\Delta_{2}) ,
\eea
in the SU(4$|$2) theory ($N_{f}=2$) where the mass $m$ in the arguments is always equal to $M_{u,u}$, and
\bea
 \tilde{\calI}_{3}(M_{a,b}) &=& \delta_{a,b} \left \{ A_{a,b} \mbox{ }\calI(M_{a,b}) + \left ( 1 - A_{a,b}\right ) 
        \mbox{ } \calI(M_{X}) + C_{a,b}\mbox{ } \calI_{\eta^{\prime}}(M_{a,b}) \right \}\nonumber\\
   &+& (1 - \delta_{a,b}) \left \{ D_{a,b}^{(a)} \mbox{ }\calI(M_{a,a}) 
      + D_{a,b}^{(b)} \mbox{ }\calI(M_{b,b}) + D_{a,b}^{(X)}\mbox{ } \calI(M_{X})\right \} ,\nonumber\\
&& \nonumber\\
  \tilde{\calH}_{3}(M_{a,b},\Delta) &=& \delta_{a,b} \left \{ A_{a,b} \mbox{ }\calH(M_{a,b},\Delta) + \left ( 1 - A_{a,b}\right ) 
        \mbox{ } \calH(M_{X},\Delta) + C_{a,b}\mbox{ } \calH_{\eta^{\prime}}(M_{a,b},\Delta) \right \}\nonumber\\
   &+& (1 - \delta_{a,b}) \left \{ D_{a,b}^{(a)} \mbox{ }\calH(M_{a,a},\Delta) 
      + D_{a,b}^{(b)} \mbox{ }\calH(M_{b,b},\Delta) + D_{a,b}^{(X)}\mbox{ } \calH(M_{X},\Delta)\right \} ,\nonumber\\
&& \nonumber\\
   \tilde{\calK}_{3}(M_{a,b},\Delta_{1},\Delta_{2}) 
   &=& \delta_{a,b} \left \{ A_{a,b} \mbox{ }\calK(M_{a,b},\Delta_{1},\Delta_{2}) + \left ( 1 - A_{a,b}\right ) 
        \mbox{ } \calK(M_{X},\Delta_{1},\Delta_{2}) + C_{a,b}\mbox{ } \calK_{\eta^{\prime}}(M_{a,b},\Delta_{1},\Delta_{2}) \right \}\nonumber\\
\label{eq:SU63_DP_IHK}
   &+& (1 - \delta_{a,b}) \left \{ D_{a,b}^{(a)} \mbox{ }\calK(M_{a,a},\Delta_{1},\Delta_{2}) 
      + D_{a,b}^{(b)} \mbox{ }\calK(M_{b,b},\Delta_{1},\Delta_{2}) + D_{a,b}^{(X)}\mbox{ } \calK(M_{X},\Delta_{1},\Delta_{2})\right \} ,
\eea
in the SU(6$|$3) theory ($N_{f} = 3$).

\section{Coefficients for wavefunction renormalisation}
\label{app:waverenorm_tables}
In this Appendix, we present the coefficients in
Eqs.~(\ref{eq:meson_waverenorm_master}) and
(\ref{eq:baryon_waverenorm_master}) relevant to the matrix elements
investigated in this work.  These coefficients are summarised in
Tables~\ref{tab:meson_waverenorm_table}, \ref{tab:T_waverenorm_table}
and \ref{tab:S_waverenorm_table}. Because of isospin symmetry, it is
not possible (or necessary) to distinguish between the $w$ and $u$ coefficients for some of the hadrons in the current study.  For such cases, we simply present $w+u$ in the tables.
\begin{table}[t]
\begin{tabular}{c|cccc}
   $\mbox{ }\mbox{ }\mbox{ }\mbox{ }a\mbox{ }\mbox{ }\mbox{ }\mbox{ }$ & 
   $\mbox{ }\mbox{ }\mbox{ }\mbox{ }u\mbox{ }\mbox{ }\mbox{ }\mbox{ }$ & 
   $\mbox{ }\mbox{ }\mbox{ }\mbox{ }s\mbox{ }\mbox{ }\mbox{ }\mbox{ }$ & 
   $\mbox{ }\mbox{ }\mbox{ }\mbox{ }u^{\prime}\mbox{ }\mbox{ }\mbox{ }\mbox{ }$ & 
   $\mbox{ }\mbox{ }\mbox{ }\mbox{ }s^{\prime}\mbox{ }\mbox{ }\mbox{ }\mbox{ }$ \\
\hline\hline
   $\mbox{ }\mbox{ }\mbox{ }\mbox{ }w_{B_{u},a}^{(2)}\mbox{ }\mbox{ }\mbox{ }\mbox{ }$  & 
   $\mbox{ }\mbox{ }\mbox{ }\mbox{ }0\mbox{ }\mbox{ }\mbox{ }\mbox{ }$ & 
   $\mbox{ }\mbox{ }\mbox{ }\mbox{ }0\mbox{ }\mbox{ }\mbox{ }\mbox{ }$ & 
   $\mbox{ }\mbox{ }\mbox{ }\mbox{ }3\mbox{ }\mbox{ }\mbox{ }\mbox{ }$ & 
   $\mbox{ }\mbox{ }\mbox{ }\mbox{ }0\mbox{ }\mbox{ }\mbox{ }\mbox{ }$ \\
\hline
   $\mbox{ }\mbox{ }\mbox{ }\mbox{ }\tilde{w}_{B_{u},a}^{(2)}\mbox{ }\mbox{ }\mbox{ }\mbox{ }$  & 
   $\mbox{ }\mbox{ }\mbox{ }\mbox{ }\frac{-3}{4}\mbox{ }\mbox{ }\mbox{ }\mbox{ }$ & 
   $\mbox{ }\mbox{ }\mbox{ }\mbox{ }0\mbox{ }\mbox{ }\mbox{ }\mbox{ }$ & 
   $\mbox{ }\mbox{ }\mbox{ }\mbox{ }0\mbox{ }\mbox{ }\mbox{ }\mbox{ }$ & 
   $\mbox{ }\mbox{ }\mbox{ }\mbox{ }0\mbox{ }\mbox{ }\mbox{ }\mbox{ }$ \\
\hline\hline
   $\mbox{ }\mbox{ }\mbox{ }\mbox{ }w_{B_{u},a}^{(3)}\mbox{ }\mbox{ }\mbox{ }\mbox{ }$  & 
   $\mbox{ }\mbox{ }\mbox{ }\mbox{ }0\mbox{ }\mbox{ }\mbox{ }\mbox{ }$ & 
   $\mbox{ }\mbox{ }\mbox{ }\mbox{ }0\mbox{ }\mbox{ }\mbox{ }\mbox{ }$ & 
   $\mbox{ }\mbox{ }\mbox{ }\mbox{ }3\mbox{ }\mbox{ }\mbox{ }\mbox{ }$ & 
   $\mbox{ }\mbox{ }\mbox{ }\mbox{ }\frac{3}{2}\mbox{ }\mbox{ }\mbox{ }\mbox{ }$ \\
\hline
   $\mbox{ }\mbox{ }\mbox{ }\mbox{ }\tilde{w}_{B_{u},a}^{(3)}\mbox{ }\mbox{ }\mbox{ }\mbox{ }$  & 
   $\mbox{ }\mbox{ }\mbox{ }\mbox{ }\frac{-1}{2}\mbox{ }\mbox{ }\mbox{ }\mbox{ }$ & 
   $\mbox{ }\mbox{ }\mbox{ }\mbox{ }0\mbox{ }\mbox{ }\mbox{ }\mbox{ }$ & 
   $\mbox{ }\mbox{ }\mbox{ }\mbox{ }0\mbox{ }\mbox{ }\mbox{ }\mbox{ }$ & 
   $\mbox{ }\mbox{ }\mbox{ }\mbox{ }0\mbox{ }\mbox{ }\mbox{ }\mbox{ }$ \\
\hline\hline
   $\mbox{ }\mbox{ }\mbox{ }\mbox{ }w_{B_{d}^{\ast},a}^{\prime (2)}\mbox{ }\mbox{ }\mbox{ }\mbox{ }$  & 
   $\mbox{ }\mbox{ }\mbox{ }\mbox{ }0\mbox{ }\mbox{ }\mbox{ }\mbox{ }$ & 
   $\mbox{ }\mbox{ }\mbox{ }\mbox{ }0\mbox{ }\mbox{ }\mbox{ }\mbox{ }$ & 
   $\mbox{ }\mbox{ }\mbox{ }\mbox{ }1\mbox{ }\mbox{ }\mbox{ }\mbox{ }$ & 
   $\mbox{ }\mbox{ }\mbox{ }\mbox{ }0\mbox{ }\mbox{ }\mbox{ }\mbox{ }$ \\
\hline
   $\mbox{ }\mbox{ }\mbox{ }\mbox{ }\tilde{w}_{B_{d}^{\ast},a}^{\prime (2)}\mbox{ }\mbox{ }\mbox{ }\mbox{ }$  & 
   $\mbox{ }\mbox{ }\mbox{ }\mbox{ }\frac{-1}{4}\mbox{ }\mbox{ }\mbox{ }\mbox{ }$ & 
   $\mbox{ }\mbox{ }\mbox{ }\mbox{ }0\mbox{ }\mbox{ }\mbox{ }\mbox{ }$ & 
   $\mbox{ }\mbox{ }\mbox{ }\mbox{ }0\mbox{ }\mbox{ }\mbox{ }\mbox{ }$ & 
   $\mbox{ }\mbox{ }\mbox{ }\mbox{ }0\mbox{ }\mbox{ }\mbox{ }\mbox{ }$ \\
\hline
   $\mbox{ }\mbox{ }\mbox{ }\mbox{ }w_{B_{d}^{\ast},a}^{(2)}\mbox{ }\mbox{ }\mbox{ }\mbox{ }$  & 
   $\mbox{ }\mbox{ }\mbox{ }\mbox{ }0\mbox{ }\mbox{ }\mbox{ }\mbox{ }$ & 
   $\mbox{ }\mbox{ }\mbox{ }\mbox{ }0\mbox{ }\mbox{ }\mbox{ }\mbox{ }$ & 
   $\mbox{ }\mbox{ }\mbox{ }\mbox{ }2\mbox{ }\mbox{ }\mbox{ }\mbox{ }$ & 
   $\mbox{ }\mbox{ }\mbox{ }\mbox{ }0\mbox{ }\mbox{ }\mbox{ }\mbox{ }$ \\
\hline
   $\mbox{ }\mbox{ }\mbox{ }\mbox{ }\tilde{w}_{B_{d}^{\ast},a}^{(2)}\mbox{ }\mbox{ }\mbox{ }\mbox{ }$  & 
   $\mbox{ }\mbox{ }\mbox{ }\mbox{ }\frac{-1}{2}\mbox{ }\mbox{ }\mbox{ }\mbox{ }$ & 
   $\mbox{ }\mbox{ }\mbox{ }\mbox{ }0\mbox{ }\mbox{ }\mbox{ }\mbox{ }$ & 
   $\mbox{ }\mbox{ }\mbox{ }\mbox{ }0\mbox{ }\mbox{ }\mbox{ }\mbox{ }$ & 
   $\mbox{ }\mbox{ }\mbox{ }\mbox{ }0\mbox{ }\mbox{ }\mbox{ }\mbox{ }$ \\
\hline\hline
   $\mbox{ }\mbox{ }\mbox{ }\mbox{ }w_{B_{d}^{\ast},a}^{\prime (3)}\mbox{ }\mbox{ }\mbox{ }\mbox{ }$  & 
   $\mbox{ }\mbox{ }\mbox{ }\mbox{ }0\mbox{ }\mbox{ }\mbox{ }\mbox{ }$ & 
   $\mbox{ }\mbox{ }\mbox{ }\mbox{ }0\mbox{ }\mbox{ }\mbox{ }\mbox{ }$ & 
   $\mbox{ }\mbox{ }\mbox{ }\mbox{ }1\mbox{ }\mbox{ }\mbox{ }\mbox{ }$ & 
   $\mbox{ }\mbox{ }\mbox{ }\mbox{ }\frac{1}{2}\mbox{ }\mbox{ }\mbox{ }\mbox{ }$ \\
\hline
   $\mbox{ }\mbox{ }\mbox{ }\mbox{ }\tilde{w}_{B_{d}^{\ast},a}^{\prime (3)}\mbox{ }\mbox{ }\mbox{ }\mbox{ }$  & 
   $\mbox{ }\mbox{ }\mbox{ }\mbox{ }\frac{-1}{6}\mbox{ }\mbox{ }\mbox{ }\mbox{ }$ & 
   $\mbox{ }\mbox{ }\mbox{ }\mbox{ }0\mbox{ }\mbox{ }\mbox{ }\mbox{ }$ & 
   $\mbox{ }\mbox{ }\mbox{ }\mbox{ }0\mbox{ }\mbox{ }\mbox{ }\mbox{ }$ & 
   $\mbox{ }\mbox{ }\mbox{ }\mbox{ }0\mbox{ }\mbox{ }\mbox{ }\mbox{ }$ \\
\hline
   $\mbox{ }\mbox{ }\mbox{ }\mbox{ }w_{B_{d}^{\ast},a}^{(3)}\mbox{ }\mbox{ }\mbox{ }\mbox{ }$  & 
   $\mbox{ }\mbox{ }\mbox{ }\mbox{ }0\mbox{ }\mbox{ }\mbox{ }\mbox{ }$ & 
   $\mbox{ }\mbox{ }\mbox{ }\mbox{ }0\mbox{ }\mbox{ }\mbox{ }\mbox{ }$ & 
   $\mbox{ }\mbox{ }\mbox{ }\mbox{ }2\mbox{ }\mbox{ }\mbox{ }\mbox{ }$ & 
   $\mbox{ }\mbox{ }\mbox{ }\mbox{ }1\mbox{ }\mbox{ }\mbox{ }\mbox{ }$ \\
\hline
   $\mbox{ }\mbox{ }\mbox{ }\mbox{ }\tilde{w}_{B_{d}^{\ast},a}^{(3)}\mbox{ }\mbox{ }\mbox{ }\mbox{ }$  & 
   $\mbox{ }\mbox{ }\mbox{ }\mbox{ }\frac{-1}{3}\mbox{ }\mbox{ }\mbox{ }\mbox{ }$ & 
   $\mbox{ }\mbox{ }\mbox{ }\mbox{ }0\mbox{ }\mbox{ }\mbox{ }\mbox{ }$ & 
   $\mbox{ }\mbox{ }\mbox{ }\mbox{ }0\mbox{ }\mbox{ }\mbox{ }\mbox{ }$ & 
   $\mbox{ }\mbox{ }\mbox{ }\mbox{ }0\mbox{ }\mbox{ }\mbox{ }\mbox{ }$ \\
\hline\hline
   $\mbox{ }\mbox{ }\mbox{ }\mbox{ }w_{B_{s}^{\ast},a}^{\prime (3)}\mbox{ }\mbox{ }\mbox{ }\mbox{ }$  & 
   $\mbox{ }\mbox{ }\mbox{ }\mbox{ }0\mbox{ }\mbox{ }\mbox{ }\mbox{ }$ & 
   $\mbox{ }\mbox{ }\mbox{ }\mbox{ }0\mbox{ }\mbox{ }\mbox{ }\mbox{ }$ & 
   $\mbox{ }\mbox{ }\mbox{ }\mbox{ }1\mbox{ }\mbox{ }\mbox{ }\mbox{ }$ & 
   $\mbox{ }\mbox{ }\mbox{ }\mbox{ }\frac{1}{2}\mbox{ }\mbox{ }\mbox{ }\mbox{ }$ \\
\hline
   $\mbox{ }\mbox{ }\mbox{ }\mbox{ }\tilde{w}_{B_{s}^{\ast},a}^{\prime (3)}\mbox{ }\mbox{ }\mbox{ }\mbox{ }$  & 
   $\mbox{ }\mbox{ }\mbox{ }\mbox{ }0\mbox{ }\mbox{ }\mbox{ }\mbox{ }$ & 
   $\mbox{ }\mbox{ }\mbox{ }\mbox{ }\frac{-1}{6}\mbox{ }\mbox{ }\mbox{ }\mbox{ }$ & 
   $\mbox{ }\mbox{ }\mbox{ }\mbox{ }0\mbox{ }\mbox{ }\mbox{ }\mbox{ }$ & 
   $\mbox{ }\mbox{ }\mbox{ }\mbox{ }0\mbox{ }\mbox{ }\mbox{ }\mbox{ }$ \\
\hline
   $\mbox{ }\mbox{ }\mbox{ }\mbox{ }w_{B_{s}^{\ast},a}^{(3)}\mbox{ }\mbox{ }\mbox{ }\mbox{ }$  & 
   $\mbox{ }\mbox{ }\mbox{ }\mbox{ }0\mbox{ }\mbox{ }\mbox{ }\mbox{ }$ & 
   $\mbox{ }\mbox{ }\mbox{ }\mbox{ }0\mbox{ }\mbox{ }\mbox{ }\mbox{ }$ & 
   $\mbox{ }\mbox{ }\mbox{ }\mbox{ }2\mbox{ }\mbox{ }\mbox{ }\mbox{ }$ & 
   $\mbox{ }\mbox{ }\mbox{ }\mbox{ }1\mbox{ }\mbox{ }\mbox{ }\mbox{ }$ \\
\hline
   $\mbox{ }\mbox{ }\mbox{ }\mbox{ }\tilde{w}_{B_{s}^{\ast},a}^{(3)}\mbox{ }\mbox{ }\mbox{ }\mbox{ }$  & 
   $\mbox{ }\mbox{ }\mbox{ }\mbox{ }0\mbox{ }\mbox{ }\mbox{ }\mbox{ }$ & 
   $\mbox{ }\mbox{ }\mbox{ }\mbox{ }\frac{-1}{3}\mbox{ }\mbox{ }\mbox{ }\mbox{ }$ & 
   $\mbox{ }\mbox{ }\mbox{ }\mbox{ }0\mbox{ }\mbox{ }\mbox{ }\mbox{ }$ & 
   $\mbox{ }\mbox{ }\mbox{ }\mbox{ }0\mbox{ }\mbox{ }\mbox{ }\mbox{ }$ \\
\hline\hline
\end{tabular}
\caption{\label{tab:meson_waverenorm_table}Coefficients for heavy-light meson wavefunction renormalisation,  in Eqs.~(\ref{eq:meson_waverenorm_master}), in the isospin limit.}
\end{table}
\begin{table}[t]
\begin{tabular}{c|cccc}
   $\mbox{ }\mbox{ }\mbox{ }\mbox{ }a\mbox{ }\mbox{ }\mbox{ }\mbox{ }$ & 
   $\mbox{ }\mbox{ }\mbox{ }\mbox{ }u\mbox{ }\mbox{ }\mbox{ }\mbox{ }$ & 
   $\mbox{ }\mbox{ }\mbox{ }\mbox{ }s\mbox{ }\mbox{ }\mbox{ }\mbox{ }$ & 
   $\mbox{ }\mbox{ }\mbox{ }\mbox{ }u^{\prime}\mbox{ }\mbox{ }\mbox{ }\mbox{ }$ & 
   $\mbox{ }\mbox{ }\mbox{ }\mbox{ }s^{\prime}\mbox{ }\mbox{ }\mbox{ }\mbox{ }$ \\
\hline\hline
   $\mbox{ }\mbox{ }\mbox{ }\mbox{ }w_{T_{du},a}^{(2)}+u_{T_{du},a}^{(2)}\mbox{ }\mbox{ }\mbox{ }\mbox{ }$  & 
   $\mbox{ }\mbox{ }\mbox{ }\mbox{ }\frac{3}{2}\mbox{ }\mbox{ }\mbox{ }\mbox{ }$ & 
   $\mbox{ }\mbox{ }\mbox{ }\mbox{ }0\mbox{ }\mbox{ }\mbox{ }\mbox{ }$ & 
   $\mbox{ }\mbox{ }\mbox{ }\mbox{ }3\mbox{ }\mbox{ }\mbox{ }\mbox{ }$ & 
   $\mbox{ }\mbox{ }\mbox{ }\mbox{ }0\mbox{ }\mbox{ }\mbox{ }\mbox{ }$ \\
\hline
   $\mbox{ }\mbox{ }\mbox{ }\mbox{ }\tilde{w}_{T_{du},a}^{(2)}+\tilde{u}_{T_{du},a}^{(2)}\mbox{ }\mbox{ }\mbox{ }\mbox{ }$  & 
   $\mbox{ }\mbox{ }\mbox{ }\mbox{ }0\mbox{ }\mbox{ }\mbox{ }\mbox{ }$ & 
   $\mbox{ }\mbox{ }\mbox{ }\mbox{ }0\mbox{ }\mbox{ }\mbox{ }\mbox{ }$ & 
   $\mbox{ }\mbox{ }\mbox{ }\mbox{ }0\mbox{ }\mbox{ }\mbox{ }\mbox{ }$ & 
   $\mbox{ }\mbox{ }\mbox{ }\mbox{ }0\mbox{ }\mbox{ }\mbox{ }\mbox{ }$ \\
\hline\hline
   $\mbox{ }\mbox{ }\mbox{ }\mbox{ }w_{T_{du},a}^{(3)}+u_{T_{du},a}^{(3)}\mbox{ }\mbox{ }\mbox{ }\mbox{ }$  & 
   $\mbox{ }\mbox{ }\mbox{ }\mbox{ }\frac{3}{2}\mbox{ }\mbox{ }\mbox{ }\mbox{ }$ & 
   $\mbox{ }\mbox{ }\mbox{ }\mbox{ }0\mbox{ }\mbox{ }\mbox{ }\mbox{ }$ & 
   $\mbox{ }\mbox{ }\mbox{ }\mbox{ }3\mbox{ }\mbox{ }\mbox{ }\mbox{ }$ & 
   $\mbox{ }\mbox{ }\mbox{ }\mbox{ }\frac{3}{2}\mbox{ }\mbox{ }\mbox{ }\mbox{ }$ \\
\hline
   $\mbox{ }\mbox{ }\mbox{ }\mbox{ }\tilde{w}_{T_{du},a}^{(3)}+\tilde{u}_{T_{du},a}^{(3)}\mbox{ }\mbox{ }\mbox{ }\mbox{ }$  & 
   $\mbox{ }\mbox{ }\mbox{ }\mbox{ }0\mbox{ }\mbox{ }\mbox{ }\mbox{ }$ & 
   $\mbox{ }\mbox{ }\mbox{ }\mbox{ }0\mbox{ }\mbox{ }\mbox{ }\mbox{ }$ & 
   $\mbox{ }\mbox{ }\mbox{ }\mbox{ }0\mbox{ }\mbox{ }\mbox{ }\mbox{ }$ & 
   $\mbox{ }\mbox{ }\mbox{ }\mbox{ }0\mbox{ }\mbox{ }\mbox{ }\mbox{ }$ \\
\hline\hline
   $\mbox{ }\mbox{ }\mbox{ }\mbox{ }w_{T_{su},a}^{(3)}\mbox{ }\mbox{ }\mbox{ }\mbox{ }$  & 
   $\mbox{ }\mbox{ }\mbox{ }\mbox{ }0\mbox{ }\mbox{ }\mbox{ }\mbox{ }$ & 
   $\mbox{ }\mbox{ }\mbox{ }\mbox{ }\frac{3}{4}\mbox{ }\mbox{ }\mbox{ }\mbox{ }$ & 
   $\mbox{ }\mbox{ }\mbox{ }\mbox{ }\frac{3}{2}\mbox{ }\mbox{ }\mbox{ }\mbox{ }$ & 
   $\mbox{ }\mbox{ }\mbox{ }\mbox{ }\frac{3}{4}\mbox{ }\mbox{ }\mbox{ }\mbox{ }$ \\
\hline
   $\mbox{ }\mbox{ }\mbox{ }\mbox{ }\tilde{w}_{T_{su},a}^{(3)}\mbox{ }\mbox{ }\mbox{ }\mbox{ }$  & 
   $\mbox{ }\mbox{ }\mbox{ }\mbox{ }\frac{-1}{4}\mbox{ }\mbox{ }\mbox{ }\mbox{ }$ & 
   $\mbox{ }\mbox{ }\mbox{ }\mbox{ }0\mbox{ }\mbox{ }\mbox{ }\mbox{ }$ & 
   $\mbox{ }\mbox{ }\mbox{ }\mbox{ }0\mbox{ }\mbox{ }\mbox{ }\mbox{ }$ & 
   $\mbox{ }\mbox{ }\mbox{ }\mbox{ }0\mbox{ }\mbox{ }\mbox{ }\mbox{ }$ \\
\hline
   $\mbox{ }\mbox{ }\mbox{ }\mbox{ }u_{T_{su},a}^{(3)}\mbox{ }\mbox{ }\mbox{ }\mbox{ }$  & 
   $\mbox{ }\mbox{ }\mbox{ }\mbox{ }\frac{3}{4}\mbox{ }\mbox{ }\mbox{ }\mbox{ }$ & 
   $\mbox{ }\mbox{ }\mbox{ }\mbox{ }0\mbox{ }\mbox{ }\mbox{ }\mbox{ }$ & 
   $\mbox{ }\mbox{ }\mbox{ }\mbox{ }\frac{3}{2}\mbox{ }\mbox{ }\mbox{ }\mbox{ }$ & 
   $\mbox{ }\mbox{ }\mbox{ }\mbox{ }\frac{3}{4}\mbox{ }\mbox{ }\mbox{ }\mbox{ }$ \\
\hline
   $\mbox{ }\mbox{ }\mbox{ }\mbox{ }\tilde{u}_{T_{su},a}^{(3)}\mbox{ }\mbox{ }\mbox{ }\mbox{ }$  & 
   $\mbox{ }\mbox{ }\mbox{ }\mbox{ }\frac{1}{2}\mbox{ }\mbox{ }\mbox{ }\mbox{ }$ & 
   $\mbox{ }\mbox{ }\mbox{ }\mbox{ }\frac{-1}{4}\mbox{ }\mbox{ }\mbox{ }\mbox{ }$ & 
   $\mbox{ }\mbox{ }\mbox{ }\mbox{ }0\mbox{ }\mbox{ }\mbox{ }\mbox{ }$ & 
   $\mbox{ }\mbox{ }\mbox{ }\mbox{ }0\mbox{ }\mbox{ }\mbox{ }\mbox{ }$ \\
\hline\hline
\end{tabular}
\caption{\label{tab:T_waverenorm_table}Coefficients for $T_{ij}$ baryon wavefunction renormalisation,  in Eq.~(\ref{eq:baryon_waverenorm_master}), in the isospin limit.}
\end{table}
\begin{table}[t]
\begin{tabular}{c|cccc}
   $\mbox{ }\mbox{ }\mbox{ }\mbox{ }a\mbox{ }\mbox{ }\mbox{ }\mbox{ }$ & 
   $\mbox{ }\mbox{ }\mbox{ }\mbox{ }u\mbox{ }\mbox{ }\mbox{ }\mbox{ }$ & 
   $\mbox{ }\mbox{ }\mbox{ }\mbox{ }s\mbox{ }\mbox{ }\mbox{ }\mbox{ }$ & 
   $\mbox{ }\mbox{ }\mbox{ }\mbox{ }u^{\prime}\mbox{ }\mbox{ }\mbox{ }\mbox{ }$ & 
   $\mbox{ }\mbox{ }\mbox{ }\mbox{ }s^{\prime}\mbox{ }\mbox{ }\mbox{ }\mbox{ }$ \\
\hline\hline
   $\mbox{ }\mbox{ }\mbox{ }\mbox{ }w_{S_{du},a}^{\prime (2)}+u_{S_{du},a}^{\prime (2)}\mbox{ }\mbox{ }\mbox{ }\mbox{ }$  & 
   $\mbox{ }\mbox{ }\mbox{ }\mbox{ }\frac{-1}{2}\mbox{ }\mbox{ }\mbox{ }\mbox{ }$ & 
   $\mbox{ }\mbox{ }\mbox{ }\mbox{ }0\mbox{ }\mbox{ }\mbox{ }\mbox{ }$ & 
   $\mbox{ }\mbox{ }\mbox{ }\mbox{ }1\mbox{ }\mbox{ }\mbox{ }\mbox{ }$ & 
   $\mbox{ }\mbox{ }\mbox{ }\mbox{ }0\mbox{ }\mbox{ }\mbox{ }\mbox{ }$ \\
\hline
   $\mbox{ }\mbox{ }\mbox{ }\mbox{ }\tilde{w}_{S_{du},a}^{\prime (2)}+\tilde{u}_{S_{du},a}^{\prime (2)}\mbox{ }\mbox{ }\mbox{ }\mbox{ }$  & 
   $\mbox{ }\mbox{ }\mbox{ }\mbox{ }0\mbox{ }\mbox{ }\mbox{ }\mbox{ }$ & 
   $\mbox{ }\mbox{ }\mbox{ }\mbox{ }0\mbox{ }\mbox{ }\mbox{ }\mbox{ }$ & 
   $\mbox{ }\mbox{ }\mbox{ }\mbox{ }0\mbox{ }\mbox{ }\mbox{ }\mbox{ }$ & 
   $\mbox{ }\mbox{ }\mbox{ }\mbox{ }0\mbox{ }\mbox{ }\mbox{ }\mbox{ }$ \\
\hline
   $\mbox{ }\mbox{ }\mbox{ }\mbox{ }w_{S_{du},a}^{(2)}+u_{S_{du},a}^{(2)}\mbox{ }\mbox{ }\mbox{ }\mbox{ }$  & 
   $\mbox{ }\mbox{ }\mbox{ }\mbox{ }\frac{1}{2}\mbox{ }\mbox{ }\mbox{ }\mbox{ }$ & 
   $\mbox{ }\mbox{ }\mbox{ }\mbox{ }0\mbox{ }\mbox{ }\mbox{ }\mbox{ }$ & 
   $\mbox{ }\mbox{ }\mbox{ }\mbox{ }1\mbox{ }\mbox{ }\mbox{ }\mbox{ }$ & 
   $\mbox{ }\mbox{ }\mbox{ }\mbox{ }0\mbox{ }\mbox{ }\mbox{ }\mbox{ }$ \\
\hline
   $\mbox{ }\mbox{ }\mbox{ }\mbox{ }\tilde{w}_{S_{du},a}^{(2)}+\tilde{u}_{S_{du},a}^{(2)}\mbox{ }\mbox{ }\mbox{ }\mbox{ }$  & 
   $\mbox{ }\mbox{ }\mbox{ }\mbox{ }\frac{-1}{2}\mbox{ }\mbox{ }\mbox{ }\mbox{ }$ & 
   $\mbox{ }\mbox{ }\mbox{ }\mbox{ }0\mbox{ }\mbox{ }\mbox{ }\mbox{ }$ & 
   $\mbox{ }\mbox{ }\mbox{ }\mbox{ }0\mbox{ }\mbox{ }\mbox{ }\mbox{ }$ & 
   $\mbox{ }\mbox{ }\mbox{ }\mbox{ }0\mbox{ }\mbox{ }\mbox{ }\mbox{ }$ \\
\hline\hline
   $\mbox{ }\mbox{ }\mbox{ }\mbox{ }w_{S_{du},a}^{\prime (3)}+u_{S_{du},a}^{\prime (3)}\mbox{ }\mbox{ }\mbox{ }\mbox{ }$  & 
   $\mbox{ }\mbox{ }\mbox{ }\mbox{ }\frac{-1}{2}\mbox{ }\mbox{ }\mbox{ }\mbox{ }$ & 
   $\mbox{ }\mbox{ }\mbox{ }\mbox{ }0\mbox{ }\mbox{ }\mbox{ }\mbox{ }$ & 
   $\mbox{ }\mbox{ }\mbox{ }\mbox{ }1\mbox{ }\mbox{ }\mbox{ }\mbox{ }$ & 
   $\mbox{ }\mbox{ }\mbox{ }\mbox{ }\frac{1}{2}\mbox{ }\mbox{ }\mbox{ }\mbox{ }$ \\
\hline
   $\mbox{ }\mbox{ }\mbox{ }\mbox{ }\tilde{w}_{S_{du},a}^{\prime (3)}+\tilde{u}_{S_{du},a}^{\prime (3)}\mbox{ }\mbox{ }\mbox{ }\mbox{ }$  & 
   $\mbox{ }\mbox{ }\mbox{ }\mbox{ }0\mbox{ }\mbox{ }\mbox{ }\mbox{ }$ & 
   $\mbox{ }\mbox{ }\mbox{ }\mbox{ }0\mbox{ }\mbox{ }\mbox{ }\mbox{ }$ & 
   $\mbox{ }\mbox{ }\mbox{ }\mbox{ }0\mbox{ }\mbox{ }\mbox{ }\mbox{ }$ & 
   $\mbox{ }\mbox{ }\mbox{ }\mbox{ }0\mbox{ }\mbox{ }\mbox{ }\mbox{ }$ \\
\hline
   $\mbox{ }\mbox{ }\mbox{ }\mbox{ }w_{S_{du},a}^{(3)}+u_{S_{du},a}^{(3)}\mbox{ }\mbox{ }\mbox{ }\mbox{ }$  & 
   $\mbox{ }\mbox{ }\mbox{ }\mbox{ }\frac{1}{2}\mbox{ }\mbox{ }\mbox{ }\mbox{ }$ & 
   $\mbox{ }\mbox{ }\mbox{ }\mbox{ }0\mbox{ }\mbox{ }\mbox{ }\mbox{ }$ & 
   $\mbox{ }\mbox{ }\mbox{ }\mbox{ }1\mbox{ }\mbox{ }\mbox{ }\mbox{ }$ & 
   $\mbox{ }\mbox{ }\mbox{ }\mbox{ }\frac{1}{2}\mbox{ }\mbox{ }\mbox{ }\mbox{ }$ \\
\hline
   $\mbox{ }\mbox{ }\mbox{ }\mbox{ }\tilde{w}_{S_{du},a}^{(3)}+\tilde{u}_{S_{du},a}^{(3)}\mbox{ }\mbox{ }\mbox{ }\mbox{ }$  & 
   $\mbox{ }\mbox{ }\mbox{ }\mbox{ }\frac{-1}{3}\mbox{ }\mbox{ }\mbox{ }\mbox{ }$ & 
   $\mbox{ }\mbox{ }\mbox{ }\mbox{ }0\mbox{ }\mbox{ }\mbox{ }\mbox{ }$ & 
   $\mbox{ }\mbox{ }\mbox{ }\mbox{ }0\mbox{ }\mbox{ }\mbox{ }\mbox{ }$ & 
   $\mbox{ }\mbox{ }\mbox{ }\mbox{ }0\mbox{ }\mbox{ }\mbox{ }\mbox{ }$ \\
\hline\hline
   $\mbox{ }\mbox{ }\mbox{ }\mbox{ }w_{S_{su},a}^{\prime (3)}\mbox{ }\mbox{ }\mbox{ }\mbox{ }$  & 
   $\mbox{ }\mbox{ }\mbox{ }\mbox{ }0\mbox{ }\mbox{ }\mbox{ }\mbox{ }$ & 
   $\mbox{ }\mbox{ }\mbox{ }\mbox{ }\frac{-1}{4}\mbox{ }\mbox{ }\mbox{ }\mbox{ }$ & 
   $\mbox{ }\mbox{ }\mbox{ }\mbox{ }\frac{1}{2}\mbox{ }\mbox{ }\mbox{ }\mbox{ }$ & 
   $\mbox{ }\mbox{ }\mbox{ }\mbox{ }\frac{1}{4}\mbox{ }\mbox{ }\mbox{ }\mbox{ }$ \\
\hline
   $\mbox{ }\mbox{ }\mbox{ }\mbox{ }\tilde{w}_{S_{su},a}^{\prime (3)}\mbox{ }\mbox{ }\mbox{ }\mbox{ }$  & 
   $\mbox{ }\mbox{ }\mbox{ }\mbox{ }\frac{-1}{12}\mbox{ }\mbox{ }\mbox{ }\mbox{ }$ & 
   $\mbox{ }\mbox{ }\mbox{ }\mbox{ }0\mbox{ }\mbox{ }\mbox{ }\mbox{ }$ & 
   $\mbox{ }\mbox{ }\mbox{ }\mbox{ }0\mbox{ }\mbox{ }\mbox{ }\mbox{ }$ & 
   $\mbox{ }\mbox{ }\mbox{ }\mbox{ }0\mbox{ }\mbox{ }\mbox{ }\mbox{ }$ \\
\hline
   $\mbox{ }\mbox{ }\mbox{ }\mbox{ }u_{S_{su},a}^{\prime (3)}\mbox{ }\mbox{ }\mbox{ }\mbox{ }$  & 
   $\mbox{ }\mbox{ }\mbox{ }\mbox{ }\frac{-1}{4}\mbox{ }\mbox{ }\mbox{ }\mbox{ }$ & 
   $\mbox{ }\mbox{ }\mbox{ }\mbox{ }0\mbox{ }\mbox{ }\mbox{ }\mbox{ }$ & 
   $\mbox{ }\mbox{ }\mbox{ }\mbox{ }\frac{1}{2}\mbox{ }\mbox{ }\mbox{ }\mbox{ }$ & 
   $\mbox{ }\mbox{ }\mbox{ }\mbox{ }\frac{1}{4}\mbox{ }\mbox{ }\mbox{ }\mbox{ }$ \\
\hline
   $\mbox{ }\mbox{ }\mbox{ }\mbox{ }\tilde{u}_{S_{su},a}^{\prime (3)}\mbox{ }\mbox{ }\mbox{ }\mbox{ }$  & 
   $\mbox{ }\mbox{ }\mbox{ }\mbox{ }\frac{1}{6}\mbox{ }\mbox{ }\mbox{ }\mbox{ }$ & 
   $\mbox{ }\mbox{ }\mbox{ }\mbox{ }\frac{-1}{12}\mbox{ }\mbox{ }\mbox{ }\mbox{ }$ & 
   $\mbox{ }\mbox{ }\mbox{ }\mbox{ }0\mbox{ }\mbox{ }\mbox{ }\mbox{ }$ & 
   $\mbox{ }\mbox{ }\mbox{ }\mbox{ }0\mbox{ }\mbox{ }\mbox{ }\mbox{ }$ \\
\hline
   $\mbox{ }\mbox{ }\mbox{ }\mbox{ }w_{S_{su},a}^{(3)}\mbox{ }\mbox{ }\mbox{ }\mbox{ }$  & 
   $\mbox{ }\mbox{ }\mbox{ }\mbox{ }0\mbox{ }\mbox{ }\mbox{ }\mbox{ }$ & 
   $\mbox{ }\mbox{ }\mbox{ }\mbox{ }\frac{1}{4}\mbox{ }\mbox{ }\mbox{ }\mbox{ }$ & 
   $\mbox{ }\mbox{ }\mbox{ }\mbox{ }\frac{1}{2}\mbox{ }\mbox{ }\mbox{ }\mbox{ }$ & 
   $\mbox{ }\mbox{ }\mbox{ }\mbox{ }\frac{1}{4}\mbox{ }\mbox{ }\mbox{ }\mbox{ }$ \\
\hline
   $\mbox{ }\mbox{ }\mbox{ }\mbox{ }\tilde{w}_{S_{su},a}^{(3)}\mbox{ }\mbox{ }\mbox{ }\mbox{ }$  & 
   $\mbox{ }\mbox{ }\mbox{ }\mbox{ }\frac{-1}{12}\mbox{ }\mbox{ }\mbox{ }\mbox{ }$ & 
   $\mbox{ }\mbox{ }\mbox{ }\mbox{ }0\mbox{ }\mbox{ }\mbox{ }\mbox{ }$ & 
   $\mbox{ }\mbox{ }\mbox{ }\mbox{ }0\mbox{ }\mbox{ }\mbox{ }\mbox{ }$ & 
   $\mbox{ }\mbox{ }\mbox{ }\mbox{ }0\mbox{ }\mbox{ }\mbox{ }\mbox{ }$ \\
\hline
   $\mbox{ }\mbox{ }\mbox{ }\mbox{ }u_{S_{su},a}^{(3)}\mbox{ }\mbox{ }\mbox{ }\mbox{ }$  & 
   $\mbox{ }\mbox{ }\mbox{ }\mbox{ }\frac{1}{4}\mbox{ }\mbox{ }\mbox{ }\mbox{ }$ & 
   $\mbox{ }\mbox{ }\mbox{ }\mbox{ }0\mbox{ }\mbox{ }\mbox{ }\mbox{ }$ & 
   $\mbox{ }\mbox{ }\mbox{ }\mbox{ }\frac{1}{2}\mbox{ }\mbox{ }\mbox{ }\mbox{ }$ & 
   $\mbox{ }\mbox{ }\mbox{ }\mbox{ }\frac{1}{4}\mbox{ }\mbox{ }\mbox{ }\mbox{ }$ \\
\hline
   $\mbox{ }\mbox{ }\mbox{ }\mbox{ }\tilde{u}_{S_{su},a}^{(3)}\mbox{ }\mbox{ }\mbox{ }\mbox{ }$  & 
   $\mbox{ }\mbox{ }\mbox{ }\mbox{ }\frac{-1}{6}\mbox{ }\mbox{ }\mbox{ }\mbox{ }$ & 
   $\mbox{ }\mbox{ }\mbox{ }\mbox{ }\frac{-1}{12}\mbox{ }\mbox{ }\mbox{ }\mbox{ }$ & 
   $\mbox{ }\mbox{ }\mbox{ }\mbox{ }0\mbox{ }\mbox{ }\mbox{ }\mbox{ }$ & 
   $\mbox{ }\mbox{ }\mbox{ }\mbox{ }0\mbox{ }\mbox{ }\mbox{ }\mbox{ }$ \\
\hline\hline
   $\mbox{ }\mbox{ }\mbox{ }\mbox{ }w_{S_{ss},a}^{\prime (3)}+u_{S_{ss},a}^{\prime (3)}\mbox{ }\mbox{ }\mbox{ }\mbox{ }$  & 
   $\mbox{ }\mbox{ }\mbox{ }\mbox{ }0\mbox{ }\mbox{ }\mbox{ }\mbox{ }$ & 
   $\mbox{ }\mbox{ }\mbox{ }\mbox{ }\frac{-1}{2}\mbox{ }\mbox{ }\mbox{ }\mbox{ }$ & 
   $\mbox{ }\mbox{ }\mbox{ }\mbox{ }1\mbox{ }\mbox{ }\mbox{ }\mbox{ }$ & 
   $\mbox{ }\mbox{ }\mbox{ }\mbox{ }\frac{1}{2}\mbox{ }\mbox{ }\mbox{ }\mbox{ }$ \\
\hline
   $\mbox{ }\mbox{ }\mbox{ }\mbox{ }\tilde{w}_{S_{ss},a}^{\prime (3)}+\tilde{u}_{S_{ss},a}^{\prime (3)}\mbox{ }\mbox{ }\mbox{ }\mbox{ }$  & 
   $\mbox{ }\mbox{ }\mbox{ }\mbox{ }0\mbox{ }\mbox{ }\mbox{ }\mbox{ }$ & 
   $\mbox{ }\mbox{ }\mbox{ }\mbox{ }0\mbox{ }\mbox{ }\mbox{ }\mbox{ }$ & 
   $\mbox{ }\mbox{ }\mbox{ }\mbox{ }0\mbox{ }\mbox{ }\mbox{ }\mbox{ }$ & 
   $\mbox{ }\mbox{ }\mbox{ }\mbox{ }0\mbox{ }\mbox{ }\mbox{ }\mbox{ }$ \\
\hline
   $\mbox{ }\mbox{ }\mbox{ }\mbox{ }w_{S_{ss},a}^{(3)}+u_{S_{ss},a}^{(3)}\mbox{ }\mbox{ }\mbox{ }\mbox{ }$  & 
   $\mbox{ }\mbox{ }\mbox{ }\mbox{ }0\mbox{ }\mbox{ }\mbox{ }\mbox{ }$ & 
   $\mbox{ }\mbox{ }\mbox{ }\mbox{ }\frac{1}{2}\mbox{ }\mbox{ }\mbox{ }\mbox{ }$ & 
   $\mbox{ }\mbox{ }\mbox{ }\mbox{ }1\mbox{ }\mbox{ }\mbox{ }\mbox{ }$ & 
   $\mbox{ }\mbox{ }\mbox{ }\mbox{ }\frac{1}{2}\mbox{ }\mbox{ }\mbox{ }\mbox{ }$ \\
\hline
   $\mbox{ }\mbox{ }\mbox{ }\mbox{ }\tilde{w}_{S_{ss},a}^{(3)}+\tilde{u}_{S_{ss},a}^{(3)}\mbox{ }\mbox{ }\mbox{ }\mbox{ }$  & 
   $\mbox{ }\mbox{ }\mbox{ }\mbox{ }0\mbox{ }\mbox{ }\mbox{ }\mbox{ }$ & 
   $\mbox{ }\mbox{ }\mbox{ }\mbox{ }\frac{-1}{3}\mbox{ }\mbox{ }\mbox{ }\mbox{ }$ & 
   $\mbox{ }\mbox{ }\mbox{ }\mbox{ }0\mbox{ }\mbox{ }\mbox{ }\mbox{ }$ & 
   $\mbox{ }\mbox{ }\mbox{ }\mbox{ }0\mbox{ }\mbox{ }\mbox{ }\mbox{ }$ \\
\hline\hline
\end{tabular}
\caption{\label{tab:S_waverenorm_table}Coefficients for $S_{ij}$ baryon wavefunction renormalisation,  in Eq.~(\ref{eq:baryon_waverenorm_master}), in the isospin limit.}
\end{table}

\section{Coefficients for the sunset diagrams}
\label{app:sunset_tables}
In this Appendix, we present the coefficients in Eqs.~(\ref{eq:meson_sunset_master}) and (\ref{eq:baryon_sunset_master}) relevant to the matrix elements investigated in this work.
Because of  the isospin symmetry, it is impossible to distinguish between some $\tilde{y}$ coefficients and their $\tilde{x}$ counterparts.  For such cases, we put the symbol $+\tilde{x}$ in Table~\ref{tab:y_tilde}, and then present $\tilde{x} + \tilde{y}$ in Tables~\ref{tab:baryon_tilde_x}, \ref{tab:baryon_tilde_x_prime} and \ref{tab:baryon_tilde_x_double_prime} in the form that
$\tilde{x}$ is written as (number - $\tilde{y}$).
\begin{table}[t]
\begin{tabular}{c|cccc}
   $\mbox{ }\mbox{ }\mbox{ }\mbox{ }a\mbox{ }\mbox{ }\mbox{ }\mbox{ }$ & 
   $\mbox{ }\mbox{ }\mbox{ }\mbox{ }\mbox{ }\mbox{ }u\mbox{ }\mbox{ }\mbox{ }\mbox{ }\mbox{ }\mbox{ }$ & 
   $\mbox{ }\mbox{ }\mbox{ }\mbox{ }\mbox{ }\mbox{ }s\mbox{ }\mbox{ }\mbox{ }\mbox{ }\mbox{ }\mbox{ }$ &
   $\mbox{ }\mbox{ }\mbox{ }\mbox{ }\mbox{ }\mbox{ }u^{\prime}\mbox{ }\mbox{ }\mbox{ }\mbox{ }\mbox{ }\mbox{ }$ & 
   $\mbox{ }\mbox{ }\mbox{ }\mbox{ }\mbox{ }\mbox{ }s^{\prime}\mbox{ }\mbox{ }\mbox{ }\mbox{ }\mbox{ }\mbox{ }$ \\
\hline\hline
   $\mbox{ }\mbox{ }\mbox{ }\mbox{ }x^{(2)}_{T_{du},S_{dd},a}\mbox{ }\mbox{ }\mbox{ }\mbox{ }$ & 
   $\mbox{ }\mbox{ }\mbox{ }\mbox{ }\mbox{ }\mbox{ }-1\mbox{ }\mbox{ }\mbox{ }\mbox{ }\mbox{ }\mbox{ }$ & 
   $\mbox{ }\mbox{ }\mbox{ }\mbox{ }\mbox{ }\mbox{ }0\mbox{ }\mbox{ }\mbox{ }\mbox{ }\mbox{ }\mbox{ }$  &
   $\mbox{ }\mbox{ }\mbox{ }\mbox{ }\mbox{ }\mbox{ }-1\mbox{ }\mbox{ }\mbox{ }\mbox{ }\mbox{ }\mbox{ }$ & 
   $\mbox{ }\mbox{ }\mbox{ }\mbox{ }\mbox{ }\mbox{ }0\mbox{ }\mbox{ }\mbox{ }\mbox{ }\mbox{ }\mbox{ }$ \\
\hline
   $\mbox{ }\mbox{ }\mbox{ }\mbox{ }x^{(3)}_{T_{du},S_{dd},a}\mbox{ }\mbox{ }\mbox{ }\mbox{ }$ & 
   $\mbox{ }\mbox{ }\mbox{ }\mbox{ }\mbox{ }\mbox{ }-1\mbox{ }\mbox{ }\mbox{ }\mbox{ }\mbox{ }\mbox{ }$ & 
   $\mbox{ }\mbox{ }\mbox{ }\mbox{ }\mbox{ }\mbox{ }0\mbox{ }\mbox{ }\mbox{ }\mbox{ }\mbox{ }\mbox{ }$  &
   $\mbox{ }\mbox{ }\mbox{ }\mbox{ }\mbox{ }\mbox{ }-1\mbox{ }\mbox{ }\mbox{ }\mbox{ }\mbox{ }\mbox{ }$ & 
   $\mbox{ }\mbox{ }\mbox{ }\mbox{ }\mbox{ }\mbox{ }\frac{-1}{2}\mbox{ }\mbox{ }\mbox{ }\mbox{ }\mbox{ }\mbox{ }$ \\
\hline
   $\mbox{ }\mbox{ }\mbox{ }\mbox{ }x^{(3)}_{T_{su},S_{sd},a}\mbox{ }\mbox{ }\mbox{ }\mbox{ }$ & 
   $\mbox{ }\mbox{ }\mbox{ }\mbox{ }\mbox{ }\mbox{ }-1\mbox{ }\mbox{ }\mbox{ }\mbox{ }\mbox{ }\mbox{ }$ & 
   $\mbox{ }\mbox{ }\mbox{ }\mbox{ }\mbox{ }\mbox{ }0\mbox{ }\mbox{ }\mbox{ }\mbox{ }\mbox{ }\mbox{ }$ &
   $\mbox{ }\mbox{ }\mbox{ }\mbox{ }\mbox{ }\mbox{ }-1\mbox{ }\mbox{ }\mbox{ }\mbox{ }\mbox{ }\mbox{ }$ & 
   $\mbox{ }\mbox{ }\mbox{ }\mbox{ }\mbox{ }\mbox{ }\frac{-1}{2}\mbox{ }\mbox{ }\mbox{ }\mbox{ }\mbox{ }\mbox{ }$ \\
\hline
   $\mbox{ }\mbox{ }\mbox{ }\mbox{ }x^{(3)}_{T_{du},S_{ds},a}\mbox{ }\mbox{ }\mbox{ }\mbox{ }$ & 
   $\mbox{ }\mbox{ }\mbox{ }\mbox{ }\mbox{ }\mbox{ }\frac{-1}{2}\mbox{ }\mbox{ }\mbox{ }\mbox{ }\mbox{ }\mbox{ }$ & 
   $\mbox{ }\mbox{ }\mbox{ }\mbox{ }\mbox{ }\mbox{ }\frac{-1}{2}\mbox{ }\mbox{ }\mbox{ }\mbox{ }\mbox{ }\mbox{ }$ &
   $\mbox{ }\mbox{ }\mbox{ }\mbox{ }\mbox{ }\mbox{ }-1\mbox{ }\mbox{ }\mbox{ }\mbox{ }\mbox{ }\mbox{ }$ & 
   $\mbox{ }\mbox{ }\mbox{ }\mbox{ }\mbox{ }\mbox{ }\frac{-1}{2}\mbox{ }\mbox{ }\mbox{ }\mbox{ }\mbox{ }\mbox{ }$ \\
\hline
   $\mbox{ }\mbox{ }\mbox{ }\mbox{ }x^{(3)}_{T_{su},S_{ss},a}\mbox{ }\mbox{ }\mbox{ }\mbox{ }$ & 
   $\mbox{ }\mbox{ }\mbox{ }\mbox{ }\mbox{ }\mbox{ }\frac{-1}{2}\mbox{ }\mbox{ }\mbox{ }\mbox{ }\mbox{ }\mbox{ }$ & 
   $\mbox{ }\mbox{ }\mbox{ }\mbox{ }\mbox{ }\mbox{ }\frac{-1}{2}\mbox{ }\mbox{ }\mbox{ }\mbox{ }\mbox{ }\mbox{ }$ &
   $\mbox{ }\mbox{ }\mbox{ }\mbox{ }\mbox{ }\mbox{ }-1\mbox{ }\mbox{ }\mbox{ }\mbox{ }\mbox{ }\mbox{ }$ & 
   $\mbox{ }\mbox{ }\mbox{ }\mbox{ }\mbox{ }\mbox{ }\frac{-1}{2}\mbox{ }\mbox{ }\mbox{ }\mbox{ }\mbox{ }\mbox{ }$ \\
\hline\hline
   $\mbox{ }\mbox{ }\mbox{ }\mbox{ }x^{(2)}_{S_{du},S_{dd},a}\mbox{ }\mbox{ }\mbox{ }\mbox{ }$ & 
   $\mbox{ }\mbox{ }\mbox{ }\mbox{ }\mbox{ }\mbox{ }\frac{-1}{2}\mbox{ }\mbox{ }\mbox{ }\mbox{ }\mbox{ }\mbox{ }$ & 
   $\mbox{ }\mbox{ }\mbox{ }\mbox{ }\mbox{ }\mbox{ }0\mbox{ }\mbox{ }\mbox{ }\mbox{ }\mbox{ }\mbox{ }$ &
   $\mbox{ }\mbox{ }\mbox{ }\mbox{ }\mbox{ }\mbox{ }\frac{-1}{2}\mbox{ }\mbox{ }\mbox{ }\mbox{ }\mbox{ }\mbox{ }$ & 
   $\mbox{ }\mbox{ }\mbox{ }\mbox{ }\mbox{ }\mbox{ }0\mbox{ }\mbox{ }\mbox{ }\mbox{ }\mbox{ }\mbox{ }$ \\
\hline
   $\mbox{ }\mbox{ }\mbox{ }\mbox{ }x^{(3)}_{S_{du},S_{dd},a}\mbox{ }\mbox{ }\mbox{ }\mbox{ }$ & 
   $\mbox{ }\mbox{ }\mbox{ }\mbox{ }\mbox{ }\mbox{ }\frac{-1}{2}\mbox{ }\mbox{ }\mbox{ }\mbox{ }\mbox{ }\mbox{ }$ & 
   $\mbox{ }\mbox{ }\mbox{ }\mbox{ }\mbox{ }\mbox{ }0\mbox{ }\mbox{ }\mbox{ }\mbox{ }\mbox{ }\mbox{ }$ &
   $\mbox{ }\mbox{ }\mbox{ }\mbox{ }\mbox{ }\mbox{ }\frac{-1}{2}\mbox{ }\mbox{ }\mbox{ }\mbox{ }\mbox{ }\mbox{ }$ & 
   $\mbox{ }\mbox{ }\mbox{ }\mbox{ }\mbox{ }\mbox{ }\frac{-1}{4}\mbox{ }\mbox{ }\mbox{ }\mbox{ }\mbox{ }\mbox{ }$ \\
\hline
   $\mbox{ }\mbox{ }\mbox{ }\mbox{ }x^{(3)}_{S_{su},S_{sd},a}\mbox{ }\mbox{ }\mbox{ }\mbox{ }$ & 
   $\mbox{ }\mbox{ }\mbox{ }\mbox{ }\mbox{ }\mbox{ }\frac{-1}{2}\mbox{ }\mbox{ }\mbox{ }\mbox{ }\mbox{ }\mbox{ }$ & 
   $\mbox{ }\mbox{ }\mbox{ }\mbox{ }\mbox{ }\mbox{ }0\mbox{ }\mbox{ }\mbox{ }\mbox{ }\mbox{ }\mbox{ }$ &
   $\mbox{ }\mbox{ }\mbox{ }\mbox{ }\mbox{ }\mbox{ }\frac{-1}{2}\mbox{ }\mbox{ }\mbox{ }\mbox{ }\mbox{ }\mbox{ }$ & 
   $\mbox{ }\mbox{ }\mbox{ }\mbox{ }\mbox{ }\mbox{ }\frac{-1}{4}\mbox{ }\mbox{ }\mbox{ }\mbox{ }\mbox{ }\mbox{ }$ \\
\hline
   $\mbox{ }\mbox{ }\mbox{ }\mbox{ }x^{(3)}_{S_{du},S_{ds},a}\mbox{ }\mbox{ }\mbox{ }\mbox{ }$ & 
   $\mbox{ }\mbox{ }\mbox{ }\mbox{ }\mbox{ }\mbox{ }\frac{-1}{4}\mbox{ }\mbox{ }\mbox{ }\mbox{ }\mbox{ }\mbox{ }$ & 
   $\mbox{ }\mbox{ }\mbox{ }\mbox{ }\mbox{ }\mbox{ }\frac{-1}{4}\mbox{ }\mbox{ }\mbox{ }\mbox{ }\mbox{ }\mbox{ }$ &
   $\mbox{ }\mbox{ }\mbox{ }\mbox{ }\mbox{ }\mbox{ }\frac{-1}{2}\mbox{ }\mbox{ }\mbox{ }\mbox{ }\mbox{ }\mbox{ }$ & 
   $\mbox{ }\mbox{ }\mbox{ }\mbox{ }\mbox{ }\mbox{ }\frac{-1}{4}\mbox{ }\mbox{ }\mbox{ }\mbox{ }\mbox{ }\mbox{ }$ \\
\hline
   $\mbox{ }\mbox{ }\mbox{ }\mbox{ }x^{(3)}_{S_{su},S_{ss},a}\mbox{ }\mbox{ }\mbox{ }\mbox{ }$ & 
   $\mbox{ }\mbox{ }\mbox{ }\mbox{ }\mbox{ }\mbox{ }\frac{-1}{4}\mbox{ }\mbox{ }\mbox{ }\mbox{ }\mbox{ }\mbox{ }$ & 
   $\mbox{ }\mbox{ }\mbox{ }\mbox{ }\mbox{ }\mbox{ }\frac{-1}{4}\mbox{ }\mbox{ }\mbox{ }\mbox{ }\mbox{ }\mbox{ }$ &
   $\mbox{ }\mbox{ }\mbox{ }\mbox{ }\mbox{ }\mbox{ }\frac{-1}{2}\mbox{ }\mbox{ }\mbox{ }\mbox{ }\mbox{ }\mbox{ }$ & 
   $\mbox{ }\mbox{ }\mbox{ }\mbox{ }\mbox{ }\mbox{ }\frac{-1}{4}\mbox{ }\mbox{ }\mbox{ }\mbox{ }\mbox{ }\mbox{ }$ \\
\hline\hline
\end{tabular}
\caption{\label{tab:baryon_x}Coefficients $x^{(N_{f})}_{H_{1},H_{2}}$ in Eq.~(\ref{eq:baryon_sunset_master}), in the isospin limit.}
\end{table}
\begin{table}[t]
\begin{tabular}{c|cccc}
   $\mbox{ }\mbox{ }\mbox{ }\mbox{ }a\mbox{ }\mbox{ }\mbox{ }\mbox{ }$ & 
   $\mbox{ }\mbox{ }\mbox{ }\mbox{ }\mbox{ }\mbox{ }u\mbox{ }\mbox{ }\mbox{ }\mbox{ }\mbox{ }\mbox{ }$ & 
   $\mbox{ }\mbox{ }\mbox{ }\mbox{ }\mbox{ }\mbox{ }s\mbox{ }\mbox{ }\mbox{ }\mbox{ }\mbox{ }\mbox{ }$ &
   $\mbox{ }\mbox{ }\mbox{ }\mbox{ }\mbox{ }\mbox{ }u^{\prime}\mbox{ }\mbox{ }\mbox{ }\mbox{ }\mbox{ }\mbox{ }$ & 
   $\mbox{ }\mbox{ }\mbox{ }\mbox{ }\mbox{ }\mbox{ }s^{\prime}\mbox{ }\mbox{ }\mbox{ }\mbox{ }\mbox{ }\mbox{ }$ \\
\hline\hline
   $\mbox{ }\mbox{ }\mbox{ }\mbox{ }x^{\prime (2)}_{S_{du},S_{dd},a}\mbox{ }\mbox{ }\mbox{ }\mbox{ }$ & 
   $\mbox{ }\mbox{ }\mbox{ }\mbox{ }\mbox{ }\mbox{ }0\mbox{ }\mbox{ }\mbox{ }\mbox{ }\mbox{ }\mbox{ }$ & 
   $\mbox{ }\mbox{ }\mbox{ }\mbox{ }\mbox{ }\mbox{ }0\mbox{ }\mbox{ }\mbox{ }\mbox{ }\mbox{ }\mbox{ }$ &
   $\mbox{ }\mbox{ }\mbox{ }\mbox{ }\mbox{ }\mbox{ }-1\mbox{ }\mbox{ }\mbox{ }\mbox{ }\mbox{ }\mbox{ }$ & 
   $\mbox{ }\mbox{ }\mbox{ }\mbox{ }\mbox{ }\mbox{ }0\mbox{ }\mbox{ }\mbox{ }\mbox{ }\mbox{ }\mbox{ }$ \\
\hline
   $\mbox{ }\mbox{ }\mbox{ }\mbox{ }x^{\prime (3)}_{S_{du},S_{dd},a}\mbox{ }\mbox{ }\mbox{ }\mbox{ }$ & 
   $\mbox{ }\mbox{ }\mbox{ }\mbox{ }\mbox{ }\mbox{ }0\mbox{ }\mbox{ }\mbox{ }\mbox{ }\mbox{ }\mbox{ }$ & 
   $\mbox{ }\mbox{ }\mbox{ }\mbox{ }\mbox{ }\mbox{ }0\mbox{ }\mbox{ }\mbox{ }\mbox{ }\mbox{ }\mbox{ }$ &
   $\mbox{ }\mbox{ }\mbox{ }\mbox{ }\mbox{ }\mbox{ }-1\mbox{ }\mbox{ }\mbox{ }\mbox{ }\mbox{ }\mbox{ }$ & 
   $\mbox{ }\mbox{ }\mbox{ }\mbox{ }\mbox{ }\mbox{ }\frac{-1}{2}\mbox{ }\mbox{ }\mbox{ }\mbox{ }\mbox{ }\mbox{ }$ \\
\hline
   $\mbox{ }\mbox{ }\mbox{ }\mbox{ }x^{\prime (3)}_{S_{su},S_{sd},a}\mbox{ }\mbox{ }\mbox{ }\mbox{ }$ & 
   $\mbox{ }\mbox{ }\mbox{ }\mbox{ }\mbox{ }\mbox{ }0\mbox{ }\mbox{ }\mbox{ }\mbox{ }\mbox{ }\mbox{ }$ & 
   $\mbox{ }\mbox{ }\mbox{ }\mbox{ }\mbox{ }\mbox{ }0\mbox{ }\mbox{ }\mbox{ }\mbox{ }\mbox{ }\mbox{ }$ &
   $\mbox{ }\mbox{ }\mbox{ }\mbox{ }\mbox{ }\mbox{ }-1\mbox{ }\mbox{ }\mbox{ }\mbox{ }\mbox{ }\mbox{ }$ & 
   $\mbox{ }\mbox{ }\mbox{ }\mbox{ }\mbox{ }\mbox{ }\frac{-1}{2}\mbox{ }\mbox{ }\mbox{ }\mbox{ }\mbox{ }\mbox{ }$ \\
\hline
   $\mbox{ }\mbox{ }\mbox{ }\mbox{ }x^{\prime (3)}_{S_{du},S_{ds},a}\mbox{ }\mbox{ }\mbox{ }\mbox{ }$ & 
   $\mbox{ }\mbox{ }\mbox{ }\mbox{ }\mbox{ }\mbox{ }\frac{1}{2}\mbox{ }\mbox{ }\mbox{ }\mbox{ }\mbox{ }\mbox{ }$ & 
   $\mbox{ }\mbox{ }\mbox{ }\mbox{ }\mbox{ }\mbox{ }\frac{-1}{2}\mbox{ }\mbox{ }\mbox{ }\mbox{ }\mbox{ }\mbox{ }$ &
   $\mbox{ }\mbox{ }\mbox{ }\mbox{ }\mbox{ }\mbox{ }-1\mbox{ }\mbox{ }\mbox{ }\mbox{ }\mbox{ }\mbox{ }$ & 
   $\mbox{ }\mbox{ }\mbox{ }\mbox{ }\mbox{ }\mbox{ }\frac{-1}{2}\mbox{ }\mbox{ }\mbox{ }\mbox{ }\mbox{ }\mbox{ }$ \\
\hline
   $\mbox{ }\mbox{ }\mbox{ }\mbox{ }x^{\prime (3)}_{S_{su},S_{ss},a}\mbox{ }\mbox{ }\mbox{ }\mbox{ }$ & 
   $\mbox{ }\mbox{ }\mbox{ }\mbox{ }\mbox{ }\mbox{ }\frac{1}{2}\mbox{ }\mbox{ }\mbox{ }\mbox{ }\mbox{ }\mbox{ }$ & 
   $\mbox{ }\mbox{ }\mbox{ }\mbox{ }\mbox{ }\mbox{ }\frac{-1}{2}\mbox{ }\mbox{ }\mbox{ }\mbox{ }\mbox{ }\mbox{ }$ &
   $\mbox{ }\mbox{ }\mbox{ }\mbox{ }\mbox{ }\mbox{ }-1\mbox{ }\mbox{ }\mbox{ }\mbox{ }\mbox{ }\mbox{ }$ & 
   $\mbox{ }\mbox{ }\mbox{ }\mbox{ }\mbox{ }\mbox{ }\frac{-1}{2}\mbox{ }\mbox{ }\mbox{ }\mbox{ }\mbox{ }\mbox{ }$ \\
\hline\hline
\end{tabular}
\caption{\label{tab:baryon_x_prime}Coefficients $x^{\prime (N_{f})}_{H_{1},H_{2}}$ in Eq.~(\ref{eq:baryon_sunset_master}), in the isospin limit.}
\end{table}
\begin{table}[t]
\begin{tabular}{c|cccc}
   $\mbox{ }\mbox{ }\mbox{ }\mbox{ }a\mbox{ }\mbox{ }\mbox{ }\mbox{ }$ & 
   $\mbox{ }\mbox{ }\mbox{ }\mbox{ }\mbox{ }\mbox{ }u\mbox{ }\mbox{ }\mbox{ }\mbox{ }\mbox{ }\mbox{ }$ & 
   $\mbox{ }\mbox{ }\mbox{ }\mbox{ }\mbox{ }\mbox{ }s\mbox{ }\mbox{ }\mbox{ }\mbox{ }\mbox{ }\mbox{ }$ &
   $\mbox{ }\mbox{ }\mbox{ }\mbox{ }\mbox{ }\mbox{ }u^{\prime}\mbox{ }\mbox{ }\mbox{ }\mbox{ }\mbox{ }\mbox{ }$ & 
   $\mbox{ }\mbox{ }\mbox{ }\mbox{ }\mbox{ }\mbox{ }s^{\prime}\mbox{ }\mbox{ }\mbox{ }\mbox{ }\mbox{ }\mbox{ }$ \\
\hline\hline
   $\mbox{ }\mbox{ }\mbox{ }\mbox{ }x^{\prime\prime (2)}_{T_{du},S_{dd},a}\mbox{ }\mbox{ }\mbox{ }\mbox{ }$ & 
   $\mbox{ }\mbox{ }\mbox{ }\mbox{ }\mbox{ }\mbox{ }0\mbox{ }\mbox{ }\mbox{ }\mbox{ }\mbox{ }\mbox{ }$ & 
   $\mbox{ }\mbox{ }\mbox{ }\mbox{ }\mbox{ }\mbox{ }0\mbox{ }\mbox{ }\mbox{ }\mbox{ }\mbox{ }\mbox{ }$  &
   $\mbox{ }\mbox{ }\mbox{ }\mbox{ }\mbox{ }\mbox{ }-1\mbox{ }\mbox{ }\mbox{ }\mbox{ }\mbox{ }\mbox{ }$ & 
   $\mbox{ }\mbox{ }\mbox{ }\mbox{ }\mbox{ }\mbox{ }0\mbox{ }\mbox{ }\mbox{ }\mbox{ }\mbox{ }\mbox{ }$ \\
\hline
   $\mbox{ }\mbox{ }\mbox{ }\mbox{ }x^{\prime\prime (3)}_{T_{du},S_{dd},a}\mbox{ }\mbox{ }\mbox{ }\mbox{ }$ & 
   $\mbox{ }\mbox{ }\mbox{ }\mbox{ }\mbox{ }\mbox{ }0\mbox{ }\mbox{ }\mbox{ }\mbox{ }\mbox{ }\mbox{ }$ & 
   $\mbox{ }\mbox{ }\mbox{ }\mbox{ }\mbox{ }\mbox{ }0\mbox{ }\mbox{ }\mbox{ }\mbox{ }\mbox{ }\mbox{ }$  &
   $\mbox{ }\mbox{ }\mbox{ }\mbox{ }\mbox{ }\mbox{ }-1\mbox{ }\mbox{ }\mbox{ }\mbox{ }\mbox{ }\mbox{ }$ & 
   $\mbox{ }\mbox{ }\mbox{ }\mbox{ }\mbox{ }\mbox{ }\frac{-1}{2}\mbox{ }\mbox{ }\mbox{ }\mbox{ }\mbox{ }\mbox{ }$ \\
\hline
   $\mbox{ }\mbox{ }\mbox{ }\mbox{ }x^{\prime\prime (3)}_{T_{su},S_{sd},a}\mbox{ }\mbox{ }\mbox{ }\mbox{ }$ & 
   $\mbox{ }\mbox{ }\mbox{ }\mbox{ }\mbox{ }\mbox{ }0\mbox{ }\mbox{ }\mbox{ }\mbox{ }\mbox{ }\mbox{ }$ & 
   $\mbox{ }\mbox{ }\mbox{ }\mbox{ }\mbox{ }\mbox{ }0\mbox{ }\mbox{ }\mbox{ }\mbox{ }\mbox{ }\mbox{ }$ &
   $\mbox{ }\mbox{ }\mbox{ }\mbox{ }\mbox{ }\mbox{ }-1\mbox{ }\mbox{ }\mbox{ }\mbox{ }\mbox{ }\mbox{ }$ & 
   $\mbox{ }\mbox{ }\mbox{ }\mbox{ }\mbox{ }\mbox{ }\frac{-1}{2}\mbox{ }\mbox{ }\mbox{ }\mbox{ }\mbox{ }\mbox{ }$ \\
\hline
   $\mbox{ }\mbox{ }\mbox{ }\mbox{ }x^{\prime\prime (3)}_{T_{du},S_{ds},a}\mbox{ }\mbox{ }\mbox{ }\mbox{ }$ & 
   $\mbox{ }\mbox{ }\mbox{ }\mbox{ }\mbox{ }\mbox{ }\frac{-1}{2}\mbox{ }\mbox{ }\mbox{ }\mbox{ }\mbox{ }\mbox{ }$ & 
   $\mbox{ }\mbox{ }\mbox{ }\mbox{ }\mbox{ }\mbox{ }\frac{1}{2}\mbox{ }\mbox{ }\mbox{ }\mbox{ }\mbox{ }\mbox{ }$ &
   $\mbox{ }\mbox{ }\mbox{ }\mbox{ }\mbox{ }\mbox{ }-1\mbox{ }\mbox{ }\mbox{ }\mbox{ }\mbox{ }\mbox{ }$ & 
   $\mbox{ }\mbox{ }\mbox{ }\mbox{ }\mbox{ }\mbox{ }\frac{-1}{2}\mbox{ }\mbox{ }\mbox{ }\mbox{ }\mbox{ }\mbox{ }$ \\
\hline
   $\mbox{ }\mbox{ }\mbox{ }\mbox{ }x^{\prime\prime (3)}_{T_{su},S_{ss},a}\mbox{ }\mbox{ }\mbox{ }\mbox{ }$ & 
   $\mbox{ }\mbox{ }\mbox{ }\mbox{ }\mbox{ }\mbox{ }\frac{-1}{2}\mbox{ }\mbox{ }\mbox{ }\mbox{ }\mbox{ }\mbox{ }$ & 
   $\mbox{ }\mbox{ }\mbox{ }\mbox{ }\mbox{ }\mbox{ }\frac{1}{2}\mbox{ }\mbox{ }\mbox{ }\mbox{ }\mbox{ }\mbox{ }$ &
   $\mbox{ }\mbox{ }\mbox{ }\mbox{ }\mbox{ }\mbox{ }-1\mbox{ }\mbox{ }\mbox{ }\mbox{ }\mbox{ }\mbox{ }$ & 
   $\mbox{ }\mbox{ }\mbox{ }\mbox{ }\mbox{ }\mbox{ }\frac{-1}{2}\mbox{ }\mbox{ }\mbox{ }\mbox{ }\mbox{ }\mbox{ }$ \\
\hline\hline
   $\mbox{ }\mbox{ }\mbox{ }\mbox{ }x^{\prime\prime (2)}_{S_{du},S_{dd},a}\mbox{ }\mbox{ }\mbox{ }\mbox{ }$ & 
   $\mbox{ }\mbox{ }\mbox{ }\mbox{ }\mbox{ }\mbox{ }0\mbox{ }\mbox{ }\mbox{ }\mbox{ }\mbox{ }\mbox{ }$ & 
   $\mbox{ }\mbox{ }\mbox{ }\mbox{ }\mbox{ }\mbox{ }0\mbox{ }\mbox{ }\mbox{ }\mbox{ }\mbox{ }\mbox{ }$ &
   $\mbox{ }\mbox{ }\mbox{ }\mbox{ }\mbox{ }\mbox{ }-1\mbox{ }\mbox{ }\mbox{ }\mbox{ }\mbox{ }\mbox{ }$ & 
   $\mbox{ }\mbox{ }\mbox{ }\mbox{ }\mbox{ }\mbox{ }0\mbox{ }\mbox{ }\mbox{ }\mbox{ }\mbox{ }\mbox{ }$ \\
\hline
   $\mbox{ }\mbox{ }\mbox{ }\mbox{ }x^{\prime\prime (3)}_{S_{du},S_{dd},a}\mbox{ }\mbox{ }\mbox{ }\mbox{ }$ & 
   $\mbox{ }\mbox{ }\mbox{ }\mbox{ }\mbox{ }\mbox{ }0\mbox{ }\mbox{ }\mbox{ }\mbox{ }\mbox{ }\mbox{ }$ & 
   $\mbox{ }\mbox{ }\mbox{ }\mbox{ }\mbox{ }\mbox{ }0\mbox{ }\mbox{ }\mbox{ }\mbox{ }\mbox{ }\mbox{ }$ &
   $\mbox{ }\mbox{ }\mbox{ }\mbox{ }\mbox{ }\mbox{ }-1\mbox{ }\mbox{ }\mbox{ }\mbox{ }\mbox{ }\mbox{ }$ & 
   $\mbox{ }\mbox{ }\mbox{ }\mbox{ }\mbox{ }\mbox{ }\frac{-1}{2}\mbox{ }\mbox{ }\mbox{ }\mbox{ }\mbox{ }\mbox{ }$ \\
\hline
   $\mbox{ }\mbox{ }\mbox{ }\mbox{ }x^{\prime\prime (3)}_{S_{su},S_{sd},a}\mbox{ }\mbox{ }\mbox{ }\mbox{ }$ & 
   $\mbox{ }\mbox{ }\mbox{ }\mbox{ }\mbox{ }\mbox{ }0\mbox{ }\mbox{ }\mbox{ }\mbox{ }\mbox{ }\mbox{ }$ & 
   $\mbox{ }\mbox{ }\mbox{ }\mbox{ }\mbox{ }\mbox{ }0\mbox{ }\mbox{ }\mbox{ }\mbox{ }\mbox{ }\mbox{ }$ &
   $\mbox{ }\mbox{ }\mbox{ }\mbox{ }\mbox{ }\mbox{ }-1\mbox{ }\mbox{ }\mbox{ }\mbox{ }\mbox{ }\mbox{ }$ & 
   $\mbox{ }\mbox{ }\mbox{ }\mbox{ }\mbox{ }\mbox{ }\frac{-1}{2}\mbox{ }\mbox{ }\mbox{ }\mbox{ }\mbox{ }\mbox{ }$ \\
\hline
   $\mbox{ }\mbox{ }\mbox{ }\mbox{ }x^{\prime\prime (3)}_{S_{du},S_{ds},a}\mbox{ }\mbox{ }\mbox{ }\mbox{ }$ & 
   $\mbox{ }\mbox{ }\mbox{ }\mbox{ }\mbox{ }\mbox{ }\frac{-1}{2}\mbox{ }\mbox{ }\mbox{ }\mbox{ }\mbox{ }\mbox{ }$ & 
   $\mbox{ }\mbox{ }\mbox{ }\mbox{ }\mbox{ }\mbox{ }\frac{1}{2}\mbox{ }\mbox{ }\mbox{ }\mbox{ }\mbox{ }\mbox{ }$ &
   $\mbox{ }\mbox{ }\mbox{ }\mbox{ }\mbox{ }\mbox{ }-1\mbox{ }\mbox{ }\mbox{ }\mbox{ }\mbox{ }\mbox{ }$ & 
   $\mbox{ }\mbox{ }\mbox{ }\mbox{ }\mbox{ }\mbox{ }\frac{-1}{2}\mbox{ }\mbox{ }\mbox{ }\mbox{ }\mbox{ }\mbox{ }$ \\
\hline
   $\mbox{ }\mbox{ }\mbox{ }\mbox{ }x^{\prime\prime (3)}_{S_{su},S_{ss},a}\mbox{ }\mbox{ }\mbox{ }\mbox{ }$ & 
   $\mbox{ }\mbox{ }\mbox{ }\mbox{ }\mbox{ }\mbox{ }\frac{-1}{2}\mbox{ }\mbox{ }\mbox{ }\mbox{ }\mbox{ }\mbox{ }$ & 
   $\mbox{ }\mbox{ }\mbox{ }\mbox{ }\mbox{ }\mbox{ }\frac{1}{2}\mbox{ }\mbox{ }\mbox{ }\mbox{ }\mbox{ }\mbox{ }$ &
   $\mbox{ }\mbox{ }\mbox{ }\mbox{ }\mbox{ }\mbox{ }-1\mbox{ }\mbox{ }\mbox{ }\mbox{ }\mbox{ }\mbox{ }$ & 
   $\mbox{ }\mbox{ }\mbox{ }\mbox{ }\mbox{ }\mbox{ }\frac{-1}{2}\mbox{ }\mbox{ }\mbox{ }\mbox{ }\mbox{ }\mbox{ }$ \\
\hline\hline
\end{tabular}
\caption{\label{tab:baryon_x_double_prime}Coefficients $x^{\prime (N_{f})}_{H_{1},H_{2}}$ in Eq.~(\ref{eq:baryon_sunset_master}), in the isospin limit.}
\end{table}
\begin{table}[t]
\begin{tabular}{ccccccccccccc}
   $\mbox{ }\tilde{y}^{(2)}_{B_{u},B^{\ast}_{d}}$ & 
   $\mbox{ }\tilde{y}^{(3)}_{B_{u},B^{\ast}_{d}}$ & 
   $\mbox{ }\tilde{y}^{(3)}_{B_{u},B^{\ast}_{s}}$ & 
   $\mbox{ }\tilde{y}^{(2)}_{T_{du},S_{dd}}$ &
   $\mbox{ }\tilde{y}^{(3)}_{T_{du},S_{dd}}$ &
   $\mbox{ }\tilde{y}^{(3)}_{T_{su},S_{sd}}$ &
   $\mbox{ }\tilde{y}^{(3)}_{T_{du},S_{ds}}$ &
   $\mbox{ }\tilde{y}^{(3)}_{T_{su},S_{ss}}$ &
   $\mbox{ }\tilde{y}^{(2)}_{S_{du},S_{dd}}$ &
   $\mbox{ }\tilde{y}^{(3)}_{S_{du},S_{dd}}$ &
   $\mbox{ }\tilde{y}^{(3)}_{S_{su},S_{sd}}$ &
   $\mbox{ }\tilde{y}^{(3)}_{S_{du},S_{ds}}$ &
   $\mbox{ }\tilde{y}^{(3)}_{S_{su},S_{ss}}$ \\
\hline\hline
   $\mbox{ }-1\mbox{ }$ & 
   $\mbox{ }\frac{-2}{3}\mbox{ }$ & 
   $\mbox{ }\frac{-2}{3}\mbox{ }$ & 
   $\mbox{ }+\tilde{x}\mbox{ }$ & 
   $\mbox{ }+\tilde{x}\mbox{ }$ &
   $\mbox{ }\frac{-1}{6}\mbox{ }$ &
   $\mbox{ }+\tilde{x}\mbox{ }$ &
   $\mbox{ }+\tilde{x}\mbox{ }$ &
   $\mbox{ }+\tilde{x}\mbox{ }$ &
   $\mbox{ }+\tilde{x}\mbox{ }$ &
   $\mbox{ }\frac{1}{12}\mbox{ }$ &
   $\mbox{ }+\tilde{x}\mbox{ }$ &
   $\mbox{ }+\tilde{x}\mbox{ }$ \\
\hline\hline
  &&&&&&&&&&&&\\
\end{tabular}
\begin{tabular}{cccccccc}
   $\mbox{ }\tilde{y}^{\prime (2)}_{B_{u},B^{\ast}_{d}}$ & 
   $\mbox{ }\tilde{y}^{\prime (3)}_{B_{u},B^{\ast}_{d}}$ & 
   $\mbox{ }\tilde{y}^{\prime (3)}_{B_{u},B^{\ast}_{s}}$ & 
   $\mbox{ }\tilde{y}^{\prime (2)}_{S_{du},S_{dd}}$ &
   $\mbox{ }\tilde{y}^{\prime (3)}_{S_{du},S_{dd}}$ &
   $\mbox{ }\tilde{y}^{\prime (3)}_{S_{su},S_{sd}}$ &
   $\mbox{ }\tilde{y}^{\prime (3)}_{S_{du},S_{ds}}$ &
   $\mbox{ }\tilde{y}^{\prime (3)}_{S_{su},S_{ss}}$ \\
\hline\hline
   $\mbox{ }\frac{1}{2}\mbox{ }$ &
   $\mbox{ }\frac{1}{3}\mbox{ }$ &
   $\mbox{ }\frac{1}{3}\mbox{ }$ &
   $\mbox{ }+\tilde{x}^{\prime}\mbox{ }$ &
   $\mbox{ }+\tilde{x}^{\prime}\mbox{ }$ &
   $\mbox{ }\frac{-1}{6}\mbox{ }$ &
   $\mbox{ }+\tilde{x}^{\prime}\mbox{ }$ &
   $\mbox{ }+\tilde{x}^{\prime}\mbox{ }$ \\
\hline\hline
   &&&&&&&\\
\end{tabular}
\begin{tabular}{cccccccccc}
   $\mbox{ }\tilde{y}^{\prime\prime(2)}_{T_{du},S_{dd}}$ &
   $\mbox{ }\tilde{y}^{\prime\prime(3)}_{T_{du},S_{dd}}$ &
   $\mbox{ }\tilde{y}^{\prime\prime(3)}_{T_{su},S_{sd}}$ &
   $\mbox{ }\tilde{y}^{\prime\prime(3)}_{T_{du},S_{ds}}$ &
   $\mbox{ }\tilde{y}^{\prime\prime(3)}_{T_{su},S_{ss}}$ &
   $\mbox{ }\tilde{y}^{\prime\prime(2)}_{S_{du},S_{dd}}$ &
   $\mbox{ }\tilde{y}^{\prime\prime(3)}_{S_{du},S_{dd}}$ &
   $\mbox{ }\tilde{y}^{\prime\prime(3)}_{S_{su},S_{sd}}$ &
   $\mbox{ }\tilde{y}^{\prime\prime(3)}_{S_{du},S_{ds}}$ &
   $\mbox{ }\tilde{y}^{\prime\prime(3)}_{S_{su},S_{ss}}$ \\
\hline\hline
   $\mbox{ }+\tilde{x}^{\prime\prime}\mbox{ }$ & 
   $\mbox{ }+\tilde{x}^{\prime\prime}\mbox{ }$ &
   $\mbox{ }\frac{1}{6}\mbox{ }$ &
   $\mbox{ }+\tilde{x}^{\prime\prime}\mbox{ }$ &
   $\mbox{ }+\tilde{x}^{\prime\prime}\mbox{ }$ &
   $\mbox{ }+\tilde{x}^{\prime\prime}\mbox{ }$ &
   $\mbox{ }+\tilde{x}^{\prime\prime}\mbox{ }$ &
   $\mbox{ }\frac{-1}{6}\mbox{ }$ &
   $\mbox{ }+\tilde{x}^{\prime\prime}\mbox{ }$ &
   $\mbox{ }+\tilde{x}^{\prime\prime}\mbox{ }$ \\
\hline\hline
\end{tabular}
\caption{\label{tab:y_tilde}Coefficients $\tilde{y}^{(N_{f})}_{H_{1},H_{2}}$, $\tilde{y}^{\prime (N_{f})}_{H_{1},H_{2}}$ and $\tilde{y}^{\prime\prime (N_{f})}_{H_{1},H_{2}}$ in Eqs.~(\ref{eq:meson_sunset_master}) and (\ref{eq:baryon_sunset_master}).  Due to the isospin symmetry, some of these coefficients cannot be distinguished from their $\tilde{x}^{(N_{f})}_{H_{1},H_{2}}$ counterparts.  For such cases, the values are denoted $+\tilde{x}$ in the table, and are presented together with the corresponding $\tilde{x}^{(N_{f})}_{H_{1},H_{2}}$.}
\end{table}
\begin{table}[t]
\begin{tabular}{c|cccc}
   $\mbox{ }\mbox{ }\mbox{ }\mbox{ }a\mbox{ }\mbox{ }\mbox{ }\mbox{ }$ & 
   $\mbox{ }\mbox{ }\mbox{ }\mbox{ }\mbox{ }\mbox{ }u\mbox{ }\mbox{ }\mbox{ }\mbox{ }\mbox{ }\mbox{ }$ & 
   $\mbox{ }\mbox{ }\mbox{ }\mbox{ }\mbox{ }\mbox{ }s\mbox{ }\mbox{ }\mbox{ }\mbox{ }\mbox{ }\mbox{ }$ &
   $\mbox{ }\mbox{ }\mbox{ }\mbox{ }\mbox{ }\mbox{ }u^{\prime}\mbox{ }\mbox{ }\mbox{ }\mbox{ }\mbox{ }\mbox{ }$ &
   $\mbox{ }\mbox{ }\mbox{ }\mbox{ }\mbox{ }\mbox{ }s^{\prime}\mbox{ }\mbox{ }\mbox{ }\mbox{ }\mbox{ }\mbox{ }$ \\
\hline\hline
   $\mbox{ }\mbox{ }\mbox{ }\mbox{ }\tilde{x}^{(2)}_{T_{du},S_{dd},a}\mbox{ }\mbox{ }\mbox{ }\mbox{ }$ & 
   $\mbox{ }\mbox{ }\mbox{ }\mbox{ }\mbox{ }\mbox{ }\mbox{ }\mbox{ }\mbox{ }\mbox{ }\mbox{ }\mbox{ }\mbox{ }0-\tilde{y}^{(2)}_{T_{du},S_{dd}}\mbox{ }\mbox{ }\mbox{ }\mbox{ }\mbox{ }\mbox{ }$ & 
   $\mbox{ }\mbox{ }\mbox{ }\mbox{ }\mbox{ }\mbox{ }0\mbox{ }\mbox{ }\mbox{ }\mbox{ }\mbox{ }\mbox{ }$ &
   $\mbox{ }\mbox{ }\mbox{ }\mbox{ }\mbox{ }\mbox{ }0\mbox{ }\mbox{ }\mbox{ }\mbox{ }\mbox{ }\mbox{ }$ &
   $\mbox{ }\mbox{ }\mbox{ }\mbox{ }\mbox{ }\mbox{ }0\mbox{ }\mbox{ }\mbox{ }\mbox{ }\mbox{ }\mbox{ }$ \\
\hline
   $\mbox{ }\mbox{ }\mbox{ }\mbox{ }\tilde{x}^{(3)}_{T_{du},S_{dd},a}\mbox{ }\mbox{ }\mbox{ }\mbox{ }$ & 
   $\mbox{ }\mbox{ }\mbox{ }\mbox{ }\mbox{ }\mbox{ }\mbox{ }\mbox{ }\mbox{ }\mbox{ }\mbox{ }\mbox{ }\mbox{ }0-\tilde{y}^{(3)}_{T_{du},S_{dd}}\mbox{ }\mbox{ }\mbox{ }\mbox{ }\mbox{ }\mbox{ }$ & 
   $\mbox{ }\mbox{ }\mbox{ }\mbox{ }\mbox{ }\mbox{ }0\mbox{ }\mbox{ }\mbox{ }\mbox{ }\mbox{ }\mbox{ }$ &
   $\mbox{ }\mbox{ }\mbox{ }\mbox{ }\mbox{ }\mbox{ }0\mbox{ }\mbox{ }\mbox{ }\mbox{ }\mbox{ }\mbox{ }$ &
   $\mbox{ }\mbox{ }\mbox{ }\mbox{ }\mbox{ }\mbox{ }0\mbox{ }\mbox{ }\mbox{ }\mbox{ }\mbox{ }\mbox{ }$ \\
\hline
   $\mbox{ }\mbox{ }\mbox{ }\mbox{ }\tilde{x}^{(3)}_{T_{su},S_{sd},a}\mbox{ }\mbox{ }\mbox{ }\mbox{ }$ & 
   $\mbox{ }\mbox{ }\mbox{ }\mbox{ }\mbox{ }\mbox{ }0\mbox{ }\mbox{ }\mbox{ }\mbox{ }\mbox{ }\mbox{ }$ & 
   $\mbox{ }\mbox{ }\mbox{ }\mbox{ }\mbox{ }\mbox{ }\frac{1}{6}\mbox{ }\mbox{ }\mbox{ }\mbox{ }\mbox{ }\mbox{ }$ &
   $\mbox{ }\mbox{ }\mbox{ }\mbox{ }\mbox{ }\mbox{ }0\mbox{ }\mbox{ }\mbox{ }\mbox{ }\mbox{ }\mbox{ }$ &
   $\mbox{ }\mbox{ }\mbox{ }\mbox{ }\mbox{ }\mbox{ }0\mbox{ }\mbox{ }\mbox{ }\mbox{ }\mbox{ }\mbox{ }$ \\
\hline
   $\mbox{ }\mbox{ }\mbox{ }\mbox{ }\tilde{x}^{(3)}_{T_{du},S_{ds},a}\mbox{ }\mbox{ }\mbox{ }\mbox{ }$ & 
   $\mbox{ }\mbox{ }\mbox{ }\mbox{ }\mbox{ }\mbox{ }0\mbox{ }\mbox{ }\mbox{ }\mbox{ }\mbox{ }\mbox{ }$ & 
   $\mbox{ }\mbox{ }\mbox{ }\mbox{ }\mbox{ }\mbox{ }\mbox{ }\mbox{ }\mbox{ }\mbox{ }\mbox{ }\mbox{ }\mbox{ }0-\tilde{y}^{(3)}_{T_{du},S_{ds}}\mbox{ }\mbox{ }\mbox{ }\mbox{ }\mbox{ }\mbox{ }$ &
   $\mbox{ }\mbox{ }\mbox{ }\mbox{ }\mbox{ }\mbox{ }0\mbox{ }\mbox{ }\mbox{ }\mbox{ }\mbox{ }\mbox{ }$ &
   $\mbox{ }\mbox{ }\mbox{ }\mbox{ }\mbox{ }\mbox{ }0\mbox{ }\mbox{ }\mbox{ }\mbox{ }\mbox{ }\mbox{ }$ \\
\hline
   $\mbox{ }\mbox{ }\mbox{ }\mbox{ }\tilde{x}^{(3)}_{T_{su},S_{ss},a}\mbox{ }\mbox{ }\mbox{ }\mbox{ }$ & 
   $\mbox{ }\mbox{ }\mbox{ }\mbox{ }\mbox{ }\mbox{ }\mbox{ }\mbox{ }\mbox{ }\mbox{ }\mbox{ }\mbox{ }\mbox{ }\frac{-1}{3}-\tilde{y}^{(3)}_{T_{su},S_{ss}}\mbox{ }\mbox{ }\mbox{ }\mbox{ }\mbox{ }\mbox{ }$ & 
   $\mbox{ }\mbox{ }\mbox{ }\mbox{ }\mbox{ }\mbox{ }\frac{1}{3}\mbox{ }\mbox{ }\mbox{ }\mbox{ }\mbox{ }\mbox{ }$ &
   $\mbox{ }\mbox{ }\mbox{ }\mbox{ }\mbox{ }\mbox{ }0\mbox{ }\mbox{ }\mbox{ }\mbox{ }\mbox{ }\mbox{ }$ &
   $\mbox{ }\mbox{ }\mbox{ }\mbox{ }\mbox{ }\mbox{ }0\mbox{ }\mbox{ }\mbox{ }\mbox{ }\mbox{ }\mbox{ }$ \\
\hline\hline
   $\mbox{ }\mbox{ }\mbox{ }\mbox{ }\tilde{x}^{(2)}_{S_{du},S_{dd},a}\mbox{ }\mbox{ }\mbox{ }\mbox{ }$ & 
   $\mbox{ }\mbox{ }\mbox{ }\mbox{ }\mbox{ }\mbox{ }\mbox{ }\mbox{ }\mbox{ }\mbox{ }\mbox{ }\mbox{ }\mbox{ }\frac{1}{2}-\tilde{y}^{(2)}_{S_{du},S_{dd}}\mbox{ }\mbox{ }\mbox{ }\mbox{ }\mbox{ }\mbox{ }$ & 
   $\mbox{ }\mbox{ }\mbox{ }\mbox{ }\mbox{ }\mbox{ }0\mbox{ }\mbox{ }\mbox{ }\mbox{ }\mbox{ }\mbox{ }$ &
   $\mbox{ }\mbox{ }\mbox{ }\mbox{ }\mbox{ }\mbox{ }0\mbox{ }\mbox{ }\mbox{ }\mbox{ }\mbox{ }\mbox{ }$ &
   $\mbox{ }\mbox{ }\mbox{ }\mbox{ }\mbox{ }\mbox{ }0\mbox{ }\mbox{ }\mbox{ }\mbox{ }\mbox{ }\mbox{ }$ \\
\hline
   $\mbox{ }\mbox{ }\mbox{ }\mbox{ }\tilde{x}^{(3)}_{S_{du},S_{dd},a}\mbox{ }\mbox{ }\mbox{ }\mbox{ }$ & 
   $\mbox{ }\mbox{ }\mbox{ }\mbox{ }\mbox{ }\mbox{ }\mbox{ }\mbox{ }\mbox{ }\mbox{ }\mbox{ }\mbox{ }\mbox{ }\frac{1}{3}-\tilde{y}^{(3)}_{S_{du},S_{dd}}\mbox{ }\mbox{ }\mbox{ }\mbox{ }\mbox{ }\mbox{ }$ & 
   $\mbox{ }\mbox{ }\mbox{ }\mbox{ }\mbox{ }\mbox{ }0\mbox{ }\mbox{ }\mbox{ }\mbox{ }\mbox{ }\mbox{ }$ &
   $\mbox{ }\mbox{ }\mbox{ }\mbox{ }\mbox{ }\mbox{ }0\mbox{ }\mbox{ }\mbox{ }\mbox{ }\mbox{ }\mbox{ }$ &
   $\mbox{ }\mbox{ }\mbox{ }\mbox{ }\mbox{ }\mbox{ }0\mbox{ }\mbox{ }\mbox{ }\mbox{ }\mbox{ }\mbox{ }$ \\
\hline
   $\mbox{ }\mbox{ }\mbox{ }\mbox{ }\tilde{x}^{(3)}_{S_{su},S_{sd},a}\mbox{ }\mbox{ }\mbox{ }\mbox{ }$ & 
   $\mbox{ }\mbox{ }\mbox{ }\mbox{ }\mbox{ }\mbox{ }\frac{1}{6}\mbox{ }\mbox{ }\mbox{ }\mbox{ }\mbox{ }\mbox{ }$ & 
   $\mbox{ }\mbox{ }\mbox{ }\mbox{ }\mbox{ }\mbox{ }\frac{1}{12}\mbox{ }\mbox{ }\mbox{ }\mbox{ }\mbox{ }\mbox{ }$ &
   $\mbox{ }\mbox{ }\mbox{ }\mbox{ }\mbox{ }\mbox{ }0\mbox{ }\mbox{ }\mbox{ }\mbox{ }\mbox{ }\mbox{ }$ &
   $\mbox{ }\mbox{ }\mbox{ }\mbox{ }\mbox{ }\mbox{ }0\mbox{ }\mbox{ }\mbox{ }\mbox{ }\mbox{ }\mbox{ }$ \\
\hline
   $\mbox{ }\mbox{ }\mbox{ }\mbox{ }\tilde{x}^{(3)}_{S_{du},S_{ds},a}\mbox{ }\mbox{ }\mbox{ }\mbox{ }$ & 
   $\mbox{ }\mbox{ }\mbox{ }\mbox{ }\mbox{ }\mbox{ }\frac{1}{6}\mbox{ }\mbox{ }\mbox{ }\mbox{ }\mbox{ }\mbox{ }$ & 
   $\mbox{ }\mbox{ }\mbox{ }\mbox{ }\mbox{ }\mbox{ }\mbox{ }\mbox{ }\mbox{ }\mbox{ }\mbox{ }\mbox{ }\mbox{ }\frac{1}{6}-\tilde{y}^{(3)}_{S_{du},S_{ds}}\mbox{ }\mbox{ }\mbox{ }\mbox{ }\mbox{ }\mbox{ }$ &
   $\mbox{ }\mbox{ }\mbox{ }\mbox{ }\mbox{ }\mbox{ }0\mbox{ }\mbox{ }\mbox{ }\mbox{ }\mbox{ }\mbox{ }$ &
   $\mbox{ }\mbox{ }\mbox{ }\mbox{ }\mbox{ }\mbox{ }0\mbox{ }\mbox{ }\mbox{ }\mbox{ }\mbox{ }\mbox{ }$ \\
\hline
   $\mbox{ }\mbox{ }\mbox{ }\mbox{ }\tilde{x}^{(3)}_{S_{su},S_{ss},a}\mbox{ }\mbox{ }\mbox{ }\mbox{ }$ & 
   $\mbox{ }\mbox{ }\mbox{ }\mbox{ }\mbox{ }\mbox{ }\mbox{ }\mbox{ }\mbox{ }\mbox{ }\mbox{ }\mbox{ }\mbox{ }\frac{1}{6}-\tilde{y}^{(3)}_{S_{su},S_{ss}}\mbox{ }\mbox{ }\mbox{ }\mbox{ }\mbox{ }\mbox{ }$ & 
   $\mbox{ }\mbox{ }\mbox{ }\mbox{ }\mbox{ }\mbox{ }\frac{1}{6}\mbox{ }\mbox{ }\mbox{ }\mbox{ }\mbox{ }\mbox{ }$ &
   $\mbox{ }\mbox{ }\mbox{ }\mbox{ }\mbox{ }\mbox{ }0\mbox{ }\mbox{ }\mbox{ }\mbox{ }\mbox{ }\mbox{ }$ &
   $\mbox{ }\mbox{ }\mbox{ }\mbox{ }\mbox{ }\mbox{ }0\mbox{ }\mbox{ }\mbox{ }\mbox{ }\mbox{ }\mbox{ }$ \\ 
\hline \hline
\end{tabular}
\caption{\label{tab:baryon_tilde_x}Coefficients $\tilde{x}^{(N_{f})}_{H_{1},H_{2}}$ in Eq.~(\ref{eq:baryon_sunset_master}).  Due to the isospin symmetry, some of them cannot be distinguished from their $\tilde{y}^{(N_{f})}_{H_{1},H_{2}}$ counterparts.  For such cases, we present $\tilde{x}^{(N_{f})}_{H_{1},H_{2}}+\tilde{y}^{(N_{f})}_{H_{1},H_{2}}$ in the table.}
\end{table}
\begin{table}[t]
\begin{tabular}{c|cccc}
   $\mbox{ }\mbox{ }\mbox{ }\mbox{ }a\mbox{ }\mbox{ }\mbox{ }\mbox{ }$ & 
   $\mbox{ }\mbox{ }\mbox{ }\mbox{ }\mbox{ }\mbox{ }u\mbox{ }\mbox{ }\mbox{ }\mbox{ }\mbox{ }\mbox{ }$ & 
   $\mbox{ }\mbox{ }\mbox{ }\mbox{ }\mbox{ }\mbox{ }s\mbox{ }\mbox{ }\mbox{ }\mbox{ }\mbox{ }\mbox{ }$ &
   $\mbox{ }\mbox{ }\mbox{ }\mbox{ }\mbox{ }\mbox{ }u^{\prime}\mbox{ }\mbox{ }\mbox{ }\mbox{ }\mbox{ }\mbox{ }$ &
   $\mbox{ }\mbox{ }\mbox{ }\mbox{ }\mbox{ }\mbox{ }s^{\prime}\mbox{ }\mbox{ }\mbox{ }\mbox{ }\mbox{ }\mbox{ }$ \\
\hline\hline
   $\mbox{ }\mbox{ }\mbox{ }\mbox{ }\tilde{x}^{\prime (2)}_{S_{du},S_{dd},a}\mbox{ }\mbox{ }\mbox{ }\mbox{ }$ & 
   $\mbox{ }\mbox{ }\mbox{ }\mbox{ }\mbox{ }\mbox{ }\mbox{ }\mbox{ }\mbox{ }\mbox{ }\mbox{ }\mbox{ }\mbox{ }0-\tilde{y}^{\prime (2)}_{S_{du},S_{dd}}\mbox{ }\mbox{ }\mbox{ }\mbox{ }\mbox{ }\mbox{ }$ & 
   $\mbox{ }\mbox{ }\mbox{ }\mbox{ }\mbox{ }\mbox{ }0\mbox{ }\mbox{ }\mbox{ }\mbox{ }\mbox{ }\mbox{ }$ &
   $\mbox{ }\mbox{ }\mbox{ }\mbox{ }\mbox{ }\mbox{ }0\mbox{ }\mbox{ }\mbox{ }\mbox{ }\mbox{ }\mbox{ }$ &
   $\mbox{ }\mbox{ }\mbox{ }\mbox{ }\mbox{ }\mbox{ }0\mbox{ }\mbox{ }\mbox{ }\mbox{ }\mbox{ }\mbox{ }$ \\
\hline
   $\mbox{ }\mbox{ }\mbox{ }\mbox{ }\tilde{x}^{\prime (3)}_{S_{du},S_{dd},a}\mbox{ }\mbox{ }\mbox{ }\mbox{ }$ & 
   $\mbox{ }\mbox{ }\mbox{ }\mbox{ }\mbox{ }\mbox{ }\mbox{ }\mbox{ }\mbox{ }\mbox{ }\mbox{ }\mbox{ }\mbox{ }0-\tilde{y}^{\prime (3)}_{S_{du},S_{dd}}\mbox{ }\mbox{ }\mbox{ }\mbox{ }\mbox{ }\mbox{ }$ & 
   $\mbox{ }\mbox{ }\mbox{ }\mbox{ }\mbox{ }\mbox{ }0\mbox{ }\mbox{ }\mbox{ }\mbox{ }\mbox{ }\mbox{ }$ &
   $\mbox{ }\mbox{ }\mbox{ }\mbox{ }\mbox{ }\mbox{ }0\mbox{ }\mbox{ }\mbox{ }\mbox{ }\mbox{ }\mbox{ }$ &
   $\mbox{ }\mbox{ }\mbox{ }\mbox{ }\mbox{ }\mbox{ }0\mbox{ }\mbox{ }\mbox{ }\mbox{ }\mbox{ }\mbox{ }$ \\
\hline
   $\mbox{ }\mbox{ }\mbox{ }\mbox{ }\tilde{x}^{\prime (3)}_{S_{su},S_{sd},a}\mbox{ }\mbox{ }\mbox{ }\mbox{ }$ & 
   $\mbox{ }\mbox{ }\mbox{ }\mbox{ }\mbox{ }\mbox{ }0\mbox{ }\mbox{ }\mbox{ }\mbox{ }\mbox{ }\mbox{ }$ & 
   $\mbox{ }\mbox{ }\mbox{ }\mbox{ }\mbox{ }\mbox{ }\frac{1}{6}\mbox{ }\mbox{ }\mbox{ }\mbox{ }\mbox{ }\mbox{ }$ &
   $\mbox{ }\mbox{ }\mbox{ }\mbox{ }\mbox{ }\mbox{ }0\mbox{ }\mbox{ }\mbox{ }\mbox{ }\mbox{ }\mbox{ }$ &
   $\mbox{ }\mbox{ }\mbox{ }\mbox{ }\mbox{ }\mbox{ }0\mbox{ }\mbox{ }\mbox{ }\mbox{ }\mbox{ }\mbox{ }$ \\
\hline
   $\mbox{ }\mbox{ }\mbox{ }\mbox{ }\tilde{x}^{\prime (3)}_{S_{du},S_{ds},a}\mbox{ }\mbox{ }\mbox{ }\mbox{ }$ & 
   $\mbox{ }\mbox{ }\mbox{ }\mbox{ }\mbox{ }\mbox{ }0\mbox{ }\mbox{ }\mbox{ }\mbox{ }\mbox{ }\mbox{ }$ & 
   $\mbox{ }\mbox{ }\mbox{ }\mbox{ }\mbox{ }\mbox{ }\mbox{ }\mbox{ }\mbox{ }\mbox{ }\mbox{ }\mbox{ }\mbox{ }0-\tilde{y}^{\prime (3)}_{S_{du},S_{ds}}\mbox{ }\mbox{ }\mbox{ }\mbox{ }\mbox{ }\mbox{ }$ &
   $\mbox{ }\mbox{ }\mbox{ }\mbox{ }\mbox{ }\mbox{ }0\mbox{ }\mbox{ }\mbox{ }\mbox{ }\mbox{ }\mbox{ }$ &
   $\mbox{ }\mbox{ }\mbox{ }\mbox{ }\mbox{ }\mbox{ }0\mbox{ }\mbox{ }\mbox{ }\mbox{ }\mbox{ }\mbox{ }$ \\
\hline
   $\mbox{ }\mbox{ }\mbox{ }\mbox{ }\tilde{x}^{\prime (3)}_{S_{su},S_{ss},a}\mbox{ }\mbox{ }\mbox{ }\mbox{ }$ & 
   $\mbox{ }\mbox{ }\mbox{ }\mbox{ }\mbox{ }\mbox{ }\mbox{ }\mbox{ }\mbox{ }\mbox{ }\mbox{ }\mbox{ }\mbox{ }\frac{-1}{3}-\tilde{y}^{\prime (3)}_{S_{su},S_{ss}}\mbox{ }\mbox{ }\mbox{ }\mbox{ }\mbox{ }\mbox{ }$ & 
   $\mbox{ }\mbox{ }\mbox{ }\mbox{ }\mbox{ }\mbox{ }\frac{1}{3}\mbox{ }\mbox{ }\mbox{ }\mbox{ }\mbox{ }\mbox{ }$ &
   $\mbox{ }\mbox{ }\mbox{ }\mbox{ }\mbox{ }\mbox{ }0\mbox{ }\mbox{ }\mbox{ }\mbox{ }\mbox{ }\mbox{ }$ &
   $\mbox{ }\mbox{ }\mbox{ }\mbox{ }\mbox{ }\mbox{ }0\mbox{ }\mbox{ }\mbox{ }\mbox{ }\mbox{ }\mbox{ }$ \\
\hline\hline
\end{tabular}
\caption{\label{tab:baryon_tilde_x_prime}Coefficients $\tilde{x}^{\prime (N_{f})}_{H_{1},H_{2}}$ in Eq.~(\ref{eq:baryon_sunset_master}).  Due to the isospin symmetry, some of them cannot be distinguished from their $\tilde{y}^{\prime (N_{f})}_{H_{1},H_{2}}$ counterparts.  For such cases, we present $\tilde{x}^{\prime (N_{f})}_{H_{1},H_{2}}+\tilde{y}^{\prime (N_{f})}_{H_{1},H_{2}}$ in the table.}
\end{table}%
\begin{table}[t]
\begin{tabular}{c|cccc}
   $\mbox{ }\mbox{ }\mbox{ }\mbox{ }a\mbox{ }\mbox{ }\mbox{ }\mbox{ }$ & 
   $\mbox{ }\mbox{ }\mbox{ }\mbox{ }\mbox{ }\mbox{ }u\mbox{ }\mbox{ }\mbox{ }\mbox{ }\mbox{ }\mbox{ }$ & 
   $\mbox{ }\mbox{ }\mbox{ }\mbox{ }\mbox{ }\mbox{ }s\mbox{ }\mbox{ }\mbox{ }\mbox{ }\mbox{ }\mbox{ }$ &
   $\mbox{ }\mbox{ }\mbox{ }\mbox{ }\mbox{ }\mbox{ }u^{\prime}\mbox{ }\mbox{ }\mbox{ }\mbox{ }\mbox{ }\mbox{ }$ &
   $\mbox{ }\mbox{ }\mbox{ }\mbox{ }\mbox{ }\mbox{ }s^{\prime}\mbox{ }\mbox{ }\mbox{ }\mbox{ }\mbox{ }\mbox{ }$ \\
\hline\hline
   $\mbox{ }\mbox{ }\mbox{ }\mbox{ }\tilde{x}^{\prime\prime (2)}_{T_{du},S_{dd},a}\mbox{ }\mbox{ }\mbox{ }\mbox{ }$ & 
   $\mbox{ }\mbox{ }\mbox{ }\mbox{ }\mbox{ }\mbox{ }\mbox{ }\mbox{ }\mbox{ }\mbox{ }\mbox{ }\mbox{ }\mbox{ }0-\tilde{y}^{\prime\prime (2)}_{T_{du},S_{dd}}\mbox{ }\mbox{ }\mbox{ }\mbox{ }\mbox{ }\mbox{ }$ & 
   $\mbox{ }\mbox{ }\mbox{ }\mbox{ }\mbox{ }\mbox{ }0\mbox{ }\mbox{ }\mbox{ }\mbox{ }\mbox{ }\mbox{ }$ &
   $\mbox{ }\mbox{ }\mbox{ }\mbox{ }\mbox{ }\mbox{ }0\mbox{ }\mbox{ }\mbox{ }\mbox{ }\mbox{ }\mbox{ }$ &
   $\mbox{ }\mbox{ }\mbox{ }\mbox{ }\mbox{ }\mbox{ }0\mbox{ }\mbox{ }\mbox{ }\mbox{ }\mbox{ }\mbox{ }$ \\
\hline
   $\mbox{ }\mbox{ }\mbox{ }\mbox{ }\tilde{x}^{\prime\prime (3)}_{T_{du},S_{dd},a}\mbox{ }\mbox{ }\mbox{ }\mbox{ }$ & 
   $\mbox{ }\mbox{ }\mbox{ }\mbox{ }\mbox{ }\mbox{ }\mbox{ }\mbox{ }\mbox{ }\mbox{ }\mbox{ }\mbox{ }\mbox{ }0-\tilde{y}^{\prime\prime (3)}_{T_{du},S_{dd}}\mbox{ }\mbox{ }\mbox{ }\mbox{ }\mbox{ }\mbox{ }$ & 
   $\mbox{ }\mbox{ }\mbox{ }\mbox{ }\mbox{ }\mbox{ }0\mbox{ }\mbox{ }\mbox{ }\mbox{ }\mbox{ }\mbox{ }$ &
   $\mbox{ }\mbox{ }\mbox{ }\mbox{ }\mbox{ }\mbox{ }0\mbox{ }\mbox{ }\mbox{ }\mbox{ }\mbox{ }\mbox{ }$ &
   $\mbox{ }\mbox{ }\mbox{ }\mbox{ }\mbox{ }\mbox{ }0\mbox{ }\mbox{ }\mbox{ }\mbox{ }\mbox{ }\mbox{ }$ \\
\hline
   $\mbox{ }\mbox{ }\mbox{ }\mbox{ }\tilde{x}^{\prime\prime (3)}_{T_{su},S_{sd},a}\mbox{ }\mbox{ }\mbox{ }\mbox{ }$ & 
   $\mbox{ }\mbox{ }\mbox{ }\mbox{ }\mbox{ }\mbox{ }\frac{-1}{3}\mbox{ }\mbox{ }\mbox{ }\mbox{ }\mbox{ }\mbox{ }$ & 
   $\mbox{ }\mbox{ }\mbox{ }\mbox{ }\mbox{ }\mbox{ }\frac{1}{6}\mbox{ }\mbox{ }\mbox{ }\mbox{ }\mbox{ }\mbox{ }$ &
   $\mbox{ }\mbox{ }\mbox{ }\mbox{ }\mbox{ }\mbox{ }0\mbox{ }\mbox{ }\mbox{ }\mbox{ }\mbox{ }\mbox{ }$ &
   $\mbox{ }\mbox{ }\mbox{ }\mbox{ }\mbox{ }\mbox{ }0\mbox{ }\mbox{ }\mbox{ }\mbox{ }\mbox{ }\mbox{ }$ \\
\hline
   $\mbox{ }\mbox{ }\mbox{ }\mbox{ }\tilde{x}^{\prime\prime (3)}_{T_{du},S_{ds},a}\mbox{ }\mbox{ }\mbox{ }\mbox{ }$ & 
   $\mbox{ }\mbox{ }\mbox{ }\mbox{ }\mbox{ }\mbox{ }0\mbox{ }\mbox{ }\mbox{ }\mbox{ }\mbox{ }\mbox{ }$ & 
   $\mbox{ }\mbox{ }\mbox{ }\mbox{ }\mbox{ }\mbox{ }\mbox{ }\mbox{ }\mbox{ }\mbox{ }\mbox{ }\mbox{ }\mbox{ }0-\tilde{y}^{\prime\prime (3)}_{T_{du},S_{ds}}\mbox{ }\mbox{ }\mbox{ }\mbox{ }\mbox{ }\mbox{ }$ &
   $\mbox{ }\mbox{ }\mbox{ }\mbox{ }\mbox{ }\mbox{ }0\mbox{ }\mbox{ }\mbox{ }\mbox{ }\mbox{ }\mbox{ }$ &
   $\mbox{ }\mbox{ }\mbox{ }\mbox{ }\mbox{ }\mbox{ }0\mbox{ }\mbox{ }\mbox{ }\mbox{ }\mbox{ }\mbox{ }$ \\
\hline
   $\mbox{ }\mbox{ }\mbox{ }\mbox{ }\tilde{x}^{\prime\prime (3)}_{T_{su},S_{ss},a}\mbox{ }\mbox{ }\mbox{ }\mbox{ }$ & 
   $\mbox{ }\mbox{ }\mbox{ }\mbox{ }\mbox{ }\mbox{ }\mbox{ }\mbox{ }\mbox{ }\mbox{ }\mbox{ }\mbox{ }\mbox{ }0-\tilde{y}^{\prime\prime (3)}_{T_{su},S_{ss}}\mbox{ }\mbox{ }\mbox{ }\mbox{ }\mbox{ }\mbox{ }$ & 
   $\mbox{ }\mbox{ }\mbox{ }\mbox{ }\mbox{ }\mbox{ }0\mbox{ }\mbox{ }\mbox{ }\mbox{ }\mbox{ }\mbox{ }$ &
   $\mbox{ }\mbox{ }\mbox{ }\mbox{ }\mbox{ }\mbox{ }0\mbox{ }\mbox{ }\mbox{ }\mbox{ }\mbox{ }\mbox{ }$ &
   $\mbox{ }\mbox{ }\mbox{ }\mbox{ }\mbox{ }\mbox{ }0\mbox{ }\mbox{ }\mbox{ }\mbox{ }\mbox{ }\mbox{ }$ \\
\hline\hline
   $\mbox{ }\mbox{ }\mbox{ }\mbox{ }\tilde{x}^{\prime\prime (2)}_{S_{du},S_{dd},a}\mbox{ }\mbox{ }\mbox{ }\mbox{ }$ & 
   $\mbox{ }\mbox{ }\mbox{ }\mbox{ }\mbox{ }\mbox{ }\mbox{ }\mbox{ }\mbox{ }\mbox{ }\mbox{ }\mbox{ }\mbox{ }0-\tilde{y}^{\prime\prime (2)}_{S_{du},S_{dd}}\mbox{ }\mbox{ }\mbox{ }\mbox{ }\mbox{ }\mbox{ }$ & 
   $\mbox{ }\mbox{ }\mbox{ }\mbox{ }\mbox{ }\mbox{ }0\mbox{ }\mbox{ }\mbox{ }\mbox{ }\mbox{ }\mbox{ }$ &
   $\mbox{ }\mbox{ }\mbox{ }\mbox{ }\mbox{ }\mbox{ }0\mbox{ }\mbox{ }\mbox{ }\mbox{ }\mbox{ }\mbox{ }$ &
   $\mbox{ }\mbox{ }\mbox{ }\mbox{ }\mbox{ }\mbox{ }0\mbox{ }\mbox{ }\mbox{ }\mbox{ }\mbox{ }\mbox{ }$ \\
\hline
   $\mbox{ }\mbox{ }\mbox{ }\mbox{ }\tilde{x}^{\prime\prime (3)}_{S_{du},S_{dd},a}\mbox{ }\mbox{ }\mbox{ }\mbox{ }$ & 
   $\mbox{ }\mbox{ }\mbox{ }\mbox{ }\mbox{ }\mbox{ }\mbox{ }\mbox{ }\mbox{ }\mbox{ }\mbox{ }\mbox{ }\mbox{ }0-\tilde{y}^{\prime\prime (3)}_{S_{du},S_{dd}}\mbox{ }\mbox{ }\mbox{ }\mbox{ }\mbox{ }\mbox{ }$ & 
   $\mbox{ }\mbox{ }\mbox{ }\mbox{ }\mbox{ }\mbox{ }0\mbox{ }\mbox{ }\mbox{ }\mbox{ }\mbox{ }\mbox{ }$ &
   $\mbox{ }\mbox{ }\mbox{ }\mbox{ }\mbox{ }\mbox{ }0\mbox{ }\mbox{ }\mbox{ }\mbox{ }\mbox{ }\mbox{ }$ &
   $\mbox{ }\mbox{ }\mbox{ }\mbox{ }\mbox{ }\mbox{ }0\mbox{ }\mbox{ }\mbox{ }\mbox{ }\mbox{ }\mbox{ }$ \\
\hline
   $\mbox{ }\mbox{ }\mbox{ }\mbox{ }\tilde{x}^{\prime\prime (3)}_{S_{su},S_{sd},a}\mbox{ }\mbox{ }\mbox{ }\mbox{ }$ & 
   $\mbox{ }\mbox{ }\mbox{ }\mbox{ }\mbox{ }\mbox{ }0\mbox{ }\mbox{ }\mbox{ }\mbox{ }\mbox{ }\mbox{ }$ & 
   $\mbox{ }\mbox{ }\mbox{ }\mbox{ }\mbox{ }\mbox{ }\frac{1}{6}\mbox{ }\mbox{ }\mbox{ }\mbox{ }\mbox{ }\mbox{ }$ &
   $\mbox{ }\mbox{ }\mbox{ }\mbox{ }\mbox{ }\mbox{ }0\mbox{ }\mbox{ }\mbox{ }\mbox{ }\mbox{ }\mbox{ }$ &
   $\mbox{ }\mbox{ }\mbox{ }\mbox{ }\mbox{ }\mbox{ }0\mbox{ }\mbox{ }\mbox{ }\mbox{ }\mbox{ }\mbox{ }$ \\
\hline
   $\mbox{ }\mbox{ }\mbox{ }\mbox{ }\tilde{x}^{\prime\prime (3)}_{S_{du},S_{ds},a}\mbox{ }\mbox{ }\mbox{ }\mbox{ }$ & 
   $\mbox{ }\mbox{ }\mbox{ }\mbox{ }\mbox{ }\mbox{ }\frac{1}{3}\mbox{ }\mbox{ }\mbox{ }\mbox{ }\mbox{ }\mbox{ }$ & 
   $\mbox{ }\mbox{ }\mbox{ }\mbox{ }\mbox{ }\mbox{ }\mbox{ }\mbox{ }\mbox{ }\mbox{ }\mbox{ }\mbox{ }\mbox{ }\frac{-1}{3}-\tilde{y}^{\prime\prime (3)}_{S_{du},S_{ds}}\mbox{ }\mbox{ }\mbox{ }\mbox{ }\mbox{ }\mbox{ }$ &
   $\mbox{ }\mbox{ }\mbox{ }\mbox{ }\mbox{ }\mbox{ }0\mbox{ }\mbox{ }\mbox{ }\mbox{ }\mbox{ }\mbox{ }$ &
   $\mbox{ }\mbox{ }\mbox{ }\mbox{ }\mbox{ }\mbox{ }0\mbox{ }\mbox{ }\mbox{ }\mbox{ }\mbox{ }\mbox{ }$ \\
\hline
   $\mbox{ }\mbox{ }\mbox{ }\mbox{ }\tilde{x}^{\prime\prime (3)}_{S_{su},S_{ss},a}\mbox{ }\mbox{ }\mbox{ }\mbox{ }$ & 
   $\mbox{ }\mbox{ }\mbox{ }\mbox{ }\mbox{ }\mbox{ }\mbox{ }\mbox{ }\mbox{ }\mbox{ }\mbox{ }\mbox{ }\mbox{ }0-\tilde{y}^{\prime\prime (3)}_{S_{su},S_{ss}}\mbox{ }\mbox{ }\mbox{ }\mbox{ }\mbox{ }\mbox{ }$ & 
   $\mbox{ }\mbox{ }\mbox{ }\mbox{ }\mbox{ }\mbox{ }0\mbox{ }\mbox{ }\mbox{ }\mbox{ }\mbox{ }\mbox{ }$ &
   $\mbox{ }\mbox{ }\mbox{ }\mbox{ }\mbox{ }\mbox{ }0\mbox{ }\mbox{ }\mbox{ }\mbox{ }\mbox{ }\mbox{ }$ &
   $\mbox{ }\mbox{ }\mbox{ }\mbox{ }\mbox{ }\mbox{ }0\mbox{ }\mbox{ }\mbox{ }\mbox{ }\mbox{ }\mbox{ }$ \\
\hline\hline
\end{tabular}
\caption{\label{tab:baryon_tilde_x_double_prime}Coefficients $\tilde{x}^{\prime\prime (N_{f})}_{H_{1},H_{2}}$ in Eq.~(\ref{eq:baryon_sunset_master}).  Due to the isospin symmetry, some of them cannot be distinguished from their $\tilde{y}^{\prime\prime (N_{f})}_{H_{1},H_{2}}$ counterparts.  For such cases, we present $\tilde{x}^{\prime\prime (N_{f})}_{H_{1},H_{2}}+\tilde{y}^{\prime\prime (N_{f})}_{H_{1},H_{2}}$ in the table.}
\end{table}

\clearpage

\bibliographystyle{apsrev.bst}
\bibliography{refs} 
 
\end{document}